\documentclass[%
 reprint,
 superscriptaddress,
nofootinbib,
 amsmath,amssymb,
 aps,
prstab,
]{revtex4-2}

\usepackage{tikz}
\usetikzlibrary{decorations.pathreplacing}
\usepackage{xcolor}
\usepackage{graphicx}
\usepackage{dcolumn}
\usepackage{bm}
\usepackage{hyperref}
\usepackage[]{algorithm2e}

\begin{document}

\preprint{APS/123-QED}

\title{Three Dimensional Theory of the Ion Channel Laser}

\author{Claire Hansel}
\email{chansel@slac.stanford.edu}
\affiliation{Center for Integrated Plasma Studies, Department of Physics, University of Colorado Boulder, Boulder, Colorado 80309, USA}
\author{Agostino Marinelli}
\affiliation{Department of Applied Physics, Stanford University, Stanford, California 94305, USA}
\affiliation{SLAC National Accelerator Laboratory, Menlo Park, California 94025, USA}
\author{Zhirong Huang}
\affiliation{Department of Applied Physics, Stanford University, Stanford, California 94305, USA}
\affiliation{SLAC National Accelerator Laboratory, Menlo Park, California 94025, USA}
\author{Michael Litos}
\affiliation{Center for Integrated Plasma Studies, Department of Physics, University of Colorado Boulder, Boulder, Colorado 80309, USA}
\date{September 13, 2026}

\begin{abstract}

The Ion Channel Laser (ICL) is a plasma-based alternative to the free-electron laser (FEL) that uses the electrostatic field of a uniform-density ion channel rather than the magnetic field of an undulator to induce transverse oscillations of relativistic electrons and thereby produce a coherent beam of short-wavelength radiation through a collective electromagnetic instability. The powerful focusing of the ion channel generally yields significantly higher gain parameters in the ICL than in a comparable FEL. This permits lasing over extremely short distances using electron bunches with an energy spread as large as a few percent, a value readily achievable with current plasma-based accelerators. ICLs, however, impose stringent transverse phase space requirements on the electron bunch beyond those required in FELs. In this work, we present a novel 3D theory of the planar off-axis configuration of the ICL that accounts for a number of effects, including diffraction, transverse radiation profile, betatron phase detuning, frequency detuning, and nonzero spread in energy and undulator parameter. We derive the ICL pendulum and field equations, which we use to write the 3D Maxwell-Klimontovich equations. After linearizing, we obtain an integro-differential equation describing the $z$-evolution of the radiation field. The 3D ICL dispersion relation is obtained using a van Kampen normal mode expansion. We numerically solve the $z$-evolution equation to compute radiation power growth rates and transverse radiation profiles over a range of ICL parameters. We also examine the reduction in gain due to 3D effects, energy spread, and emittance. Electron bunch phase space and emittance requirements for lasing are then derived for simple Gaussian distributions. Finally, we make general observations about the performance and feasibility of the ICL and discuss future prospects.
\end{abstract}

\maketitle

\section{Introduction \label{sec:introduction}}

Free-electron lasers (FELs)~\cite{kim2017, pellegrini2016, huang2007} are capable of producing ultrafast, high-brightness, coherent pulses of short-wavelength radiation, making them an invaluable tool across a diverse range of scientific disciplines~\cite{bostedt2016}. However, the significant cost and scale of UV and X-ray FEL facilities limit their practical utility. The creation of compact laser plasma accelerator- (LPA)~\cite{esarey2009} driven FELs has garnered significant research interest~\cite{nakajima2008,corde2013,albert2014,albert2016,albert2021,Emma2021b}, with substantial progress being made in recent years~\cite{wang2021,labat2023,barber2025}. LPA-driven FELs would represent enormous cost savings over conventional FELs, enabling them to be deployed at scale in academia and industry, potentially accelerating the pace of scientific progress across a wide range of fields. Even at wavelengths longer than X-rays, such devices could be used for impactful applications, including angle-resolved photoemission spectroscopy (ARPES) \cite{lv2019} for $\sim$100 nm light and semiconductor metrology \cite{holler2017} for $\sim$10 nm light.

Plasma-based electron injectors have been able to produce ultra-bright beams of order 100 nm-rad emittance, sub-percent energy spread, and kiloamp-scale peak current capable of achieving lasing in an FEL undulator~\cite{wang2021,barber2025}. In the case of Ref.~\cite{barber2025}, stable lasing was maintained over an 8-hour period with reliable day-to-day repeatability. Generation of even higher brightness bunches of order 10 nm-rad emittance via multiple plasma-based injection schemes has been simulated in detail~\cite{li2013,hu2016,manahan2017,habib2023,xiang2024,tomassini2025}. Despite this, the stringent beam requirements for lasing in an FEL, particularly the constraint on energy spread, have made developing LPA-driven short-wavelength FELs extremely challenging~\cite{nakajima2008,corde2013,albert2014,albert2016,albert2021}. A number of approaches to overcoming this limitation have been proposed, including beam loading~\cite{tzoufras2008}, decompression~\cite{maier2012,seggebrock2013}, transverse gradient~\cite{huang2012} and electromagnetic wave~\cite{xu2024} undulators, as well as non-FEL coherent radiation sources~\cite{emma2021, malaca2023}. In this work, we focus on a different approach: the Ion Channel Laser (ICL)~\cite{whittum1990a, whittum1990b, chen1990, chen1991, whittum1992, whittum1993, kupershmidt1994, esarey2002, kostyukov2003, bosch2005, liu2007, ersfeld2014, chen2016, shpakov2017, davoine2018, litos2018}, a plasma-based alternative to the FEL.

In an ICL, an electron bunch or laser pulse drives a blowout regime~\cite{rosenzweig1991} plasma wake in which the light plasma electrons are completely expelled from a region trailing the driver. Inside this region, the stationary heavy plasma ions form a uniform-density ion channel. A trailing electron bunch inside this ion channel experiences a linear transverse focusing force due to the electrostatic field of the ions, causing the electrons in the bunch to undergo betatron oscillations and emit betatron radiation~\cite{esarey2002,kostyukov2003,corde2013}. The interaction between the trailing bunch and its own betatron radiation causes its structure to change and the radiation to increase exponentially in power as it becomes increasingly coherent through a collective electromagnetic instability. While the ICL instability is broadly analogous to the FEL instability, there are a number of subtle differences. For example, the microbunching process in the ICL is more complicated than that in the FEL, responding to the sum of the ponderomotive and betatron phases rather than just the ponderomotive phase. 

In order for the ICL instability to grow, the longitudinal and transverse wakefields (barring the electrostatic field of the ion column) must be negligible or, preferably, zero. This condition can be met by operating in the so-called ``wakeless'' regime, wherein a narrow plasma column is pre-formed in a high ionization threshold gas (e.g., He) via laser-induced field ionization \cite{diederichs2019, litos2018}. With a sufficiently narrow plasma column and a strong (dense, high-current, high-energy) drive electron bunch, the plasma electrons can be expelled with enough force such that they do not return to the ion column and produce no appreciable wakefield for many picoseconds. This leaves a large window of time in which to place the trailing witness bunch in a bare ion column in the absence of plasma electron-induced wakefields. The high ionization threshold of the background gas ensures there is no undesired beam field induced ionization beyond the periphery of the pre-ionized narrow plasma channel that could spoil the wakeless ion column.

While in principle either a drive laser or drive electron bunch could be used to blow out electrons from the plasma and create the bare ion channel, diffraction would make it challenging for a laser driver to produce an ion channel with sufficient length to reach ICL saturation. An appropriate drive electron bunch could be produced from a conventional radio-frequency (RF) linear accelerator or from an LPA \cite{hidding2019,foerster2022,foerster2026}, and a trailing bunch could be created by a conventional RF injector and accelerator, or by a variety of different plasma injection schemes~\cite{wang2021,barber2025,li2013,hu2016,manahan2017,habib2023,xiang2024,tomassini2025}. The most compact scheme would utilize plasma-based injection and acceleration methods for both the drive and trailing bunches.

The trailing bunch must satisfy certain constraints on its emittance (or, more accurately, on its transverse phase space distribution) and its energy spread in order to meet the ICL lasing criteria. Compared to the FEL, the ICL has less stringent energy-spread requirements but more stringent emittance requirements to achieve lasing. These tradeoffs make the ICL well suited to the beams produced by plasma injectors, which struggle to achieve energy spreads as low as those of conventional RF accelerators, but can offer orders of magnitude improvement in transverse emittance while maintaining a small laboratory footprint. An additional potential advantage of the ICL over both conventional RF-driven and LPA-driven FELs is that the ICL radiator (i.e., plasma column) need only be meter-scale, whereas FEL magnetic undulators are many tens of meters in length.

Multiple configurations of the ICL are possible depending on the characteristics of the bunch's phase space distribution. The original literature on the ICL~\cite{whittum1990a, whittum1990b, chen1990, chen1991, whittum1992, whittum1993, kupershmidt1994, esarey2002, bosch2005, liu2007, shpakov2017} considered an ``on-axis'' configuration wherein the transverse centroid position of the trailing bunch is aligned to the center of the ion channel and each particle has a small transverse normalized momentum ($\gamma \bm{\beta}_{\perp} \ll 1$); this was shown by Ersfeld {\it et al.}~\cite{ersfeld2014} to require an impractically small emittance to lase. We focus on the more feasible ``planar off-axis'' configuration~\cite{ersfeld2014,davoine2018,litos2018}, wherein the transverse centroid position of the trailing bunch is initially offset from the ion channel axis about which it oscillates in a single plane, and where the bunch spot size is small compared to the maximum oscillation amplitude. While other ICL configurations are possible, such as those with circular or elliptical motion of the trailing bunch leading to correspondingly circular or elliptical radiation polarization, these are beyond the scope of this paper.

Despite a modest amount of prior literature, much of the fundamental physics of the ICL is poorly understood. This is in large part due to the challenging nature of simulating the ICL. The unique physics of the ICL precludes the use of conventional FEL codes such as \texttt{GENESIS}~\cite{reiche1999}. While particle-in-cell (PIC) codes are able to correctly model the physics, the computational expense of such simulations is generally prohibitive for the ICL. This is because the ICL is a multiscale problem in both the longitudinal (radiation wavelength vs. bunch length) and transverse (bunch spot size vs. bunch offset) dimensions. In addition, PIC simulations can be vulnerable to numerical space-charge instabilities without special solvers~\cite{xu2020} unless performed in the Lorentz-boosted frame~\cite{vay2007,fawley2009}, which introduces its own challenges related to the initialization of fields and particles as well as the process of ``unboosting'' the numerical diagnostics back to the lab frame for practical interpretation.

To date, only Davoine {\it et al.}~\cite{davoine2018} have managed to simulate the high-gain ICL in 3D. However, these simulations are of limited use as they simulated an unrealizable ICL configuration. In that simulation, circularly polarized radiation interacts with an on-axis cylindrical bunch where particles are initialized with no azimuthal momentum or spread in radial momentum. The existing literature also lacks a comprehensive analytic theory of the ICL. Only Davoine {\it et al.} consider 3D effects in their model of ICL dynamics, which represented a significant step forward at the time of publication that has, until now, stood as the most thorough description of ICL theory. However, their work is built on assumptions that limit the theory's accuracy. In particular, they assume an arbitrary form and width of the transverse radiation profile, allowing them to ignore the effect of the transverse radiation profile on electron motion. Further, they do not derive the ICL dispersion relation or radiation field $z$-evolution equation.

In our work, we take a theoretical approach to understanding the 3D physics of the planar off-axis ICL. In Section~\ref{sec:analytictheory}, we present a novel analytic theory of the ICL instability and derive equations describing lasing. We introduce numerical methods to solve these equations in Section~\ref{sec:numericalalgorithm}, which we use in Section~\ref{sec:numericalresults} to obtain radiation power growth rates and transverse radiation profiles. Next, in Section~\ref{sec:phasespace}, we derive the transverse phase space requirements for lasing. Finally, we discuss the most salient features of our results and their real-world implications, and draw general conclusions about the ICL in Section~\ref{sec:discussion}. Throughout our work, we use units of $\omega_p^{-1}$ for time, $c / \omega_p$ for distance, $m_e$ for mass, $m_e c^2$ for energy, and $e$ for charge, where $\omega_p = \left(e^2 n_0/m_e \epsilon_0\right)^{1/2}$ is the plasma frequency and $n_0$ is the plasma density.

\section{\label{sec:analytictheory} Analytic Theory}

\subsection{\label{sec:hamiltonian} Hamiltonian}

The Hamiltonian describing a relativistic bunch of $N_e$ electrons traveling down an ion channel is

\begin{equation} \label{eq:thamiltonian}
H = \sum_{j = 1}^{N_e} \sqrt{1 + (\bm{p}_{c,j} + \bm{a}(\bm{x}_j, t))^2} + \frac{1}{4} \bm{x}_{\perp,j}^2 - \phi(\bm{x}_j, t)
\end{equation}

\noindent where $j$ is an integer index labeling each electron, $\bm{p}_{c,j} = \bm{p}_j - \bm{a}(\bm{x}_j, t)$ is the canonical momentum, $\bm{a}(\bm{x},t)$ is the magnetic vector potential, $\phi(\bm{x}, t)$ is the electric scalar potential, excluding the $\bm{x}_{\perp,j}^2/4$ potential~\cite{kostyukov2003} of the ion channel, and $\perp$ refers to the transverse ($\hat{\bm{x}}$ and $\bm{\hat{y}}$) directions as opposed to the longitudinal ($\bm{\hat{z}}$) direction. We can transform Eq.~(\ref{eq:thamiltonian}) into a more convenient Hamiltonian $\mathcal{H} \equiv H - \sum_{j = 1}^{N_e} p_{c,z,j}$ that uses $z$ as the new time variable. Making the paraxial approximation $1 + \bm{p}_{\perp,j}^2 \ll \gamma_j^2$, we can write the new Hamiltonian as 

\begin{equation} \label{eq:zhamiltonian}
\begin{split}
\mathcal{H} = \sum_{j = 1}^{N_e} &\frac{1 + (\bm{p}_{c,\perp,j} + \bm{a}_{\perp}(\bm{x}_{\perp,j}, \zeta_j, z))^2}{2 \gamma_j}  \\
&+ \frac{1}{4} \bm{x}_{\perp,j}^2 - \psi(\bm{x}_{\perp,j}, \zeta_j, z)
\end{split}
\end{equation}

\noindent where $\gamma_j = \sqrt{1 + \bm{p}_j^2}$ is the Lorentz factor and $\psi \equiv \phi - a_z$ is the wake pseudopotential. The canonical position variables of this new Hamiltonian are $x_j$, $y_j$, and $\zeta_j \equiv z - t_j(z)$, and the canonical momentum variables are $p_{c,x,j}$, $p_{c,y,j}$, and $H_j$. Next, we make two key assumptions. First, we assume $\phi_j \ll \gamma_j$, which is equivalent to saying that the electrostatic potential energy due to the bunch charge is much smaller than the electron's total energy. This is an excellent approximation in practice; for a uniform-density cylindrical bunch with current $I$, the maximum normalized potential is $|I| / I_A$ times a geometric factor of at most a few, where $I_A$ is the Alfv\'en current, which in SI units is $I_A = e c / r_e \approx 17 \, \mathrm{kA}$, where $r_e$ is the classical electron radius. Second, we assume $\bm{x}_{\perp,j}^2 / 4 \ll \gamma_j$, implying that the potential energy due to the Coulomb force of the ion column is much smaller than the electron's total energy. As discussed in Section~\ref{sec:electronmotion}, this is equivalent to assuming the motion is paraxial and is true for virtually all ICLs. With these assumptions we can use $\gamma_j = H_j - \bm{x}_{\perp,j}^2/4 + \phi_j \simeq H_j$ instead of $H_j$ as the third canonical momentum variable in Eq.~(\ref{eq:zhamiltonian}).

\subsection{\label{sec:electronmotion} Electron Motion}

Next we consider the motion of the electrons in the ion channel in the absence of electromagnetic or bunch space-charge fields ($\bm{a}_{\perp} = \psi = 0$). The equations of motion can be derived from the Hamiltonian in Eq.~(\ref{eq:zhamiltonian}). For now we consider motion entirely in the $x$-$z$ plane. Solving for the transverse motion gives
\begin{equation} \label{eq:transversemotion}
\begin{split}
x_j(z) &= a_{\beta,j} \cos(k_{\beta,j} z + \varphi_j) \\
p_{x,j}(z) &= -K_j \sin(k_{\beta,j} z + \varphi_j)
\end{split}
\end{equation}

\noindent where $a_{\beta,j}$ is the betatron oscillation amplitude, $k_{\beta,j}=1/\sqrt{2\gamma_j}$ is the betatron angular wavenumber, $\varphi_j$ is the initial phase at $z=0$, and 
\begin{equation}
K_j = \gamma_j k_{\beta,j} a_{\beta,j}
\end{equation}

\noindent is the undulator parameter. In Section~\ref{sec:hamiltonian} we assumed $\bm{x}_{\perp,j}^2 / 4 \ll \gamma_j$. Here we see this is equivalent to the assumption $K_j^2 \ll \gamma_j^2$, which is equivalent to the paraxial assumption $p_{x,j}^2 \ll \gamma_j^2$. For ICLs with very low energies and large transverse offsets, this assumption can be violated, leading to relativistic effects causing the betatron oscillations to become anharmonic~\cite{kostyukov2003}. While we consider this extreme regime to be beyond the scope of this work, we note it has been investigated by Frazzitta {\it et al.}~\cite{frazzitta2024}. The two longitudinal equations can be solved to give

\begin{equation} \label{eq:longitudinalmotion}
\begin{split}
\zeta_j(z) &= \zeta_{0,j} - \frac{1 + \frac{K_j^2}{2}}{2 \gamma_j^2} z -\frac{K_j^2}{8 \gamma_j^2 k_{\beta,j}} \sin(2 \varphi_j) \\
&+ \frac{K_j^2}{8 \gamma_j^2 k_{\beta,j}} \sin(2(k_{\beta,j} z + \varphi_j)) \\
\gamma_j(z) &= \gamma_j
\end{split}
\end{equation}

\noindent where $\zeta_{0,j}$ and $\gamma_j$ are constants. Unlike in an undulator, where $\gamma_j(z)$ is truly constant since static magnetic fields do no work, in an ion channel $\gamma_j(z)$ is only approximately constant; the paraxial approximation allows us to ignore the $z$-dependent subleading term which is oscillatory with amplitude of order $K_j^2 / \gamma_j$. Each electron in an ICL emits $\bm{\hat{z}}$-directed betatron radiation at a fundamental wavelength $\lambda_{1,j}$ equal to the distance behind a particle moving at the speed of light the electron slips during one oscillation period. Mathematically, $-\lambda_{1,j} = \zeta_j(z + \lambda_{\beta,j}) - \zeta_j(z)$, where $\lambda_{\beta,j} = 2\pi/k_{\beta,j}$ is the betatron wavelength. Using Eq.~(\ref{eq:longitudinalmotion}), this yields the ICL resonance condition
\begin{equation} \label{eq:resonancecondition}
\lambda_{1, j} = \lambda_{\beta,j} \frac{1 + \frac{K_j^2}{2}}{2 \gamma_j^2}.
\end{equation}

\noindent For $K_j \gtrsim 1$, radiation is also emitted at odd harmonics $\lambda_{h,j} = \lambda_{1,j} / h$ where $h$ is an odd integer.

As discussed in the introduction, in this paper we consider only the planar off-axis ICL configuration. In this configuration, the fictitious ideal ``reference particle'' oscillates entirely in the $x$-$z$ plane with motion given by Eqs.~(\ref{eq:transversemotion}) and (\ref{eq:longitudinalmotion}). However, because we aim to include the effect of nonzero emittance in our theory, we must contend with the fact that, in general, particles in the bunch do not oscillate strictly in this plane. For general non-planar motion, the trajectories of electrons in the ion channel are elliptical helices. We call the ellipse traced out in the $x$-$y$ plane during this motion the orbit ellipse. These electrons have oscillation amplitudes in both dimensions ($a_{\beta,x,j}$ and $a_{\beta,y,j}$) as well as undulator parameters in both dimensions ($K_{x,j}$ and $K_{y,j}$). By computing $\zeta_j(z)$ it can be shown that the resonance condition Eq.~(\ref{eq:resonancecondition}) also holds in the non-planar case, provided the ``radial'' undulator parameter $K_j \equiv \sqrt{K_{x,j}^2 + K_{y,j}^2}$ is used. While planar transverse motion is parameterized by only two quantities, the undulator parameter $K_j$ and the oscillation phase $\varphi_j$, non-planar transverse motion is parameterized by four quantities: the undulator parameter $K_j$, the oscillation phase $\varphi_j$, the orientation angle of the orbit ellipse (angle between the $x$ axis and the semimajor axis), and the eccentricity angle of the orbit ellipse (arctangent of the aspect ratio). When a nonzero emittance bunch enters an ion channel, the spread in $(x_j, y_j, p_{x,j}, p_{y,j})$ can be expressed as a spread in these four parameters. Of these four parameters, only the undulator parameter $K_j$ appears in Eq.~(\ref{eq:resonancecondition}) for the fundamental wavelength $\lambda_{1,j}$. The spread in $\lambda_{1,j}$, which is determined by the spread in $K_j$ due to nonzero emittance along with the energy spread, must satisfy $\Delta \lambda_1 / \lambda_1 \lesssim \rho$ for lasing as discussed in Section~\ref{sec:phasespace}, where $\rho$ is the gain parameter, including any 3D or finite emittance or energy spread effects, defined in terms of the power gain length, i.e., the propagation length required to produce an $e$-fold increase in radiation power:

\begin{equation} \label{eq:gainlength}
L_G = \frac{\lambda_{\beta,r}}{4 \pi \sqrt{3} \rho}.
\end{equation}

As $\Delta \lambda_1 / \lambda_1$ approaches $\rho$, the ICL radiation power growth rate decreases until eventually the ICL instability is stabilized. This is the fundamental constraint from which the ICL emittance and energy spread requirements arise. While spread in orientation and eccentricity angle of the orbit ellipses in the bunch does not lead to spread in fundamental wavelength, it does still decrease the ICL gain. Electrons with very different orbit ellipses couple less efficiently to each other's emitted radiation and produce radiation with different polarizations, leading to some destructive interference. These effects directly decrease $\rho$. However, because we consider bunches with spot size much smaller than their transverse offset, the amount these effects decrease the ICL gain is negligible. Moving forward, we ignore these effects by approximating the motion of the bunch electrons as being entirely in the $x$-$z$ plane and described by Eqs.~(\ref{eq:transversemotion}) and (\ref{eq:longitudinalmotion}), with the caveat that $K_j$ is understood to be the {\it radial} undulator parameter rather than simply $K_{x,j}$. This distinction is important, for example, in Section~\ref{sec:phasespace} when computing $\Delta K$.

It is illuminating to make a comparison here to the FEL. In both the FEL and ICL, $\Delta \lambda_1 / \lambda_1 \lesssim \rho$ is the fundamental constraint from which the emittance and energy spread constraints arise \cite{bonifacio1992}. However, the way in which a bunch's emittance causes a spread in radiation wavelength $\Delta \lambda_1$ differs between the two. In the FEL, the undulator period-averaged electron motion is simply a constant velocity motion at some small angle $\theta_{\beta,j}$ with respect to the $\bm{\hat{z}}$ direction. Since the undulator radiation is emitted at this angle, the component of this radiation projected along the $\bm{\hat{z}}$-direction has a shifted frequency. The FEL resonance condition is $\lambda_{1,j} = (\lambda_u / (2 \gamma_j^2)) (1 + K_u^2 / 2 + \gamma_j^2 \theta_{\beta,j}^2)$ (SI) where $\lambda_u$ is the undulator period, $K_u = e \lambda_u B_0 / (2 \pi m_e c)$ (SI) is the undulator parameter, and $B_0$ is the undulator magnetic field strength. When a nonzero emittance bunch enters an FEL, the spread in $(x_j, y_j, p_{x,j}, p_{y,j})$ due to its emittance leads to a spread in the angle $\theta_{\beta,j}$ at which radiation is emitted, leading to a spread in fundamental wavelength. The undulator parameter $K_u$ is the same for each electron and makes no contribution to the spread in fundamental wavelength. In contrast, in the ICL the betatron period-averaged motion of every electron points exactly along the $\bm{\hat{z}}$ axis of the ion channel. There is no angular spread in the direction of the emitted radiation, and the spread in fundamental wavelengths $\Delta \lambda_1$ is entirely due to the spread in $K_j$.

\subsection{\label{sec:pendulumequations} Pendulum Equations}

To derive the pendulum equations we adopt the following approach broadly similar to existing FEL theory \cite{kim2017}. Previously we solved for the electron motion in the absence of the radiation fields, obtaining Eqs.~(\ref{eq:transversemotion}) and (\ref{eq:longitudinalmotion}) for the evolution of the four coordinates $(x_j, p_{x,j}, \zeta_j, \gamma_j)$ in terms of $z$ and four constants $(K_j, \varphi_j, \zeta_{0,j}, \gamma_j)$. In this subsection we turn those fields back on and let the motion be described by these same equations, except with the four constants now allowed to vary with $z$. Next we construct four ``pendulum quantities'' defined in terms of $(K_j, \varphi_j, \zeta_{0,j}, \gamma_j)$ and $z$ that vary slowly over a single betatron oscillation. After inverting Eqs.~(\ref{eq:transversemotion}) and (\ref{eq:longitudinalmotion}) and rewriting the pendulum quantities in terms of $(x_j, p_{x,j}, \zeta_j, \gamma_j)$ and $z$, we use the equations of motion derived from the Hamiltonian in Eq.~(\ref{eq:zhamiltonian}) to compute the (unaveraged) pendulum equations describing the $z$ evolution of the pendulum quantities. After simplifying these equations by ignoring terms we can show are small, we average these equations over the betatron oscillation period to obtain the period-averaged pendulum equations, which are the main result of this subsection.

The four pendulum quantities we choose are 
\begin{equation} \label{eq:pendulumquantities}
\begin{split}
\eta_j &\equiv \frac{\gamma_j - \gamma_r}{\gamma_r} \\
\delta_j &\equiv \frac{K_j - K_r}{K_r} \\
\theta_j &\equiv k_{\beta,r} z + k_{1,r} \Bigg(\zeta_{0,j}  - \frac{1 + \frac{K_j^2}{2}}{2 \gamma_j^2} z \\
&-\frac{K_j^2}{8 \gamma_j^2 k_{\beta,j}} \sin(2 \varphi_j)\Bigg) \\
\vartheta_j &\equiv \varphi_j + k_{\beta,j} z - k_{\beta,r} z
\end{split}
\end{equation}

\noindent where the $z$-dependence of $(\eta_j, \delta_j, \theta_j, \vartheta_j)$ and $(K_j, \varphi_j, \zeta_{0,j}, \gamma_j)$ is implicit and the subscript $r$ refers to the reference value of a quantity which does not depend on $z$ or the particle index $j$. The parameters $\eta_j$ and $\theta_j$ are the energy detuning and ponderomotive phase, which are defined in the same way as in FEL theory \cite{kim2017}. It is important to note that in the definition of $\theta_j$ the reference value---rather than the individual particle values---of $k_{\beta}$ and $k_1$ are used, since $\theta_j$ is not only a quantity defined for each particle, but also a field coordinate $\theta(z, \zeta) \equiv k_{\beta,r} z + k_{1,r} \zeta$ used in the field equation in Section~\ref{sec:fieldequation}. The parameters $\delta_j$ and $\vartheta_j$ are the undulator parameter detuning and the betatron phase detuning, respectively. These are unique to ICL theory and not found in FEL theory. In the off-axis configuration of the ICL, the bunch spot size is small compared to the centroid offset and so the spread in $K_j$ is small. Since we also assume small energy spread, $\eta_j, \delta_j \ll 1$. It is worth mentioning that the pendulum quantities in Eq.~(\ref{eq:pendulumquantities}) are chosen somewhat arbitrarily. We could have, for example, instead used $a_{\beta,j}$, $K_j^2$, or $\lambda_{1,j}$ detuning.

Having defined the pendulum quantities, we now derive the pendulum equations. By inverting Eqs.~(\ref{eq:transversemotion}) and (\ref{eq:longitudinalmotion}) to give $(K_j, \varphi_j, \zeta_{0,j}, \gamma_j)$ in terms of $(x_j, p_{x,j}, \zeta_j, \gamma_j)$ and substituting these into Eq.~(\ref{eq:pendulumquantities}), we can write the pendulum quantities in terms of the four coordinates $(x_j, p_{x,j}, \zeta_j, \gamma_j)$ and $z$

\begin{equation} \label{eq:pendulumquantitiescoordinates}
\begin{split}
\eta_j &= \frac{\gamma_j - \gamma_r}{\gamma_r} \\
\delta_j &= \frac{\sqrt{p_{x,j}^2 + \gamma_j^2 k_{\beta,j}^2 x_j^2} - K_r}{K_r} \\
\theta_j &= k_{\beta,r} z + k_{1,r} \zeta_j + \frac{k_{1,r} x_j p_{x,j}}{4 \gamma_j} \\
\vartheta_j &= \mathrm{atan2}\left(-p_{x,j}, \gamma_j k_{\beta,j} x_j\right) - k_{\beta,r} z.
\end{split}
\end{equation}

Next, we take the $z$ derivative of the pendulum parameters in Eq.~(\ref{eq:pendulumquantitiescoordinates}) and substitute first the equations of motion obtained from the Hamiltonian in Eq.~(\ref{eq:zhamiltonian}) and then Eqs.~(\ref{eq:transversemotion}) and (\ref{eq:longitudinalmotion}) for the four coordinates to obtain the unaveraged pendulum equations

\begin{widetext}
\begin{equation} \label{eq:fullunaveragedpendulum}
\begin{split}
\eta_j' &= \frac{K_j}{\gamma_r \gamma_j} \sin(k_{\beta,j} z + \varphi_j) \left . \frac{\partial a}{\partial \zeta} \right \vert_j + \frac{1}{\gamma_r} \left. \frac{\partial \psi}{\partial \zeta} \right \vert_j\\
\delta_j' &= \frac{1 + K_j^2}{ 2 \gamma_j^2 K_r} \sin(k_{\beta,j} z + \varphi_j) \left . \frac{\partial a}{\partial \zeta} \right \vert_j - \frac{1}{K_r} \sin(k_{\beta,j} z + \varphi_j) \left . \frac{\partial a}{\partial z} \right \vert_j + \frac{K_j}{2 \gamma_j K_r} \cos^2(k_{\beta,j} z + \varphi_j) \left. \frac{\partial \psi}{\partial \zeta} \right \vert_j\\
\theta_j' &= k_{\beta,r} - k_{1,r} \frac{1 + \frac{K_j^2}{2}}{2 \gamma_j^2} - \frac{k_{1,r} k_{\beta,j} K_j \cos(k_{\beta,j} z + \varphi_j)}{2 \gamma_j} \left(\frac{1 - K_j^2 \sin^2(k_{\beta,j} z + \varphi_j)}{2 \gamma_j^2} \left . \frac{\partial a}{\partial \zeta} \right \vert_j - \left . \frac{\partial a}{\partial z} \right \vert_j\right) \\
&+ \frac{k_{1,r} k_{\beta,j} K_j^2}{4 \gamma_j^2} \sin(2 (k_{\beta,j} z + \varphi_j))\left. \frac{\partial \psi}{\partial \zeta} \right \vert_j \\ 
\vartheta_j' &= k_{\beta,j} - k_{\beta,r} + \frac{1}{2 \gamma_j^2 K_j} \cos(k_{\beta,j} z + \varphi_j) \left . \frac{\partial a}{\partial \zeta} \right \vert_j - \frac{1}{K_j} \cos(k_{\beta,j} z + \varphi_j)  \left . \frac{\partial a}{\partial z} \right \vert_j - \frac{1}{4 \gamma_j} \sin(2 (k_{\beta,j} z + \varphi_j))\left. \frac{\partial \psi}{\partial \zeta} \right \vert_j 
\end{split}
\end{equation}
\end{widetext}

\noindent where $a$ refers to the $\hat{\bm{x}}$ component of the field $\bm{a}(\bm{x}_{\perp}, \zeta, z)$, and $|_j$ refers to evaluation at $\bm{x}_{\perp} = \bm{x}_{\perp,j}(z)$, $\zeta = \zeta_j(z)$. Additionally, in Eq.~(\ref{eq:fullunaveragedpendulum}) we neglected transverse space-charge fields (which are a factor of $\gamma$ smaller than the longitudinal space-charge fields) by ignoring the transverse derivatives of $\psi$.

Equation~(\ref{eq:fullunaveragedpendulum}) can be simplified considerably, and the four pendulum parameters can be shown to vary slowly over one betatron period by making order of magnitude arguments about the various terms. The orders of magnitude of the various quantities can be derived from the normalization used later on in Section~\ref{sec:maxwellklimontovich} and are $\partial_{\zeta} \sim h k_{1,r}$, $\partial_z \sim L_G^{-1}$, $\delta_j, \eta_j \lesssim \rho \ll 1$, $L_p \lesssim 40 L_G$, and $a \lesssim \rho^2 / (2 \xi_r [\mathrm{JJ}]_h)$ where $L_p$ is the plasma length, $h$ is the integer harmonic number, $[\mathrm{JJ}]_n \equiv J_{(n - 1) / 2}(h \xi_r) - J_{(n + 1) / 2}(h \xi_r)$ where $J$ is the Bessel function of the first kind, and $\xi_r \equiv K_r^2 / (2(2 + K_r^2))$. The second term in the $\delta_j'$ equation can be neglected, since it is subleading, and the terms in the $\theta_j'$ and $\vartheta_j'$ equations that depend on $a$ can be shown to change $\theta_j$ and $\varphi_j$, respectively, by an amount much less than unity over the length of the plasma. While making these order-of-magnitude arguments we assume the harmonic number is no larger than a few and $K \gtrsim 1$. As discussed further in Section~\ref{sec:maxwellklimontovich}, an ICL with $K \lesssim 1$ is impractical due to diffraction. 

Finally, while we have included the effect of longitudinal space-charge up to this point, we ignore it going forward by setting longitudinal derivatives of $\psi$ to zero. This is also done in Davoine {\it et al.} \cite{davoine2018}, and we refer to the justification used there, which says longitudinal space-charge can be ignored when $k_{p,b} L_{G,0} \lesssim 1$ where $k_{p,b} = \sqrt{n_b / (\gamma^3 n_0)}$ is the beam's plasma angular wavenumber and $L_{G,0}$ is the cold 1D gain length defined in Section~\ref{sec:maxwellklimontovich}. After writing $\gamma_j$ and $K_j$ in terms of their respective reference values and detunings, and ignoring subleading order terms in $\eta_j$ and $\delta_j$, we arrive at

\begin{equation} \label{eq:simplifiedunaveragedpendulum}
\begin{split}
\eta_j' &= \frac{K_r}{\gamma_r^2} \sin(k_{\beta,r} z + \vartheta_j) \left . \frac{\partial a}{\partial \zeta} \right \vert_j \\
\delta_j' &= \frac{1 + K_r^2}{2 K_r^2} \eta_j' \\
\theta_j' &= 2 k_{\beta,r} \left(\eta_j - \frac{K_r^2}{2 + K_r^2} \delta_j\right)\\
\vartheta_j' &= -\frac{1}{2} k_{\beta,r} \eta_j
\end{split}
\end{equation}

\noindent where we have also rewritten the argument of the sine function in the first equation using the definition of $\vartheta_j$ from Eq.~(\ref{eq:pendulumquantities}).

The next step is to period average Eq.~(\ref{eq:simplifiedunaveragedpendulum}) according to the definition of the period average in Appendix~\ref{app:periodaveraging}. The $\delta_j'$, $\theta_j'$, and $\vartheta_j'$ pendulum equations do not change when period-averaged since they contain only slowly varying quantities, but the $\eta_j'$ equation is nontrivial. We make the slowly varying envelope approximation as described in Appendix~\ref{app:slowlyvaryingenvelopeapproximation}. Using Eqs.~(\ref{eq:fieldfouriertransformreal}) and (\ref{eq:mathcaladefinition}), the period-averaged $\eta_j'$ equation is
\begin{equation} \label{eq:etaprimeunaveraged}
\begin{split}
\eta_j' &= \sum_{h \in \mathbb{N}^{+}} \int_{\nu \approx h}  d\nu \, \frac{i \nu k_{1,r} K_r}{\gamma_r^2} \overline{\mathcal{A}_h(\bm{x}_{\perp,j}(z), \nu, z) } \\
&\overline{\times \sin(k_{\beta,r} z + \vartheta_j) e^{i \nu k_{1,r} \zeta_j(z)} e^{i \Delta \nu k_{\beta,r} z}} + \mathrm{c.c.}.
\end{split}
\end{equation}

In the FEL, the transverse undulator oscillation amplitude $a_{u,j} \equiv K_u / (\gamma_j k_u)$ is always small compared to the transverse radiation size, and so particles oscillate locally in a small part of the radiation field. This means that the field can be treated as effectively constant over an oscillation period and when computing the period average of the $\eta_j'$ equation, $\mathcal{A}_h(\bm{x}_{\perp,j}(z), \nu, z)$ is slowly varying and can be taken out of the period average. In the ICL, however, $a_{\beta, j}$ is the same order as the transverse radiation size, which means electrons oscillate in and out of the core of the radiation field every oscillation. Even though $\mathcal{A}_h(\bm{x}_{\perp}, \nu, z)$ is slowly varying, $\mathcal{A}_h(\bm{x}_{\perp,j}(z), \nu, z)$ is not, and it cannot necessarily be taken out of the period average in Eq.~(\ref{eq:etaprimeunaveraged}). Despite this, it is still possible to compute the period average for arbitrary transverse mode shape by performing a Fourier transform of the field in the transverse plane, period averaging, and then performing an inverse Fourier transform. This is equivalent to using a delta function to write $\mathcal{A}_h(\bm{x}_{\perp,j}(z), \nu, z) = \int d^2\bm{x}_{\perp} \mathcal{A}_h(\bm{x}_{\perp}, \nu, z) \delta^2(\bm{x}_{\perp} - \bm{x}_{\perp,j}(z))$ and then taking the slowly varying $\mathcal{A}_h(\bm{x}_{\perp}, \nu, z)$ out of the period average.

Using this procedure along with the result Eq.~(\ref{eq:fancyperiodaverage}) from Appendix~\ref{app:periodaveraging}, we can period average Eq.~(\ref{eq:etaprimeunaveraged}). Combining this with the other three period-averaged pendulum equations, which are the same as their unaveraged forms from Eq.~(\ref{eq:simplifiedunaveragedpendulum}), we can write down the period-averaged pendulum equations 
\begin{equation} \label{eq:pendulumequationsperiodaveraged}
\begin{split}
\eta_j' &= \sum_{h \in \mathbb{N}^{+}} \int_{\nu \approx h} d\nu \, \frac{\nu k_{1,r} K_r [\mathrm{JJ}]_h}{2 \gamma_r^2} e^{i h \vartheta_j} e^{i \nu \theta_j} \\
&\times \int d^2\bm{x}_{\perp} \mathcal{W}_h(\bm{x}_{\perp}) \mathcal{A}_h(\bm{x}_{\perp}, \nu, z) + \mathrm{c.c.} \\
\delta_j' &= \frac{1 + K_r^2}{2 K_r^2} \eta_j' \\
\theta_j' &= 2 k_{\beta,r} \left(\eta_j - \frac{K_r^2}{2 + K_r^2} \delta_j\right)\\
\vartheta_j' &= -\frac{1}{2} k_{\beta,r} \eta_j
\end{split}
\end{equation}

\noindent where $\mathcal{W}_h(\bm{x}_{\perp})$ is defined in Eq.~(\ref{eq:wdefinition}) and $[\mathrm{JJ}]_h$ is defined in Eq.~(\ref{eq:jjdefinition}).

\subsection{\label{sec:fieldequation} Field Equation}

In this section, we derive the period-averaged field equation for $\mathcal{A}_h(\bm{x}_{\perp}, \nu, z)$. The field equation in Lorentz gauge for the $\hat{\bm{x}}$ component of the magnetic vector potential with a source term due to the $N_e$ electrons in the bunch is
\begin{equation}
\begin{split}
&\left[\frac{\partial^2}{\partial t^2} - \bm{\nabla}^2\right] a(\bm{x}, t) = \\
& -\frac{4 \pi}{I_A}
\sum_{j = 1}^{N_e} \frac{d x_j(t)}{dt} \delta^3(\bm{x} - \bm{x}_j(t)).
\end{split}
\end{equation}

\noindent We transform coordinates from $(z,t)$ to $(\zeta,z)$ and make the slowly varying envelope approximation ($\partial_z \ll \partial_{\zeta}$) as discussed in Appendix~\ref{app:slowlyvaryingenvelopeapproximation} to obtain
\begin{equation}
\begin{split}
&\left[\frac{\partial^2}{\partial z \partial \zeta} + \frac{1}{2} \bm{\nabla}_{\perp}^2\right] a(\bm{x}_{\perp}, \zeta, z) = \\
&\frac{2 \pi}{I_A} \sum_{j = 1}^{N_e} x_j'(z) \delta^2(\bm{x}_{\perp} - \bm{x}_{\perp,j}(z)) \delta(\zeta - \zeta_j(z)).
\end{split}
\end{equation}

\noindent Next, we Fourier transform using Eq.~(\ref{eq:fieldfouriertransform}) and use the definition of $\mathcal{A}_h(\bm{x}_{\perp}, \nu, z)$ from Eq.~(\ref{eq:mathcaladefinition}) to get
\begin{equation}
\begin{split}
&\left[\frac{\partial}{\partial z} + i \Delta \nu k_{\beta,r} - \frac{i}{2 \nu k_{1,r}} \bm{\nabla}_{\perp}^2\right] \mathcal{A}_h(\bm{x}_{\perp}, \nu, z) = \\
&\frac{e^{-i \Delta \nu k_{\beta,r} z}}{i \nu I_A} \sum_{j = 1}^{N_e} x_j'(z) \delta^2(\bm{x}_{\perp} - \bm{x}_{\perp,j}(z)) e^{-i \nu k_{1,r} \zeta_j}.
\end{split}
\end{equation}

\noindent The left-hand side of the above equation is unchanged by period averaging. The period average of the source term of the above equation can be computed by substituting in $x_j'(z)$ from Eq.~(\ref{eq:transversemotion}) and then using Eq.~(\ref{eq:fancyperiodaverage}), yielding 
\begin{equation} \label{eq:periodaveragedwaveequation}
\begin{split}
&\left[\frac{\partial}{\partial z} + i \Delta \nu k_{\beta,r} - \frac{i}{2 \nu k_{1,r}} \bm{\nabla}_{\perp}^2\right] \mathcal{A}_h(\bm{x}_{\perp}, \nu, z) = \\
&-\frac{K_r [\mathrm{JJ}]_{h,r}}{2 \nu I_A \gamma_r} \mathcal{W}_h(\bm{x}_{\perp}) \sum_{j = 1}^{N_e} e^{-i \nu \theta_j} e^{-i h \vartheta_j},
\end{split}
\end{equation}

\noindent which is the period-averaged field equation and the main result of this section.

The source term on the right-hand side of Eq.~(\ref{eq:periodaveragedwaveequation}) contains a bunching factor $\sum_{j=1}^{N_e} e^{-i (\nu \theta_j + h \vartheta_j)}$ which includes the ponderomotive phase and betatron phase detuning. This is different from the analogous FEL equation \cite[(3.68)]{kim2017} which has a bunching factor $\sum_{j=1}^{N_e} e^{-i \nu \theta_j}$ that includes only the ponderomotive phase. In both cases the radiation power is proportional to the absolute value of this factor squared. If the arguments of the complex exponential are randomly distributed, which is the case for a bunch with no microscopic structure, many of the terms in the sum will cancel each other out causing the power to scale as $P \propto N_e$. What is physically happening in this case is that the radiation emitted by the particles have random phases and destructively interfere, or in other words the bunch is radiating incoherently. If the bunch instead has microscopic structure such that the complex exponential arguments are all equal (modulo $2\pi$), the terms in the sum add, causing the power to scale as $P \propto N_e^2$. In this case, the radiation is emitted with the same phases and adds constructively, or in other words the bunch is radiating coherently.

The difference between $e^{-i (\nu \theta_j + h \vartheta_j)}$ in the ICL equation and the $e^{-i \nu \theta_j}$ in the FEL equation is the mathematical manifestation of a subtle but crucial difference between the ICL and FEL. A bunch in an FEL radiates coherently when $\nu \theta_j \;\mathrm{mod}\;2\pi$ are all equal, which occurs only when the bunch is microbunched at the radiation wavelength $\lambda$. A bunch in an ICL, however, radiates coherently when $\nu \theta_j + h \vartheta_j \;\mathrm{mod}\;2\pi$ are all equal, which can occur in different ways. The bunch could have $\nu \theta_j\;\mathrm{mod}\;2\pi$ the same for each particle and $h\vartheta_j\;\mathrm{mod}\;2\pi$ the same for each particle, in which case it would be microbunched at $\lambda_1$. The bunch could also have a uniform random distribution of both $\theta_j$ and $\vartheta_j$ but with a correlation between the two such that $\nu \theta_j + h \vartheta_j \;\mathrm{mod}\;2\pi$ are equal for each particle. This would be a bunch with no microbunching ``microsnaked'' in space at wavelength $\lambda$. Bunches with a combination of microbunching and microsnaking could also radiate coherently. While we note the complexity of the microscopic physics of the ICL, a full investigation of it is beyond the scope of this paper.

\subsection{\label{sec:maxwellklimontovich} Maxwell-Klimontovich Equations}

\begin{figure}[b]
    \centering
    \includegraphics[width=\columnwidth]{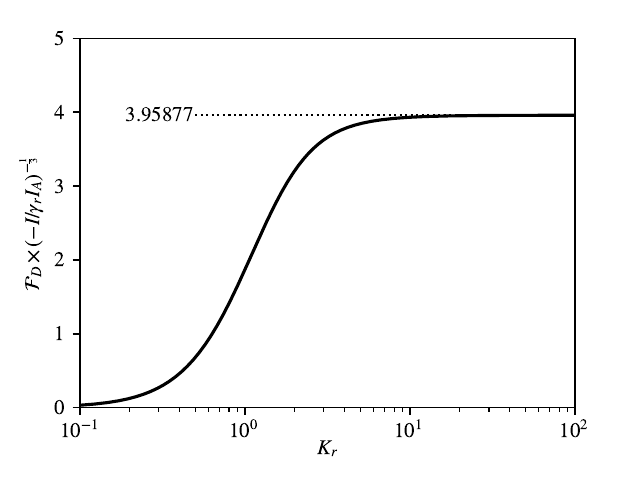}
    \caption{Dependence of the Fresnel parameter $\mathcal{F}_D$ defined in Eq.~(\ref{eq:fresnel}) on the undulator parameter $K_r$, assuming $|\Delta \nu| \ll 1$. $\mathcal{F}_D$ quantifies the level of diffraction in an ICL with smaller values corresponding to a greater degree of diffraction. Note that $\mathcal{F}_D \times (-I / \gamma_r I_A)^{-1/3}$ is plotted instead of $\mathcal{F}_D$ directly to pull out the dependence on $I$ and $\gamma_r$.
    }
    \label{fig:fresnel}
\end{figure}

The period-averaged pendulum equations Eq.~(\ref{eq:pendulumequationsperiodaveraged}) and field equation Eq.~(\ref{eq:periodaveragedwaveequation}) form a system of $4N_e + 1$ equations that describe the physics of the ICL instability. As in FEL theory \cite{kim2017}, we can reduce the number of equations to only two by introducing a distribution function and writing down the Maxwell-Klimontovich equations. For the ICL, the distribution function is
\begin{equation} \label{eq:distributionfunction}
\begin{split}
f(\eta, \delta, \vartheta, \theta, z) &= \frac{2 \pi}{N_{\lambda_{1,r}}} \sum_{j = 1}^{N_e} \delta(\eta - \eta_j(z)) \delta(\delta - \delta_j(z)) \\
&\times \delta(\vartheta - \vartheta_j(z)) \delta(\theta - \theta_j(z))    
\end{split}
\end{equation}

\noindent where $N_{\lambda_{1,r}} = |I| \lambda_{1,r}$ is the number of particles in one longitudinal slice of length $\lambda_{1,r}$. We have chosen this normalization arbitrarily. The right-hand side of the field Eq.~(\ref{eq:periodaveragedwaveequation}) can be rewritten as an integral over $f$ and the Klimontovich equation can be computed from $df/dz = 0$ using the pendulum equations Eq.~(\ref{eq:pendulumequationsperiodaveraged}). The resulting system of equations is
\begin{widetext}
\begin{equation} \label{eq:mkharmonic}
\begin{split}
&\left[\frac{\partial}{\partial z} + i \Delta \nu k_{\beta,r} - \frac{i}{2 \nu k_{1,r}} \bm{\nabla}_{\perp}^2\right] \mathcal{A}_h(\bm{x}_{\perp}, \nu, z) = -\frac{K_r [\mathrm{JJ}]_h N_{\lambda_{1,r}}}{4 \pi \nu I_A \gamma_r} \mathcal{W}_h(\bm{x}_{\perp}) \iiiint d\eta \,d\delta  \,d\vartheta \,d\theta  \, e^{-i \nu \theta} e^{-i h \vartheta} f(\eta, \delta, \vartheta, \theta, z) \\
&\Bigg[\frac{\partial}{\partial z} + \left[\sum_{h \in \mathbb{N}^{+}} \int_{\nu \approx h} d\nu \, \frac{\nu k_{1,r} K_r [\mathrm{JJ}]_h}{2 \gamma_r^2} e^{i h \vartheta} e^{i \nu \theta} \int d^2\bm{x}_{\perp} \mathcal{W}_h(\bm{x}_{\perp}) \mathcal{A}_h(\bm{x}_{\perp}, \nu, z) + \mathrm{c.c.} \right] \left(\frac{\partial}{\partial \eta} + \frac{1 + K_r^2}{2 K_r^2} \frac{\partial}{\partial \delta}\right) \\
&+ 2 k_{\beta,r} \left(\eta - \frac{K_r^2}{2 + K_r^2} \delta\right) \frac{\partial}{\partial \theta} - \frac{1}{2} k_{\beta,r} \eta \frac{\partial}{\partial \vartheta}\Bigg] f(\eta, \delta, \vartheta, \theta, z) = 0. \\
\end{split}
\end{equation}

\noindent If not explicitly specified, the bounds of $\theta$ integrals are from $0$ to $k_{1,r} L_b$ where $L_b$ is the bunch length, and all other integrals are from $-\infty$ to $\infty$. In Appendix~\ref{app:energy}, we verify that Eq.~(\ref{eq:mkharmonic}) obeys conservation of energy. Introducing normalized quantities, Eq.~(\ref{eq:mkharmonic}) becomes

\begin{equation} \label{eq:mk}
\begin{split}
&\left[\frac{\partial}{\partial \hat{z}} + i \hat{\Delta \nu} - i \mathcal{F}_D^{-1} \hat{\bm{\nabla}}_{\perp}^2\right] \hat{\mathcal{A}}_h(\hat{\bm{x}}_{\perp}, \nu, \hat{z}) = -\frac{1}{2 \nu \mathcal{I}} \frac{[\mathrm{JJ}]_h^2}{[\mathrm{JJ}]_1^2} \hat{\mathcal{W}}_h(\hat{\bm{x}}_{\perp})  \iiiint d\hat{\eta} \,d\hat{\delta} \,d\vartheta \,d\theta \, e^{-i \nu \theta} e^{-i h \vartheta} \hat{f}(\hat{\eta}, \hat{\delta}, \vartheta, \theta, \hat{z}) \\
&\left[\frac{\partial}{\partial \hat{z}} + \frac{d\hat{\eta}}{d\hat{z}} \left(\frac{\partial}{\partial \hat{\eta}} + \frac{1 + K_r^2}{2 K_r^2} \frac{\partial}{\partial \hat{\delta}}\right) + \left(\hat{\eta} - \frac{K_r^2}{2 + K_r^2} \hat{\delta}\right) \frac{\partial}{\partial \theta} - \frac{\hat{\eta}}{4} \frac{\partial}{\partial \vartheta}\right] \hat{f}(\hat{\eta}, \hat{\delta}, \vartheta, \theta, \hat{z}) = 0 \\
\end{split}
\end{equation}

\noindent where 

\begin{equation}
\frac{d\hat{\eta}}{d\hat{z}} = \sum_{h \in \mathbb{N}^{+}} \int_{\nu \approx h} d\nu \, \nu  e^{i h \vartheta}  e^{i \nu \theta}  \int d^2\hat{\bm{x}}_{\perp} \hat{\mathcal{W}}_h(\hat{\bm{x}}_{\perp})\hat{\mathcal{A}}_h(\hat{\bm{x}}_{\perp}, \nu, \hat{z}) + \mathrm{c.c.}
\end{equation}

\end{widetext}

\noindent and where we have switched to normalized coordinates $\hat{z} = 2 k_{\beta,r} \rho_0 z$, $\hat{\bm{x}}_{\perp} = \bm{x}_{\perp} / a_{\beta,r}$, $\hat{\Delta \nu} = \Delta \nu / (2 \rho_0)$, $\hat{\eta} = \eta / \rho_0$, $\hat{\delta} = \delta / \rho_0$, $\hat{\mathcal{W}} = \mathcal{W} a_{\beta,r}^2$, $\hat{\mathcal{A}}_h = 2 \xi_r [\mathrm{JJ}]_h / (K_r \rho_0^2) \mathcal{A}_h$, $\hat{f} = f \rho_0^2$, and where we have defined the Fresnel parameter
\begin{equation} \label{eq:fresnel}
\mathcal{F}_D \equiv 4 \nu k_{1,r} k_{\beta,r} \rho_0 a_{\beta,r}^2 = 32 \nu \xi_r \rho_0,
\end{equation}

\noindent the cold 1D gain parameter 
\begin{equation} \label{eq:rho1d}
\rho_0 = \left(\frac{|I|}{I_A} \frac{\mathcal{I} [\mathrm{JJ}]_1^2}{8 \gamma_r}\right)^{\frac{1}{3}},
\end{equation}

\noindent and the ``ICL factor''
\begin{equation} \label{eq:iclparameter}
\mathcal{I} \equiv \frac{4 + K_r^2}{4 (2 + K_r^2)}.
\end{equation}

\noindent The interpretation of these three quantities is discussed in the following three paragraphs.

The Fresnel parameter Eq.~(\ref{eq:fresnel}) is defined analogously to that of Pellegrini {\it et al.} \cite{pellegrini2016} and quantifies the effect of diffraction in the ICL. It can be written as $\mathcal{F}_D = (4 / \sqrt{3}) z_R / L_{G, 0} \approx 2.3 z_R / L_{G, 0}$ where $L_{G,0} = \lambda_{\beta, r} / (4 \pi \sqrt{3} \rho_0)$ is the cold 1D gain length, and $z_R = (1/2) \nu k_{1,r} a_{\beta,r}^2$ is the Rayleigh length of a Gaussian mode of waist $a_{\beta,r}$, which is an approximation of the actual radiation waist. The impact of diffraction is greatest when $\mathcal{F}_D$ is smallest. A plot of the dependence of $\mathcal{F}_D$ on $K_r$ is shown in Fig.~\ref{fig:fresnel}. Diffraction is much more significant at lower $K_r$ and plateaus at $K_r \gtrsim 2$. There is little reason to operate an ICL at $K \lesssim 2$. In the on-axis configuration of the ICL~\cite{whittum1990a}, the particles' normalized transverse momenta are assumed to be much less than one, i.e., $K_j \ll 1$. In this case, $K_j$ drops out of Eq.~(\ref{eq:resonancecondition}) for $\lambda_{1,j}$ and so the spread in $K_j$ due to nonzero emittance cannot induce a spread in $\lambda_{1,j}$ that damps the instability. While this circumvents the stringent emittance requirements of the off-axis ICL, we can see from Fig.~\ref{fig:fresnel} that it introduces a new problem as the degree of diffraction at $K_j \ll 1$ becomes extremely large, which is what makes the on-axis configuration infeasible.

The cold 1D gain parameter is defined in Eq.~(\ref{eq:rho1d}). While this definition was made arbitrarily, we show later on in Section~\ref{sec:dispersion} that in the 1D limit with no detuning or spread in energy or undulator parameter and for $h = 1$, $\rho$ as defined in terms of the power gain length in Eq.~(\ref{eq:gainlength}) reduces to $\rho_0$ as defined in Eq.~(\ref{eq:rho1d}). Unlike in the FEL, $\rho_0$ has no dependence on lasing wavelength or bunch size and only a very weak dependence on $K_r$ (although $\rho$, which, unlike $\rho_0$, includes effects of diffraction, becomes very small for $K_r \lesssim 2$ due to the corresponding increase in diffraction). There is also no dependence of $\rho_0$ on plasma density. Very high values of $\rho_0$ can be achieved in the ICL compared with the FEL. If we consider, for example, the parameters of the LPA-driven XUV FEL demonstrated in Wang {\it et al.} \cite{wang2021} ($K = 1.41$, $\gamma m_e c^2 = 490 \; \mathrm{MeV}$, $I = 5.7 \; \mathrm{kA}$, $\sigma_x = 50 \; \mathrm{\mu m}$, $\sigma_y = 70 \; \mathrm{\mu m}$, $\lambda_u = 25 \; \mathrm{mm}$), $\rho_{0, \mathrm{ICL}} = 0.023$ while $\rho_{0, \mathrm{FEL}} = 0.0054$. For slightly longer wavelengths where $\gamma$ is typically low, it could be possible to reach regimes where $\rho$ becomes close enough to unity that the physics of the instability begins to deviate from our theoretical results, which assume $\rho \ll 1$, although this regime is beyond the scope of this paper.

The ICL factor Eq.~(\ref{eq:iclparameter}) quantifies some of the subtle differences between the physics of the FEL and the ICL in 1D. If we start with the equation for the 1D FEL gain parameter \cite[(3.87)]{kim2017}, take $\pi a_{\beta,r}^2$ to be the transverse area, and use the equation $K_r = \gamma_r k_{\beta,r} a_{\beta,r}$ and the resonance condition Eq.~(\ref{eq:resonancecondition}), we can derive an equation identical to Eq.~(\ref{eq:rho1d}) save for the factor of $\mathcal{I}^{1/3}$. This extra factor arises due to the $z$ evolution of $K_j$ and $\vartheta_j$ and the dependence of $k_{\beta}$ on energy. It is accounted for when properly deriving the ICL equations, but not when making simple arguments ignoring the differences between FELs and ICLs. $\mathcal{I}$ is of order unity for all values of $K$, with $1/4 \leq \mathcal{I} \leq 1/2$.

The Maxwell-Klimontovich equations in Eq.~(\ref{eq:mk}) describe many aspects of the physics of the ICL, including the startup process, exponential gain, and saturation. Moving forward, we split the distribution function into a $\theta$-independent background plus a perturbation: $\hat{f}(\hat{\eta}, \hat{\delta}, \vartheta, \theta, \hat{z}) = \hat{f}_0(\hat{\eta}, \hat{\delta}, \vartheta, \hat{z}) + \hat{f}_1(\hat{\eta}, \hat{\delta}, \vartheta, \theta, \hat{z})$ where the bunch has no microscopic structure at $\hat{z}=0$, i.e., $\hat{f}_1(\hat{\eta}, \hat{\delta}, \vartheta, \theta, 0)=0$. Linearizing the Maxwell-Klimontovich equations in this way still allows them to describe the physics of the exponential gain, though they no longer conserve energy or accurately capture the saturation physics.

From Eq.~(\ref{eq:distributionfunction}) and the definitions for the normalized units, $\int d\hat{\eta} \,d\hat{\delta} \,d\vartheta \,d\theta \hat{f}(\hat{\eta}, \hat{\delta}, \vartheta, \theta, \hat{z}) = 2 \pi N_e / N_{\lambda_{1,r}}$. Since the $\theta$ integration is performed over an interval $k_1 L_b = 2 \pi N_e / N_{\lambda_{1,r}}$, the normalization condition for $\hat{f}_0$ is
\begin{equation} \label{eq:f0normalization}
\iiint d\hat{\eta} d\hat{\delta} d\vartheta \hat{f}_0(\hat{\eta}, \hat{\delta}, \vartheta, \hat{z}) = 1.
\end{equation}

The linearized Klimontovich equation is  
\begin{widetext} 
\begin{equation} \label{eq:mklinearized}
\left[\frac{\partial}{\partial \hat{z}}  + \left(\hat{\eta} - \frac{K_r^2}{2 + K_r^2} \hat{\delta}\right) \frac{\partial}{\partial \theta} - \frac{\hat{\eta}}{4} \frac{\partial}{\partial \vartheta}\right] \hat{f}_1(\hat{\eta}, \hat{\delta}, \vartheta, \theta, \hat{z}) = -\left[\frac{\partial}{\partial \hat{z}} + \hat{\eta}' \left(\frac{\partial}{\partial \hat{\eta}} + \frac{1 + K_r^2}{2 K_r^2} \frac{\partial}{\partial \hat{\delta}}\right) - \frac{\hat{\eta}}{4} \frac{\partial}{\partial \vartheta}\right] \hat{f}_0(\hat{\eta}, \hat{\delta}, \vartheta, \hat{z})
\end{equation}

\noindent where the higher order $\hat{\mathcal{A}} \hat{f}_1$ term is neglected. We introduce the transformed distribution function
\begin{equation}
\begin{split}
F(\hat{\eta}, \hat{\delta}, \vartheta, \nu, \hat{z}) &= e^{-i h \vartheta} \int d\theta \, e^{-i \nu \theta} \hat{f}(\hat{\eta}, \hat{\delta}, \vartheta, \theta, \hat{z}) \\
\hat{f}_1(\hat{\eta}, \hat{\delta}, \vartheta, \theta, \hat{z}) &= \frac{1}{2 \pi} \int d \nu \, e^{i \nu \theta} e^{i h \vartheta} F(\hat{\eta}, \hat{\delta}, \vartheta, \nu, \hat{z}) - \hat{f}_0(\hat{\eta}, \hat{\delta}, \vartheta, \hat{z})
\end{split}
\end{equation}

\noindent which allows Eq.~(\ref{eq:mklinearized}) to be rewritten as 
\begin{equation} \label{eq:mklinearized2}
\begin{split}
&\left[\frac{\partial}{\partial \hat{z}} + i \hat{\Delta \nu} - i \mathcal{F}_D^{-1} \hat{\bm{\nabla}}_{\perp}^2\right] \hat{\mathcal{A}}_h(\hat{\bm{x}}_{\perp}, \nu, \hat{z}) = -\frac{1}{2 \nu \mathcal{I}} \frac{[\mathrm{JJ}]_h^2}{[\mathrm{JJ}]_1^2}\hat{\mathcal{W}}_h(\hat{\bm{x}}_{\perp})  \iiint d\hat{\eta} \,d\hat{\delta} \,d\vartheta F(\hat{\eta}, \hat{\delta}, \vartheta, \nu, \hat{z})\\
&\left[\frac{\partial}{\partial \hat{z}} + i \left(\left(\nu - \frac{h}{4}\right)\hat{\eta} - \nu \frac{K_r^2}{2 + K_r^2} \hat{\delta}\right) - \frac{\hat{\eta}}{4} \frac{\partial}{\partial \vartheta}\right] F(\hat{\eta}, \hat{\delta}, \vartheta, \nu, \hat{z}) = \\
&- 2 \pi \nu \left(\frac{\partial \hat{f}_0}{\partial \hat{\eta}} + \frac{1 + K_r^2}{2 K_r^2} \frac{\partial \hat{f}_0}{\partial \hat{\delta}}\right)  \int d^2\hat{\bm{x}}_{\perp} \hat{\mathcal{W}}_h(\hat{\bm{x}}_{\perp})\hat{\mathcal{A}}_h(\hat{\bm{x}}_{\perp}, \nu, \hat{z}).\\
\end{split}
\end{equation}

\noindent This can be further simplified by integrating the Klimontovich equation in Eq.~(\ref{eq:mklinearized2}) with respect to $\vartheta$ and using the fact that $F(\hat{\eta}, \hat{\delta}, \vartheta, \nu, \hat{z})$ vanishes at $\vartheta \rightarrow \pm \infty$ to get 
\begin{equation} \label{eq:simplifiedmv}
\begin{split}
&\left[\frac{\partial}{\partial \hat{z}} + i \hat{\Delta \nu} - i \mathcal{F}_D^{-1} \hat{\bm{\nabla}}_{\perp}^2\right] \hat{\mathcal{A}}_h(\hat{\bm{x}}_{\perp}, \nu, \hat{z}) = -\frac{1}{2 \nu \mathcal{I}} \frac{[\mathrm{JJ}]_h^2}{[\mathrm{JJ}]_1^2} \hat{\mathcal{W}}_h(\hat{\bm{x}}_{\perp})  \iint d \hat{\eta} \,d\hat{\delta} \, \mathcal{F}(\hat{\eta}, \hat{\delta}, \nu, \hat{z})\\
&\left[\frac{\partial}{\partial \hat{z}} + i \left(\left(\nu - \frac{h}{4}\right)\hat{\eta} - \nu \frac{K_r^2}{2 + K_r^2} \hat{\delta}\right)\right] \mathcal{F}(\hat{\eta}, \hat{\delta}, \nu, \hat{z}) = \\
&- 2 \pi \nu \left(\int d\vartheta \left(\frac{\partial \hat{f}_0}{\partial \hat{\eta}} + \frac{1 + K_r^2}{2 K_r^2} \frac{\partial \hat{f}_0}{\partial \hat{\delta}}\right)\right)  \int d^2\hat{\bm{x}}_{\perp} \hat{\mathcal{W}}_h(\hat{\bm{x}}_{\perp})\hat{\mathcal{A}}_h(\hat{\bm{x}}_{\perp}, \nu, \hat{z}) \\
\end{split}
\end{equation}

\noindent where $\mathcal{F}(\hat{\eta}, \hat{\delta}, \nu, \hat{z}) \equiv \int d\vartheta \, F(\hat{\eta}, \hat{\delta}, \vartheta, \nu, \hat{z})$. These are the linearized Maxwell-Klimontovich equations that describe the growth of the ICL instability. The second equation of Eq.~(\ref{eq:simplifiedmv}) can be integrated to give an expression for $\mathcal{F}(\hat{\eta}, \hat{\delta}, \nu, \hat{z})$, which can be substituted into the first equation to yield the following integro-differential equation describing the $\hat{z}$-evolution of $\hat{\mathcal{A}}_h$
\begin{equation} \label{eq:ivp}
\left[\frac{\partial}{\partial \hat{z}} + i \hat{\Delta \nu} - i \mathcal{F}_D^{-1} \hat{\bm{\nabla}}_{\perp}^2\right] \hat{\mathcal{A}}_h(\hat{\bm{x}}_{\perp}, \nu, \hat{z}) = \hat{\mathcal{W}}_h(\hat{\bm{x}}_{\perp}) \int_0^{\hat{z}} d\hat{z}' \mathcal{X}_h(\hat{z}, \hat{z}') \int d^2 \hat{\bm{x}}_{\perp}' \hat{\mathcal{W}}_h(\hat{\bm{x}}_{\perp}') \hat{\mathcal{A}}_h(\hat{\bm{x}}_{\perp}', \nu, \hat{z}')
\end{equation}

\noindent where we have defined
\begin{equation} \label{eq:Xdefinition}
\mathcal{X}_h(\hat{z}, \hat{z}') = \frac{\pi}{\mathcal{I}} \frac{[\mathrm{JJ}]_h^2}{[\mathrm{JJ}]_1^2} \iiint d\hat{\eta} \,d\hat{\delta} \,d\vartheta \, e^{-i \left(\left(\nu - \frac{h}{4}\right)\hat{\eta} - \nu \frac{K_r^2}{2 + K_r^2} \hat{\delta}\right) (\hat{z} - \hat{z}')} \times \left[\frac{\partial}{\partial \hat{\eta}} + \frac{1 + K_r^2}{2 K_r^2} \frac{\partial}{\partial \hat{\delta}}\right] \hat{f}_0(\hat{\eta}, \hat{\delta}, \vartheta, \hat{z}').
\end{equation}

\noindent In Appendix~\ref{app:computingVX}, we compute Eq.~(\ref{eq:Xdefinition}) for bunches with Gaussian spread in $\hat{\eta}$ and $\hat{\delta}$ and a bunch with no spread in these. Equation~(\ref{eq:ivp}) is the equation for the ICL initial value problem, which is solved numerically in Sections~\ref{sec:numericalalgorithm} and \ref{sec:numericalresults}.

\end{widetext}

\subsection{\label{sec:dispersion} Dispersion Relation}

While Eqs.~(\ref{eq:ivp}) and (\ref{eq:Xdefinition}) of the previous subsection are the basis for the numerical algorithm used throughout the rest of this paper, for completeness we derive the dispersion relation for the ICL in this subsection. The bulk of this derivation is done in Appendix~\ref{app:vankampennormalmodes}, where the linearized Maxwell-Klimontovich equations Eq.~(\ref{eq:simplifiedmv}) are solved using a van Kampen normal mode expansion to give a general solution to the ICL initial value problem in the form of a series expansion Eq.~(\ref{eq:vankampenexpansion}). To use this expansion to compute the main quantities of interest, the complex growth rate $\mu_m$ and the transverse radiation mode profile $\hat{\mathcal{A}}_m(\bm{x}_{\perp})$ of the fastest growing mode $m$, one must solve the dispersion relation Eq.~(\ref{eq:as}) for the set of solution pairs $\{(\mu_{\ell}, \hat{\mathcal{A}}_{\ell}(\bm{x}_{\perp}))\}$ for the different modes indexed by $\ell$. Written out without the shorthands used in Appendix~\ref{app:vankampennormalmodes}, Eq.~(\ref{eq:as}) is

\begin{equation} \label{eq:dispersion}
\begin{split}
&\left[\mu_{\ell} - \hat{\Delta \nu} + \mathcal{F}_D^{-1} \hat{\bm{\nabla}}_{\perp}^2\right] \hat{\mathcal{A}}_{\ell}(\hat{\bm{x}}_{\perp}) \\
&- \pi \mathcal{V}(\mu_{\ell}) \hat{\mathcal{W}} (\hat{\bm{x}}_{\perp}) \int d^2 \hat{\bm{x}}_{\perp}' \hat{\mathcal{W}}(\hat{\bm{x}}_{\perp}') \hat{\mathcal{A}}_{\ell}(\hat{\bm{x}}_{\perp}') = 0.
\end{split}
\end{equation}

\noindent The function $\mathcal{V}(\mu)$ appearing in this equation is defined in Eq.~(\ref{eq:vdefinition}); using the definitions in Eq.~(\ref{eq:variousdefinitions}) and integrating by parts, it can be rewritten as

\begin{equation} \label{eq:vdefinition2}
\begin{split}
\mathcal{V}(\mu) &= \left(h + \frac{2 (3 + K_r^2)}{4 + K_r^2} \Delta \nu\right) \frac{[\mathrm{JJ}]_h^2 }{[\mathrm{JJ}]_1^2 }\iiint d\hat{\eta} \,d\hat{\delta} \,d\vartheta \\
&\frac{\hat{f}_0(\hat{\eta}, \hat{\delta}, \vartheta)}{\left(\mu - \left(\frac{3}{4} h + \Delta \nu\right)\hat{\eta} +  (h + \Delta \nu) \frac{K_r^2}{2 + K_r^2} \hat{\delta}\right)^2}.
\end{split}
\end{equation}

In Appendix~\ref{app:computingVX}, we compute $\mathcal{V}(\mu)$ for bunches with Gaussian spread in $\hat{\eta}$ and $\hat{\delta}$ and a bunch with no spread in these quantities. We denote the solution pair with the complex growth rate that has the largest imaginary part by the subscript $m$. At large enough $\hat{z}$, this solution pair should become the dominant term in the series Eq.~(\ref{eq:vankampenexpansion}) and $\hat{\mathcal{A}}(\hat{\bm{x}}_{\perp}, \hat{z}) \propto \hat{\mathcal{A}}_m(\hat{\bm{x}}_{\perp}) e^{-i \mu_m \hat{z}}$. The 3D gain parameter is defined through the equation for the power gain length and is, in terms of the complex growth rate,
\begin{equation} \label{eq:rho}
\rho = \rho_0  \frac{2}{\sqrt{3}} \mathrm{im}(\mu_m).
\end{equation}

This method of computing growth rates using the dispersion relation Eq.~(\ref{eq:dispersion}) works well for the FEL. However, for the ICL it is much more difficult in practice. One reason for this is that ICL radiation profiles are asymmetric. This means that one cannot ignore asymmetric radiation modes, which increases the dimensionality of the problem. Another reason is the higher $\rho$ in the ICL compared to the FEL, which means there is a greater contribution of higher order modes to the radiation. It is for these reasons that we opt to use the algorithm described in Section~\ref{sec:numericalalgorithm}, which is based instead on Eqs.~(\ref{eq:ivp}) and (\ref{eq:Xdefinition}). 

The 1D limit of the theory can be taken by assuming $\hat{\mathcal{A}}_{\ell}$ has no transverse dependence and then integrating Eq.~(\ref{eq:dispersion}) over a transverse circle of radius $a_{\beta, r}$. The resulting dispersion relation is $\mu_{\ell} - \hat{\Delta \nu} - \mathcal{V}(\mu_{\ell}) = 0$. For a cold bunch with no wavelength detuning and $h = 1$, the dispersion relation reduces to $\mu_{\ell}^3 = 1$, and from this the growth rate can be shown to satisfy $\rho = \rho_0$. This result agrees with Davoine {\it et al.}~\cite{davoine2018}, although there is a factor of $\mathcal{I}$ difference between the definition of $\rho$ here and in Ref.~\cite{davoine2018}. 

\section{\label{sec:numericalalgorithm} Numerical Algorithm}

A method to compute gain lengths and transverse radiation profiles from Eqs.~(\ref{eq:ivp}) or (\ref{eq:dispersion}) is needed in order to obtain useful results from the theory presented in Section~\ref{sec:analytictheory}. As in FEL theory, there are a number of possibilities for approaching this problem. One is the variational principle~\cite{yu1990,xie2000,kim2017}, in which a functional is constructed from the dispersion relation and the transverse radiation profile is approximated by a trial family of functions. Approximate complex growth rates and radiation profiles are obtained by varying the trial function's parameters to minimize the functional. While relatively quick to compute, this method is only an approximation, and whereas a Gaussian trial function is a good approximation for the radially symmetric fundamental mode of the FEL, a more complicated trial function is required for the ICL, which has an asymmetric source term $\hat{\mathcal{W}}(\hat{\bm{x}}_{\perp})$ and therefore an asymmetric radiation profile.

Matrix methods \cite{xie2000,kim2017} are another approach. In Refs.~\cite{xie2000,kim2017}, the FEL dispersion relation is converted to radial coordinates and Hankel transformed. The resulting equation is discretized over a 1D radial spatial frequency grid, yielding a matrix equation, which is solved iteratively. While this method could be adapted to the ICL, the radial non-uniformity of $\hat{\mathcal{W}}(\hat{\bm{x}}_{\perp})$ in Eq.~(\ref{eq:dispersion}) means that a 2D Fourier transform rather than a Hankel transform would be required, and the transformed equation would be discretized over a 2D---rather than 1D---grid of spatial frequencies. This implies that significantly larger matrices would be required for the ICL as compared to the FEL when using the matrix method, drastically increasing the computational cost.

While other more complicated approaches exist, such as that of Baxevanis {\it et al.}~\cite{baxevanis2013}, rather than attempt to adapt these to the ICL, we choose to use a novel, albeit relatively simple, Crank-Nicolson finite-difference approach to efficiently solve the ICL initial value problem Eq.~(\ref{eq:ivp}). The method works as follows. First, we introduce $\mathcal{B}(\hat{\bm{x}}_{\perp}, \hat{z}) \equiv e^{i \hat{\Delta \nu} \hat{z}} \hat{\mathcal{A}}(\hat{\bm{x}}_{\perp}, \nu, \hat{z})$ and $\Gamma(\hat{z}, \hat{z}') \equiv e^{i \hat{\Delta \nu} (\hat{z} - \hat{z}')} \mathcal{X}(\hat{z}, \hat{z}')$, which allows Eq.~(\ref{eq:ivp}) to be rewritten as
\begin{widetext}
\begin{equation} \label{eq:ivpb}
\left[\frac{\partial}{\partial \hat{z}} - i \mathcal{F}_D^{-1} \hat{\bm{\nabla}}_{\perp}^2\right] \mathcal{B}(\hat{\bm{x}}_{\perp}, \hat{z}) = \hat{\mathcal{W}}(\hat{\bm{x}}_{\perp}) \int_0^{\hat{z}} d\hat{z}' \, \Gamma(\hat{z}, \hat{z}')  \int d^2 \hat{\bm{x}}_{\perp}' \hat{\mathcal{W}}(\hat{\bm{x}}_{\perp}') \mathcal{B}(\hat{\bm{x}}_{\perp}', \hat{z}').
\end{equation}

We discretize this equation over a uniform transverse spatial grid. On the left-hand side we discretize using the Crank-Nicolson method in order to ensure numerical stability and speed up the code by allowing larger step sizes. For the source term on the right-hand side we discretize by replacing the integrals with Riemann sums. We use the Euler method (evaluating at $\hat{z}$) rather than the Crank-Nicolson method (evaluating at $\hat{z}$ and $\hat{z} + d\hat{z}$ and averaging) for the right-hand side, as the difference between the two is negligible and the former is simpler. The discretized equation is 

\begin{equation} \label{eq:discretization1}
\begin{split}
&\frac{\mathcal{B}^{n + 1}_{i,j} - \mathcal{B}^{n}_{i,j}}{d\hat{z}} - \frac{i \mathcal{F}_D^{-1}}{2} \Bigg[\frac{1}{d\hat{x}^2} \left(\mathcal{B}^{n}_{i - 1,j} - 2 \mathcal{B}^{n}_{i,j} + \mathcal{B}^{n}_{i + 1,j} + \mathcal{B}^{n+1}_{i - 1,j} - 2 \mathcal{B}^{n+1}_{i,j} + \mathcal{B}^{n+1}_{i + 1,j}\right) \\
&+\frac{1}{d\hat{y}^2} \left(\mathcal{B}^{n}_{i,j - 1} - 2 \mathcal{B}^{n}_{i,j} + \mathcal{B}^{n}_{i,j + 1} + \mathcal{B}^{n+1}_{i,j - 1} - 2 \mathcal{B}^{n+1}_{i,j} + \mathcal{B}^{n+1}_{i,j + 1}\right)\Bigg] = \hat{\mathcal{W}}_{i,j} d\hat{z} \sum_{m = 0}^{n - 1} \Gamma^{n, m} d\hat{x} d\hat{y} \sum_{p,q} \hat{\mathcal{W}}_{p,q} \mathcal{B}^m_{p,q}
\end{split}
\end{equation}

\end{widetext}

\begin{figure}[b]
    \centering
    \includegraphics[width=\columnwidth]{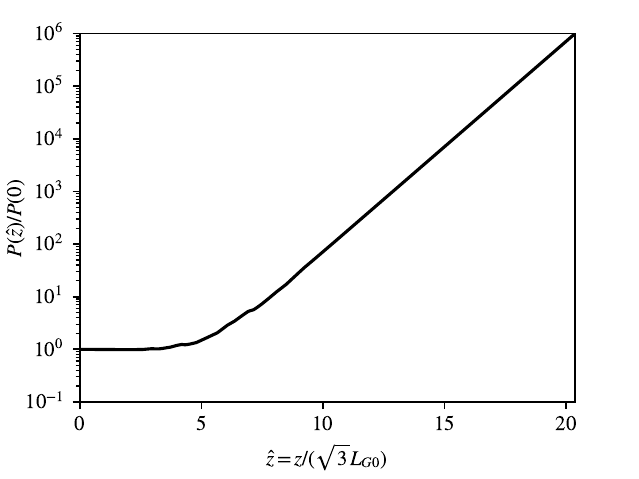}
    \caption{Simulated radiation power gain of an ICL with the ``base'' parameters described in Section~\ref{sec:numericalresults}.}
    \label{fig:prototype_power}
\end{figure}

\noindent where $\mathcal{B}_{i,j}^n = \mathcal{B}(\hat{x}_i, \hat{y}_j, \hat{z}_n)$ and $\Gamma^{n, m} = \Gamma(\hat{z}_n, \hat{z}_m)$, and where, in order to discretize $\hat{\mathcal{W}}(\hat{x},\hat{y})$, we approximated the delta function in Eq.~(\ref{eq:wdefinition}) with a step function of width $d\hat{y}$ and height $1 / d\hat{y}$ centered at $\hat{y} = 0$: $\hat{\mathcal{W}}_{i,j} = \hat{\mathcal{W}}_{x,i}  \delta_{j, j_0} / d\hat{y} = \hat{\mathcal{W}}_x(\hat{x}_i)  \delta_{j, j_0} / d\hat{y}$ where $\hat{\mathcal{W}}_x(\hat{x}) = \int d\hat{y}\, \hat{\mathcal{W}}(\hat{x},\hat{y})$ and $j_0$ is the index where $\hat{y}_{j_0} = 0$. Equation~(\ref{eq:discretization1}) can be rearranged and converted to the matrix equation
\begin{equation} \label{eq:discretization2}
\begin{split}
&(1 - i \mu_x \bm{\Delta}_x - i \mu_y \bm{\Delta}_y) \bm{\mathcal{B}}^{n + 1} \\
&= (1 + i \mu_x \bm{\Delta}_x + i \mu_y \bm{\Delta}_y) \bm{\mathcal{B}}^n + \bm{\mathcal{C}}^n
\end{split}
\end{equation}

\noindent where 
\begin{equation} \label{eq:discretization2definitions}
\begin{split}
(\bm{\Delta}_x)_{i,j, i', j'} &\equiv (\delta_{i - 1, i'} - 2 \delta_{i,i'} + \delta_{i + 1, i'}) \delta_{j,j'}\\
(\bm{\Delta}_y)_{i,j, i', j'} &\equiv \delta_{i,i'} (\delta_{j - 1, j'} - 2 \delta_{j,j'} + \delta_{j + 1, j'})\\
(\bm{\mathcal{C}}^n)_{i,j} &\equiv \delta_{j, j_0} \hat{\mathcal{W}}_{x,i}  \frac{d\hat{z}^2 d\hat{x}}{d\hat{y}} \sum_{m = 0}^{n - 1} \Gamma^{n, m} \left(\sum_p \hat{\mathcal{W}}_{x,p} \mathcal{B}^m_{p,j_0}\right) \\
\mu_x &= d\hat{z} / (2 \mathcal{F}_D d\hat{x}^2) \\
\mu_y &= d\hat{z} / (2 \mathcal{F}_D d\hat{y}^2). \\
\end{split}
\end{equation}

\begin{figure*}[t]
    \centering
    {
        \centering
        \includegraphics[trim={2.7cm 0 5.0cm 0}, clip, width=2\columnwidth]{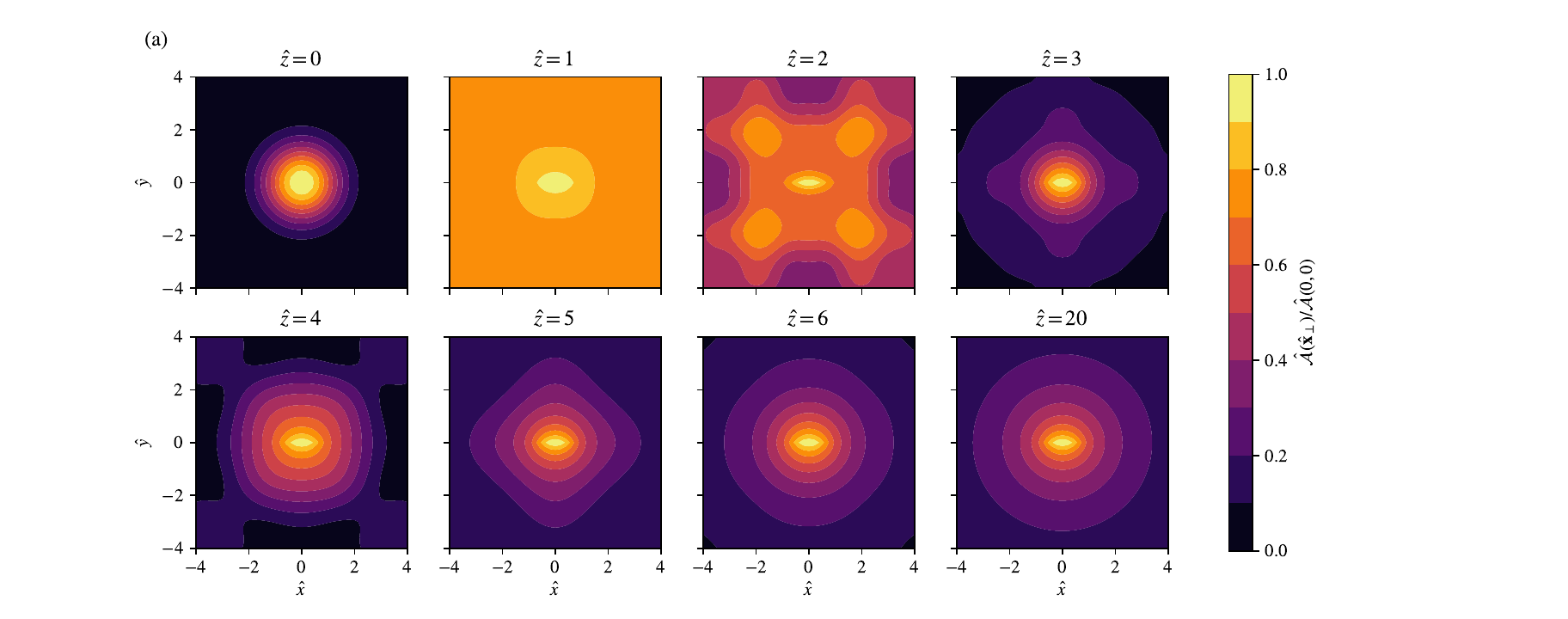}
    }
    \hfill
    {
        \centering
        \includegraphics[trim={2.7cm 0 5.0cm 0}, clip, width=2\columnwidth]{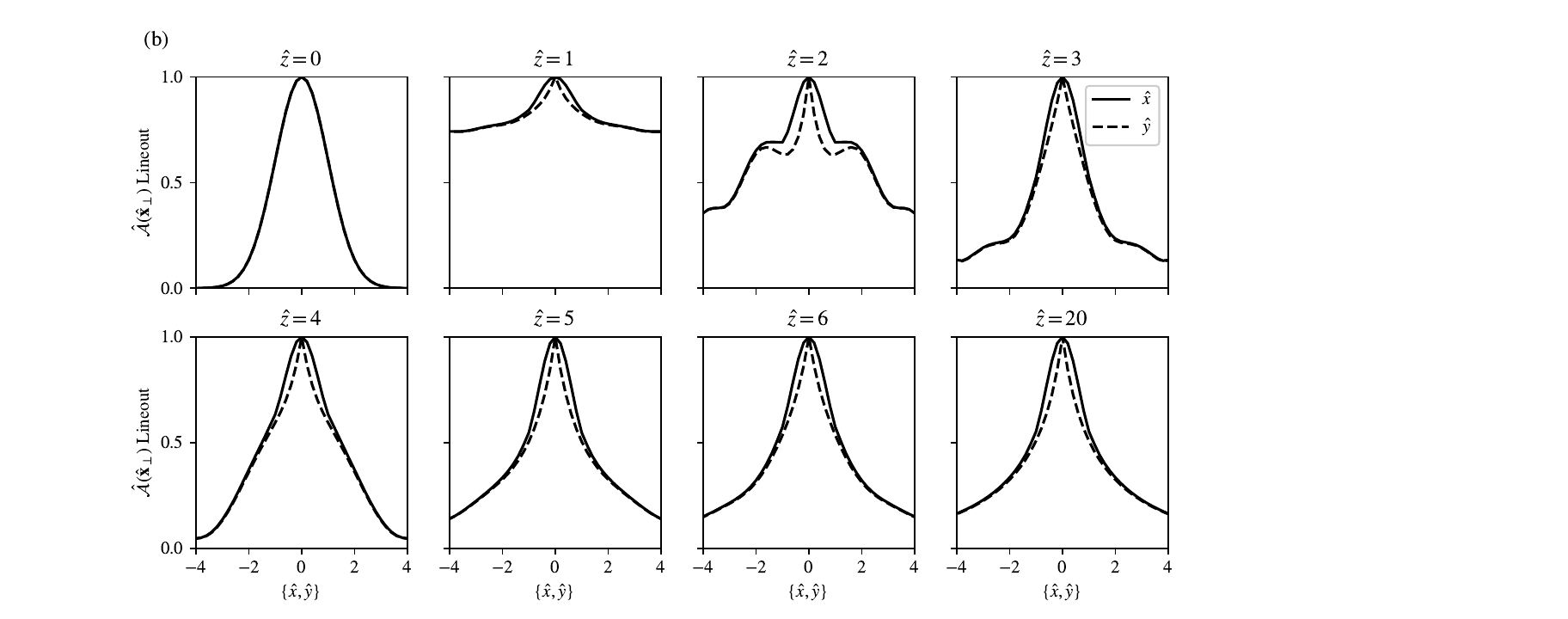}
    }

    \caption{Simulated transverse radiation profiles at different $\hat{z}$ positions for an ICL with the ``base'' parameters described in Section~\ref{sec:numericalresults}. (a) Normalized two-dimensional field profile. (b) Lineouts at $\hat{y}=0$ (solid) and $\hat{x}=0$ (dashed). Panels show the evolution of the radiation profile from the initial seed, through the startup process, to the final steady state.}
\label{fig:prototype_modes}
\end{figure*}

\noindent and where vectors and matrices are constructed by flattening the 2D transverse position indices into 1D using an ordering scheme such as column-major (e.g., $\bm{\mathcal{B}}^n = \begin{pmatrix}
\mathcal{B}^n_{11} & \mathcal{B}^n_{21} & \dots & \mathcal{B}^n_{12} & \mathcal{B}^n_{22} & \dots 
\end{pmatrix}^{\intercal}$) or row-major (e.g., $\bm{\mathcal{B}}^n = \begin{pmatrix}
\mathcal{B}^n_{11} & \mathcal{B}^n_{12} & \dots & \mathcal{B}^n_{21} & \mathcal{B}^n_{22} & \dots 
\end{pmatrix}^{\intercal}$). We note that $\bm{\mathcal{C}}^n$ depends on the time ($\hat{z}$) history $\bm{\mathcal{B}}^0, \bm{\mathcal{B}}^1, \cdots, \bm{\mathcal{B}}^{n - 1}$.

\begin{figure*}[t]
    \centering
    
    \includegraphics[width=\columnwidth]{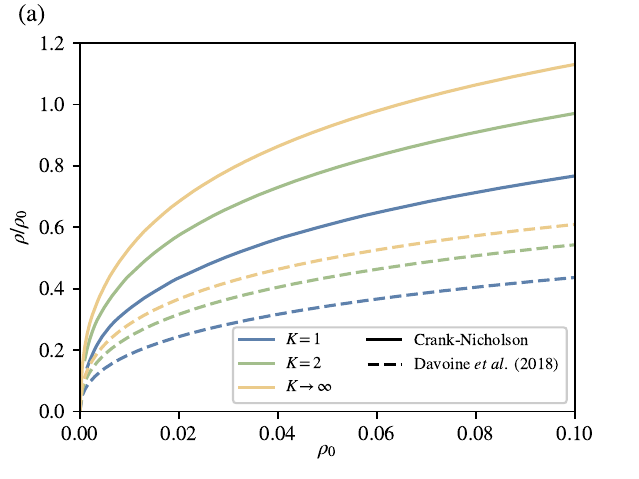}    \includegraphics[width=\columnwidth]{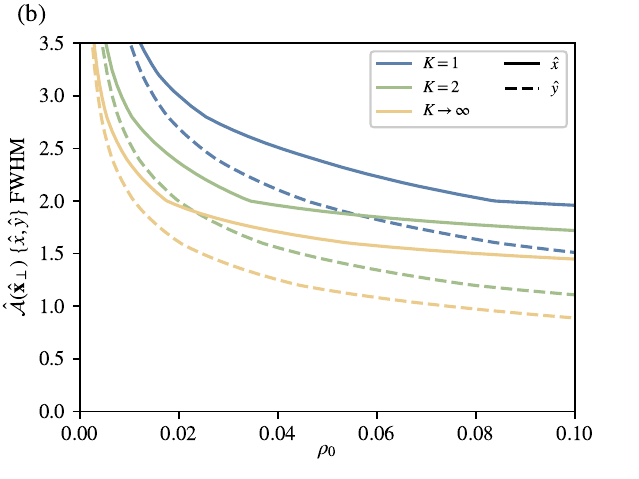}

    \caption{Dependence of the relative 3D gain parameter (a) and the transverse radiation sizes in $\hat{x}$ (solid lines) and $\hat{y}$ (dashed lines) (b) on the cold 1D gain parameter $\rho_0$ for various values of the undulator parameter: $K_r=1$ (blue), $K_r=2$ (green), and $K_r\rightarrow\infty$ (yellow). Results are shown for a cold electron bunch with no frequency detuning ($\Delta \nu = 0$) and are computed using the algorithm described in Section~\ref{sec:numericalalgorithm}, except for the dashed lines in panel (a), which are computed using the method from Davoine {\it et al.}~\cite{davoine2018}.}
    \label{fig:no_detuning_cold}
\end{figure*}

Due to the large number of grid points and thus the dimension of the vectors and matrices involved, it is inefficient to directly solve Eq.~(\ref{eq:discretization2}) numerically at every step. Were the problem one (transverse) dimensional, the matrices in this equation would be tridiagonal, enabling it to be solved using fast and space-efficient tridiagonal matrix algorithms. However, in two transverse dimensions this is not the case. If column-major ordering is used, $\bm{\Delta}_x$ is tridiagonal but $\bm{\Delta}_y$ is not, and vice versa for row-major ordering. To circumvent this, we use the alternating direction implicit method \cite{peaceman1955}, which instead works by solving the approximate equation

\begin{equation} \label{eq:discretization3}
\begin{split}
&(1 - i \mu_x \bm{\Delta}_x)(1 - i \mu_y \bm{\Delta}_y) \bm{\mathcal{B}}^{n + 1} \\
&=(1 + i \mu_x \bm{\Delta}_x)(1 + i \mu_y \bm{\Delta}_y) \bm{\mathcal{B}}^n \\
&+ (1 + i \mu_x \bm{\Delta}_x)(1 - i \mu_x \bm{\Delta}_x) \bm{\mathcal{C}}^n
\end{split}
\end{equation}

\noindent which is equal to Eq.~(\ref{eq:discretization2}) to first order in $\mu_x$ and $\mu_y$, which are small, and differs only in the second order terms. Equation~(\ref{eq:discretization3}) can be broken into two simpler equations by introducing a new vector $\bm{\mathcal{B}}^{*}$, which is solved for as an intermediate step:

\begin{equation} \label{eq:discretization4}
\begin{split}
&(1 - i \mu_x \bm{\Delta}_x) \bm{\mathcal{B}}^{*} = (1 + i \mu_y \bm{\Delta}_y) \bm{\mathcal{B}}^n \\
&(1 - i \mu_y \bm{\Delta}_y) \bm{\mathcal{B}}^{n+1} = (1 + i \mu_x \bm{\Delta}_x) (\bm{\mathcal{B}}^{*} + \bm{\mathcal{C}}^n). \\
\end{split}
\end{equation}

Unlike Eq.~(\ref{eq:discretization2}), Eq.~(\ref{eq:discretization4}) can be solved entirely using tridiagonal algorithms, which vastly speed up the computation. This requires reordering the entries of the vectors to switch between row-major and column-major at various points during the computation.

While in principle this algorithm can be applied to the FEL, in practice it is likely significantly less efficient for the FEL than for the ICL. This is due to different definitions of $\bm{\mathcal{C}}^n$, which arise from the source term on the right-hand side of the $\hat{z}$-evolution equation (Eq.~(\ref{eq:ivp}) for the ICL and \cite[(18)]{baxevanis2013} for the FEL). In the case of the FEL, to calculate $\bm{\mathcal{C}}^n$ at each step, a 3D numerical integration over $\bm{x}_{\perp}'$ and $z'<z$ must be performed for each transverse grid point $\bm{x}_{\perp}$. In the case of the ICL, due to the separability of the integral kernel in Eq.~(\ref{eq:ivp}) and the fact that $\hat{\mathcal{W}}(\hat{x}, \hat{y}) = 0$ when $\hat{y} \neq 0$, only two 1D numerical integrations are needed per step. The first is computing $a_n \equiv \sum_p \hat{\mathcal{W}}_{x,p} \mathcal{B}^n_{p,j_0}$ which is added to a list $\{a_0, a_1, \cdots, a_{n-1}\}$, which can be stored in memory. The second is computing $\sum_{m = 0}^{n - 1} \Gamma^{n, m} a_m$. This significant added computational complexity in the FEL case explains why other previously mentioned approaches, such as the variational principle or matrix methods, are used for FELs instead of the algorithm outlined here.

We developed a Python code implementing the algorithm described in this section. This code computes the $\hat{z}$-evolution of the transverse radiation profile as well as $L_G$ and $\rho$ by fitting the total radiation power growth to an exponential. For simplicity, we use a square transverse grid with square transverse cells so $\hat{x}_{\mathrm{max}} = \hat{y}_{\mathrm{max}} = -\hat{x}_{\mathrm{min}} = -\hat{y}_{\mathrm{min}}$, $d\hat{x} = d\hat{y}$, and $\mu = \mu_x = \mu_y$. A Gaussian initial radiation field is used to seed the interaction. Convergence of the simulation with respect to transverse box size, transverse cell size, and $\mu$ as well as insensitivity of the simulation to the initial radiation shape is demonstrated in Appendix~\ref{app:convergence}. This code can simulate a cold bunch as well as a bunch with Gaussian spread in energy or undulator parameter, and can be run with nonzero frequency detuning $\Delta \nu$.

\section{\label{sec:numericalresults} Numerical Results}

Having introduced an algorithm in the previous section to solve the ICL initial value problem Eq.~(\ref{eq:ivp}), we turn to using it to examine the behavior of the ICL in a number of different cases. First we ran a ``base'' simulation with parameters $h = 1$, $K_r \rightarrow \infty$, $\hat{\Delta \nu} = 0$, $\Sigma = 0$, and $\rho_0 = 0.01$. The nonphysical parameters $\hat{x}_{\mathrm{max}} = 20$, $d\hat{x} = 0.2$, $\mu = 0.5$, and $\sigma_{\hat{x}, \mathrm{initial}} = \sigma_{\hat{y}, \mathrm{initial}} = 1$ were selected based on convergence scans performed in Appendix~\ref{app:convergence}. The simulation showed exponential growth in the radiation power after an initial startup period, which can be seen in Fig.~\ref{fig:prototype_power}. The evolution of the radiation profile from the initial bi-Gaussian distribution used to seed the interaction to a steady state during this exponential growth regime is shown in Fig.~\ref{fig:prototype_modes}. The steady state transverse radiation profile is asymmetric near the axis---dropping off faster in $\hat{y}$ than in $\hat{x}$. However, at $\hat{r} \gtrsim 1$, this asymmetry quickly disappears and the field becomes symmetric. This shape is due to the competing effects of the asymmetric source term in Eq.~(\ref{eq:ivp}) and the effect of diffraction which causes higher order (asymmetric) modes to diffract faster. In Appendix~\ref{app:convergence} we demonstrate that, as expected, the ICL radiation power growth rate is insensitive to the size of the initial seed radiation field.

We now consider variations to this base case. The dependence of the gain parameter and the (steady state) transverse radiation sizes on $K_r$ and $\rho_0$ is shown in Fig.~\ref{fig:no_detuning_cold}. Previously, in Section~\ref{sec:maxwellklimontovich}, we introduced the Fresnel parameter $\mathcal{F}_D$ defined in Eq.~(\ref{eq:fresnel}), which quantifies diffraction. The results in Fig.~\ref{fig:no_detuning_cold} agree with what would be expected from that discussion: for larger $\rho_0$ and $K_r$, $\mathcal{F}_D$ is larger, meaning diffraction is less severe and $\rho / \rho_0$ is larger; for smaller $\rho_0$ and $K_r$, $\mathcal{F}_D$ is smaller, meaning diffraction is more severe and $\rho / \rho_0$ is smaller. Since $\mathcal{F}_D \propto K_r^2 / (2(2 + K_r^2))$, the ability of larger $K_r$ to reduce diffraction plateaus for $K_r \gg 1$. 

From Fig.~\ref{fig:no_detuning_cold}, we also see that the radiation field is smaller and more asymmetric when diffraction is less significant, and larger and more symmetric when diffraction is more significant. Additionally, in this figure we compare the $\rho$ predicted by our theory to that predicted by the theory in Ref.~\cite{davoine2018}. We find Ref.~\cite{davoine2018} gives radiation power growth rates a factor of a few smaller than the method presented in this work. We speculate that this is because in Ref.~\cite{davoine2018}, the transverse size of the radiation was assumed to be equal to that of the emitter, which was well approximated by a Gaussian with $\sigma = \frac{3}{8} a_{\beta,r}  = 0.375 a_{\beta,r}$, i.e., a FWHM of $\frac{3}{4}\sqrt{2\ln 2} a_{\beta,r} \approx 0.883 a_{\beta,r}$. In contrast, our theory predicts a larger radiation field, which would have larger gain due to its longer Rayleigh length and thus lower diffraction.

Next we consider a bunch with a Gaussian spread in energy and undulator parameter. As can be seen mathematically in Eq.~(\ref{eq:xGaussian}), the dependence of the ICL physics in this case on energy spread $\sigma_{\gamma}$ and undulator parameter spread $\sigma_K$ is exclusively through the parameter 

\begin{equation} \label{eq:bigsigmadef}
\begin{split}
\Sigma &= \frac{1}{\rho_0} \Bigg [\left((h + \Delta \nu) \frac{K_r^2}{2 + K_r^2} \frac{\sigma_K}{K_r}\right)^2  \\
&+ \left(\left(\frac{3}{4} h + \Delta \nu\right) \frac{\sigma_{\gamma}}{\gamma_r} \right)^2 \Bigg]^{\frac{1}{2}}.
\end{split}
\end{equation}

\begin{figure}[t]
    \centering
    \includegraphics[width=\columnwidth]{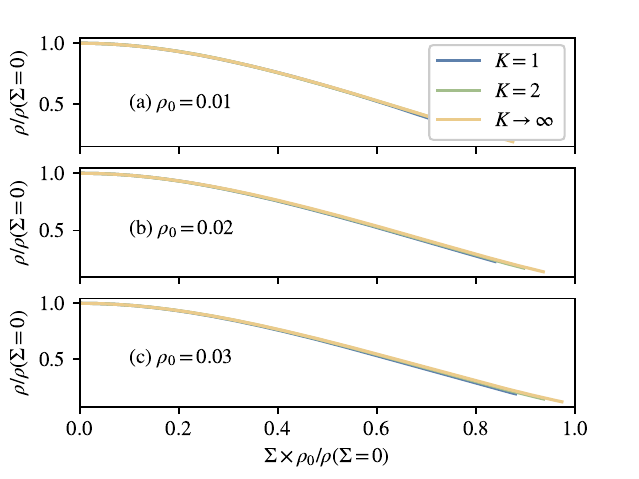}
   \caption{Effect of finite Gaussian energy/undulator spread on the gain parameter $\rho$ for $K_r=1$ (blue), $K_r=2$ (green), and $K_r\rightarrow\infty$ (yellow), and for $\rho_0=0.01$ (a), $\rho_0=0.02$ (b), and $\rho_0 =0.03$ (c). $\Sigma$ is defined in terms of $\sigma_K$, $\sigma_{\gamma}$, and $\rho_0$ in Eq.~(\ref{eq:bigsigmadef}).}
    \label{fig:no_detuning_warm}
\end{figure}

\noindent This parameter $\Sigma$ essentially quantifies the degree to which $\sigma_K$ and $\sigma_{\gamma}$ violate the constraints on them required for lasing \cite[(1.1)]{davoine2018}, although $\rho_0$ instead of $\rho$ appears in Eq.~(\ref{eq:bigsigmadef}). $\Sigma \rho_0 / \rho \lesssim 1/2$ is equivalent to these constraints \cite[(1.1)]{davoine2018}. In Fig.~\ref{fig:no_detuning_warm} we show the effect of energy and undulator parameter spread on the ICL by plotting the relative change in gain parameter as a function of $\Sigma \rho_0 / \rho(\Sigma=0)$. We see that $\rho$ decreases as $\Sigma$ is increased, falling essentially to zero by the point $\Sigma = \rho / \rho_0$. This result is important because it quantifies how ICL gain decreases with large bunch emittance and energy spread near the beam quality requirement limit. This makes this limit, which is often stated simply as an approximate inequality such as in Section~\ref{sec:phasespace}, much more specific.

\begin{figure}[t]
\centering
\includegraphics[width=\columnwidth]{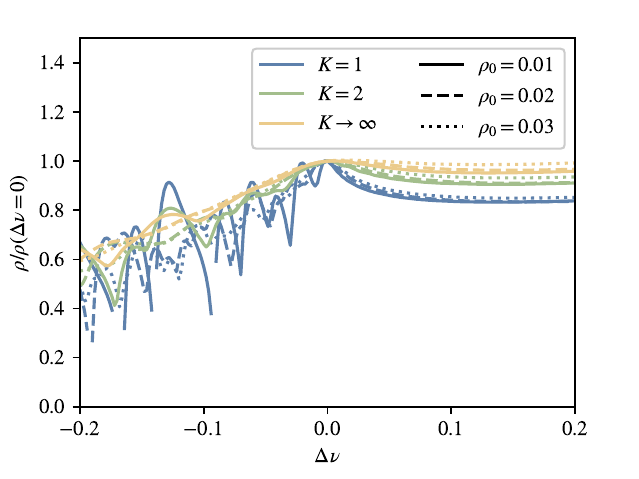}
   \caption{ICL detuning curve. Proportional change in gain parameter due to nonzero detuning is plotted for different values of $\Delta \nu$. Curves are shown for a range of parameters: $K_r = 1$ (blue), $K_r = 2$ (green), $K_r \rightarrow \infty$ (yellow), $\rho_0 = 0.01$ (solid), $\rho_0 = 0.02$ (dashed), $\rho_0 = 0.03$ (dotted).}
    \label{fig:detuning}
\end{figure}

Finally, we see the ICL detuning curve in Fig.~\ref{fig:detuning}. This curve appears to have a similar shape to that of the FEL but with a significantly larger width of order $1$ in the ICL compared to order $\rho$ in the FEL. This could be related to the fact that $\sigma_{\perp} \ll a_{\beta,r}$ in an ICL while $\sigma_{\perp} \gg a_{u,r}$ in an FEL. However, we hesitate to draw any firm conclusions, one reason being that the theory presented in this work depends on the assumption $|\Delta \nu| \ll 1$ made during the period averaging procedure. We believe the effect of frequency detuning in the ICL merits further study.

\section{\label{sec:phasespace} Phase Space Requirements}

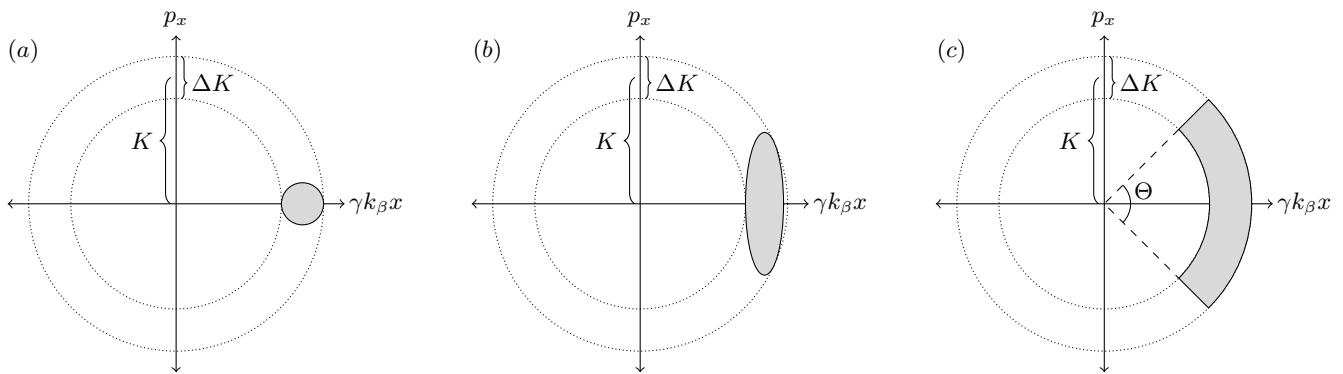
\begin{figure*}[t]
    \centering
    {
        \centering
        \begin{tikzpicture}[scale=0.7]
        \node at (-2.9, 2.9) {$(a)$};
        \draw[<->] (-3.2,0) -- (3.2,0);
        \draw[<->] (0,-3.2) -- (0,3.2);
        \node at (3.8,0) {$\gamma k_{\beta} x$};
        \node at (0,3.5) {$p_x$};
        \draw[densely dotted] (0,0) circle (2.0);
        \draw[densely dotted] (0,0) circle (2.8);
        \draw[fill=gray!30] (2.4, 0) circle (0.4);
        \draw [decorate,decoration={brace,amplitude=2,mirror,raise=2,aspect=0.375}] (0,2.0) -- (0,2.8);
        \node at (0.7, 2.3) {$\Delta K$};
        \draw [decorate,decoration={brace,amplitude=4,mirror,raise=2}] (0,2.4) -- (0,0);
        \node[rotate=0] at (-0.65, 1.19) {$K$};
        \end{tikzpicture}
    }
    \hfill
    {
        \centering
        \begin{tikzpicture}[scale=0.7]
        \node at (-2.9, 2.9) {$(b)$};
        \draw[<->] (-3.2,0) -- (3.2,0);
        \draw[<->] (0,-3.2) -- (0,3.2);
        \node at (3.8,0) {$\gamma k_{\beta} x$};
        \node at (0,3.5) {$p_x$};
        \draw[densely dotted] (0,0) circle (2.0);
        \draw[densely dotted] (0,0) circle (2.8);
        \draw[fill=gray!30] (2.36137, 0) ellipse (0.361372 and 1.35545);
        \draw [decorate,decoration={brace,amplitude=2,mirror,raise=2,aspect=0.375}] (0,2.0) -- (0,2.8);
        \node at (0.7, 2.3) {$\Delta K$};
        \draw [decorate,decoration={brace,amplitude=4,mirror,raise=2}] (0,2.4) -- (0,0);
        \node[rotate=0] at (-0.65, 1.19) {$K$};
        \end{tikzpicture}
    }
    \hfill
    {
        \centering
        \begin{tikzpicture}[scale=0.7]
        \node at (-2.9, 2.9) {$(c)$};
        \draw[<->] (-3.2,0) -- (3.2,0);
        \draw[<->] (0,-3.2) -- (0,3.2);
        \node at (3.8,0) {$\gamma k_{\beta} x$};
        \node at (0,3.5) {$p_x$};
        \draw[densely dotted] (0,0) circle (2.0);
        \draw[densely dotted] (0,0) circle (2.8);
        \filldraw[fill=gray!30] (-45:2) arc (-45:45:2) -- (45:2.8) arc (45:-45:2.8) -- cycle;
        \draw[dashed] (1.41421,1.41421) -- (0,0) -- (1.41421, -1.41421);
        \draw[solid] (-45:0.5) arc (-45:45:0.5);
        \node at (0.75,0.25) {$\Theta$};
        \draw [decorate,decoration={brace,amplitude=2,mirror,raise=2,aspect=0.375}] (0,2.0) -- (0,2.8);
        \node at (0.7, 2.3) {$\Delta K$};
        \draw [decorate,decoration={brace,amplitude=4,mirror,raise=2}] (0,2.4) -- (0,0);
        \node[rotate=0] at (-0.65, 1.19) {$K$};
        \end{tikzpicture}
    }
    \caption{Conceptual diagram showing (a) matched offset Gaussian, (b) optimally mismatched offset Gaussian, and (c) annular sector bunch distributions shaded in gray in normalized $\gamma k_{\beta} x-p_x$ phase space. The radii of the inner and outer dotted circles give the minimum and maximum tolerable values of $K$, respectively. Each bunch will rotate clockwise about the origin on a circular path at the betatron frequency. The difference in radii of the dotted circles corresponds to the bunch's $K$-spread. This figure shows how different bunch phase space distributions can maintain the same spread in $K$ while occupying a larger or smaller volume in phase space, corresponding to a larger or smaller emittance.}
    \label{fig:phasespace}
\end{figure*}

Elucidating the beam quality requirements for lasing in the ICL is crucial to judging the feasibility of the concept as a whole. In the ICL, as in the FEL, the spread in resonant wavelengths of particles in each longitudinal slice must obey the condition $\Delta \lambda_1 / \lambda_{1,r} \lesssim  \rho$ for lasing to occur. As discussed previously in Section~\ref{sec:electronmotion}, while for the FEL this implies a constraint on the beam's angular spread and thus its emittance, differences in the physics mean that for the ICL this condition instead implies a constraint on the beam's spread in $K$:
\begin{equation} \label{eq:kspreadcondition}
\frac{\Delta K}{K} \lesssim \frac{2 + K^2}{2 K^2} \rho. 
\end{equation}

While there is no {\it intrinsic} constraint on the emittance in order to lase in an ICL, for different families of phase space distributions there is a maximum emittance a beam can have for a given $\Delta K$. Along with Eq.~(\ref{eq:kspreadcondition}), this allows a maximum emittance for lasing to be written. In this section we derive emittance constraints for three families of phase space distributions: matched offset Gaussian beams, ``optimally mismatched'' offset Gaussian beams, and ``phase space annular sector'' beams.

To illustrate these three phase space distribution families and the relationship between $K$-spread and emittance, consider a bunch in an ion channel containing resonant ($\gamma_j = \gamma_r$), paraxial ($K_j / \gamma_j \ll 1$) electrons oscillating entirely in the $x$-$z$ plane ($y_j = p_{y,j} = 0$). As discussed in Section~\ref{sec:electronmotion}, the electrons will exhibit simple harmonic betatron motion. In normalized $\gamma k_{\beta} x$-$p_x$ phase space, they move clockwise along circular trajectories of radii $K_j$. Figure~\ref{fig:phasespace} shows diagrams of each of the three previously mentioned distribution families in this normalized phase space. Figure~\ref{fig:phasespace}(a) illustrates how a matched beam, while optimal for the on-axis ICL configuration, demands a relatively small emittance to achieve a given $\Delta K$, making it suboptimal for the off-axis configuration. Note that while matching prevents projected emittance growth due to chromatic phase mixing for an on-axis beam, the same is not true for an off-axis beam, although this projected emittance growth will not cause an increase in $\Delta K$.

As originally pointed out by Ersfeld {\it et al.} \cite{ersfeld2014}, using a mismatched bunch distribution makes it possible to achieve the same $\Delta K$ with a larger emittance. This can be seen in Fig.~\ref{fig:phasespace}(b), which shows an ``optimally mismatched'' offset Gaussian phase space distribution with spot size and angular spread chosen to maximize emittance for a given $K$-spread. Finally, a non-Gaussian ``phase space annular sector'' distribution is shown in Fig.~\ref{fig:phasespace}(c). Such a phase space distribution would enable lasing with even larger emittances but would be challenging to prepare. One potential method could be to send an underfocused, on-axis Gaussian beam through a bending magnet with a stronger kick at smaller $|x|$ than at larger $|x|$. This would produce a beam resembling the one shown in Fig.~\ref{fig:phasespace}(c).

We now move on to a quantitative analysis of these three cases. In Appendix~\ref{app:optimization} we write the distribution functions for each case and derive the optimal parameters that maximize emittance for a given wavelength spread subject to the resonance condition. For each case we write the emittance of the optimized distribution as a function of wavelength spread and require $\Delta \lambda_1 / \lambda_{1,r} \lesssim \rho$ to obtain the emittance constraint. These calculations consider the full 4D transverse phase space, rather than enforcing the restriction $y = p_y = 0$. The emittance constraints are 
\begin{equation} \label{eq:matchedemittanceconstraintbody}
\epsilon_{n,x}, \epsilon_{n,y} \lesssim \frac{\gamma \lambda_1}{\pi} \frac{1 + \frac{K^2}{2}}{K^2} \rho^2
\end{equation}

\noindent for the matched offset Gaussian bunch,
\begin{equation} \label{eq:mismatchedemittanceconstraintbody}
\begin{split}
\epsilon_{n,x} &\lesssim \frac{\gamma \lambda_1}{\pi} \left(\frac{2}{5}\right)^{\frac{3}{4}} \left(\frac{1 + \frac{K^2}{2}}{K^2}\right)^{\frac{1}{2}} \rho^{\frac{3}{2}} \\
\epsilon_{n,y} &\lesssim \frac{\gamma \lambda_1}{\pi} \sqrt{\frac{2}{5}} \rho
\end{split}
\end{equation}

\noindent for the optimally mismatched offset Gaussian bunch, and

\begin{equation} \label{eq:annularsectoremittanceconstraintbody}
\begin{split}
\epsilon_{n,x} &\lesssim \frac{\gamma \lambda_1}{\pi} \frac{\sqrt{6} \Theta}{\pi} \rho \\
\epsilon_{n,y} &\lesssim \frac{\gamma \lambda_1}{\pi} \frac{1}{\sqrt{2}}\rho
\end{split}
\end{equation}

\noindent for the phase space annular sector bunch, where $\Theta$ is the total angle subtended by the annular sector in phase space.

As expected, the emittance constraint Eq.~(\ref{eq:matchedemittanceconstraintbody}) on the matched offset Gaussian bunch is the most stringent, followed by the constraint Eq.~(\ref{eq:mismatchedemittanceconstraintbody}) on the optimally mismatched offset Gaussian bunch, and finally by the constraint Eq.~(\ref{eq:annularsectoremittanceconstraintbody}) on the phase space annular sector bunch, which is the least stringent. The emittance constraints in all of these cases are, however, more stringent than those for an FEL \cite{kim2017}
\begin{equation}
\epsilon_{n,x}, \epsilon_{n,y} \lesssim \frac{\gamma \lambda_1}{\pi}  \frac{\bar{\beta}}{4 L_G}
\end{equation}

\noindent where $\bar{\beta} \approx L_G$ is the beta function.

Before we continue, there is a fourth phase space distribution family that bears mentioning: optimally mismatched, offset Gaussian bunches subject to the constraint $\epsilon_{n, x} = \epsilon_{n,y}$. The emittance constraint in this case is more stringent than Eq.~(\ref{eq:mismatchedemittanceconstraintbody}) for the optimally mismatched case where $\epsilon_{n,x} \neq \epsilon_{n,y}$, but less stringent than Eq.~(\ref{eq:matchedemittanceconstraintbody}) for the matched case. While we don't include a quantitative analysis of this fourth case due to its analytic complexity, numerically computing the emittance constraint and optimal parameters is relatively straightforward.

There are a number of caveats to the discussion in this section and the emittance constraints Eqs.~(\ref{eq:matchedemittanceconstraintbody})-(\ref{eq:annularsectoremittanceconstraintbody}). First, the betatron phases of the phase space distributions shown in Fig.~\ref{fig:phasespace} and the phase space distribution functions in Eq.~(\ref{eq:distributionfunctions}) are arbitrary. These distributions can be rotated about the origin in normalized $\gamma k_{\beta} x$-$p_x$ phase space without changing the emittance constraint. This is important because it may, for instance, be easier to inject a bunch into a plasma with an offset in $p_x$ (an underfocused beam centered in the ion channel but with mean velocity at an angle with respect to the ion channel axis) than with an offset in $x$ (an overfocused beam traveling parallel to---but with a transversely displaced centroid position from---the axis defining the center of the ion channel).

Second, we have assumed a monoenergetic bunch in this section. Both the $K$-spread due to nonzero emittance and that due to energy spread contribute to the resonant wavelength spread that ultimately damps the instability. A bunch with a correlation between $\gamma$ and $K$ due to nonzero dispersion could have a lower resonant wavelength spread than one with comparable emittance and no such correlation, and could lase despite exceeding the emittance constraints derived in this section. This is an important point, as linear $x$ dispersion can easily be added to a beam using a magnetic chicane, thereby relaxing the emittance constraints derived in this section, perhaps significantly so. This is in contrast to beam conditioning methods for FELs, which are more complicated \cite{sessler1992}.

Third, the emittance constraints Eqs.~(\ref{eq:matchedemittanceconstraintbody})-(\ref{eq:annularsectoremittanceconstraintbody}) are not binary: as these limits are approached, $\rho$ decreases gradually. This was discussed in Section~\ref{sec:numericalresults} with the gradual fall-off computed numerically and shown in Fig.~\ref{fig:no_detuning_warm}, although it bears mentioning that the bunch distributions in this section do not have Gaussian $K$ distributions as assumed in Section~\ref{sec:numericalresults}, and thus these curves are only approximately applicable in this section.

\section{\label{sec:discussion} Discussion}

\begin{figure*}[t]
    \centering
    \includegraphics[width=2\columnwidth]{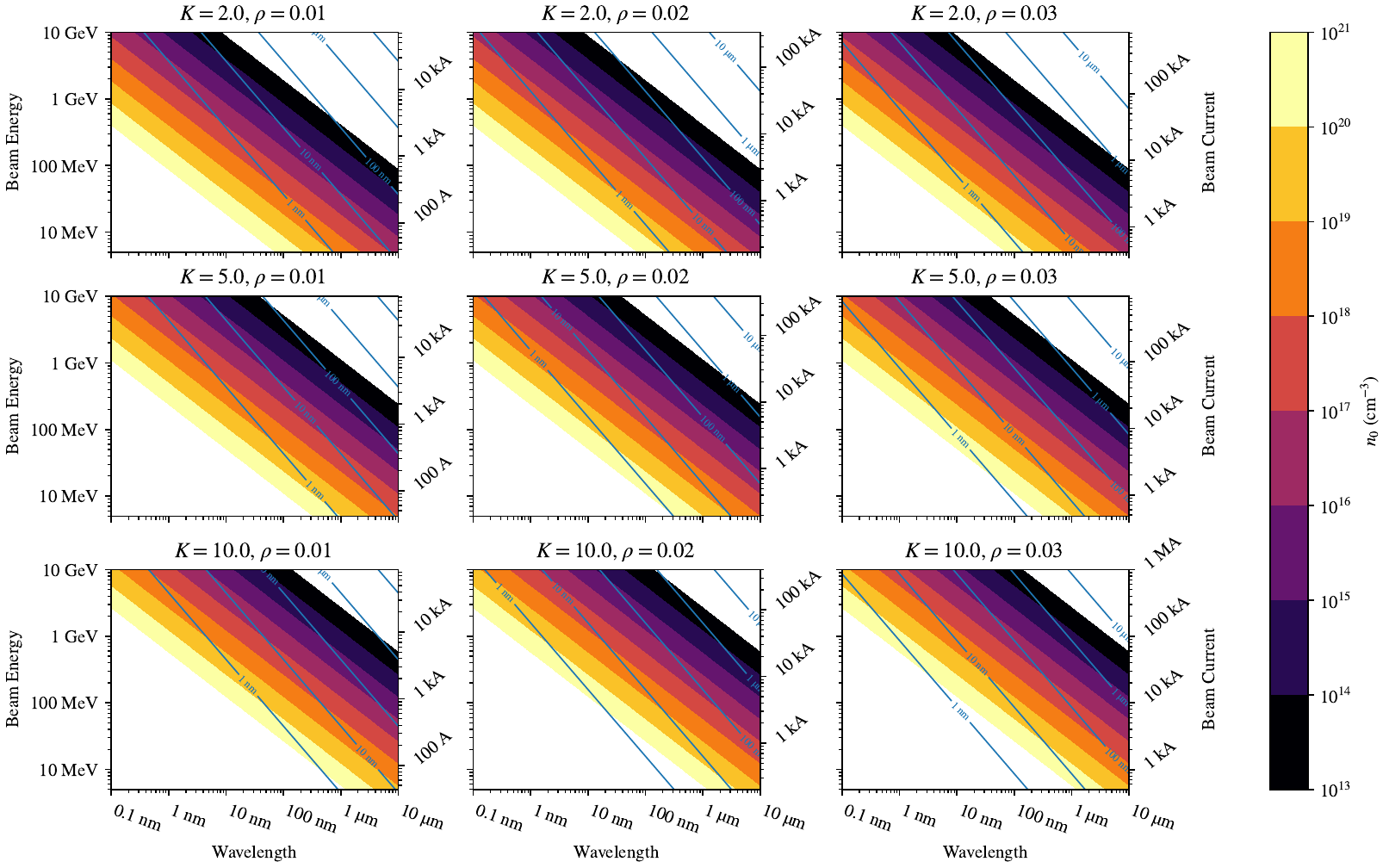}
    \caption{Electron beam and plasma parameter requirements for ICL lasing as a function of fundamental wavelength for various combinations of undulator parameter $K$ and 3D gain parameter $\rho$. $\rho$ is computed numerically using the method in Section~\ref{sec:numericalalgorithm}. Colors show plasma density required to satisfy the ICL resonance condition Eq.~(\ref{eq:resonancecondition}) for a given radiation wavelength ($x$ axis) and beam energy (left $y$ axis). The right $y$ axis shows the beam current required to achieve the given $\rho$ for the plot, which is independent of radiation wavelength. Blue contours show the maximum allowed $x$ normalized emittance that satisfies the lasing condition Eq.~(\ref{eq:mismatchedemittanceconstraintbody}). The $y$ normalized emittance constraint is not shown but is significantly less stringent.
    }
    \label{fig:parameter_space}
\end{figure*}

We have presented the most complete theory of the ICL to date, accounting for 3D effects, diffraction, transverse radiation profile, frequency detuning, betatron phase detuning, and nonzero emittance and energy spread. Additionally, we have developed a novel numerical algorithm for solving the ICL equations, which we have used to compute results for a range of parameters. Finally, we have discussed in detail the nature of the ICL's beam quality requirements, deriving emittance constraints and the optimal Gaussian beam distribution for lasing in an ICL. Still, there are many topics we have left to future work. We focused solely on the planar off-axis configuration of the ICL, leaving out discussion of circular and elliptical configurations. Longitudinal space-charge effects---which are important in some regimes---are ignored in our theory. While we include some discussion of the microscopic physics of the ICL and the effect of detuning on ICL performance, we believe both of these topics warrant further investigation. What we believe to be the two topics of highest priority for future research, however, are running 3D PIC simulations of physical ICL configurations, and investigating how electron beam dispersion could be leveraged to relax the emittance constraints of the ICL.

As discussed in Section~\ref{sec:introduction}, compared to the FEL, the ICL trades a less stringent energy spread constraint for a more stringent emittance constraint. In Section~\ref{sec:phasespace} we showed how the right transverse phase space distribution can somewhat relax these constraints for Gaussian bunches. These constraints could be relaxed further by using a bunch with a linear $x$ dispersion or other form of beam conditioning \cite{penn2007}, although we did not examine this case in detail. 

Figure~\ref{fig:parameter_space} shows the parameters needed to achieve ICL lasing as a function of the fundamental wavelength for different combinations of undulator parameter $K$ and 3D gain parameter $\rho$. The plots in Fig.~\ref{fig:parameter_space} assume an optimally mismatched Gaussian beam, as depicted in Fig.~\ref{fig:phasespace}(b). We make the general observation that multi-kiloampere beams with a low emittance of at most 10 nm-rad are required to reach a fundamental wavelength of 10 nm, representing the EUV scale of interest to the advanced chip manufacturing industry. Meanwhile, wavelengths of order 100 nm relevant for ARPES could be reached with 100 nm-rad emittance beams.

While these parameters are extreme, they appear to be within reach of plasma-injected and accelerated beams. Multiple plasma injection schemes have been demonstrated in PIC simulations to generate high peak current bunches with emittance of order 10 nm-rad \cite{li2013,manahan2017,habib2023,xiang2024,tomassini2025}, and experimental demonstrations are not far behind~\cite{wang2021,barber2025}. If an appropriate low-emittance, high-current plasma-injected bunch could be paired with a strong drive bunch to produce the ion column, the first experimental demonstration of the short-wavelength ICL could be achieved in the immediate future. SLAC National Accelerator Laboratory's Facility for Advanced Accelerator Experimental Tests II (FACET-II) \cite{yakimenko2019,joshi2018} is particularly well suited to carry out such an experiment, though with some effort, a hybrid laser-plasma wakefield accelerator (L/PWFA) \cite{hidding2010,foerster2026} may also be capable of producing a drive and trailing beam pair with appropriate parameters to demonstrate and study the unique and promising high-brightness coherent radiation source that is the ICL.

\begin{acknowledgments}

We graciously thank Joanna Piper Morgan (LLNL) for her assistance running simulations.

This work is supported by the National Science Foundation under grant NSF-2047083 and the US Department of Energy under grant DE-SC0017906.

This research used resources of the National Energy Research Scientific Computing Center (NERSC), a Department of Energy User Facility (project m306-2024).


\end{acknowledgments}

\appendix

\section{\label{app:periodaveraging} Period Averaging}

In this Appendix we define the period average and derive Eq.~(\ref{eq:fancyperiodaverage}) which computes a useful period average appearing in both the pendulum and field equation derivations. During the course of this derivation we define a new function $\mathcal{W}_h(\bm{x}_{\perp})$ which we plot and whose properties we discuss. 

For some function $f(z)$, the period average is defined as the average of the quantity over one betatron period

\begin{equation} \label{eq:periodaverage}
\overline{f(z)} \equiv \frac{1}{\lambda_{\beta,r}} \int_{z - \frac{\lambda_{\beta,r}}{2}}^{z + \frac{\lambda_{\beta,r}}{2}} dz\, f(z).
\end{equation}

\noindent We make heavy use of the fact that if a function $g(z)$ is slowly varying over one betatron period ($g'(z) \ll k_{\beta,r} g(z)$), then for any $h(z)$, $\overline{g(z) h(z)} \simeq g(z) \overline{h(z)}$.

We begin the derivation of Eq.~(\ref{eq:fancyperiodaverage}) by noting a number of identities. First we note that $\forall n \in \mathbb{Z}$,

\begin{equation} \label{eq:periodaverageexp}
\overline{e^{i n k_{\beta,r} z}} = \delta_{n,0}
\end{equation}

\noindent which can be shown by directly computing the integral in Eq.~(\ref{eq:periodaverage}). Second we note two Jacobi-Anger identities 

\begin{equation}  \label{eq:jacobianger}
\begin{split}
e^{i x \sin(\phi)} &= \sum_{n \in \mathbb{Z}} J_n(x) e^{i n \phi} \\
e^{-i x \cos(\phi)} &= \sum_{n \in \mathbb{Z}} (-i)^n J_n(x) e^{\pm i n \phi} \\
\end{split}
\end{equation}

\noindent where the second identity is true for both $\pm = +$ and $\pm = -$. Third we note that from Eqs.~(\ref{eq:longitudinalmotion}) and (\ref{eq:pendulumquantities}),

\begin{equation} \label{eq:ponderomotiveidentity}
\begin{split}
\nu k_{1,r} \zeta_j &= \nu \theta_j - \Delta \nu k_{\beta,r} z \\
&- h k_{\beta,r} z + \nu \xi_j \sin(2 (k_{\beta,r} z + \vartheta_j)) 
\end{split}
\end{equation}

\noindent where $\xi_j \equiv k_{1,r} K_j^2 / (8 \gamma_j^2 k_{\beta,j})$. The reference value of this new quantity is $\xi_r = k_{1,r} K_r^2 / (8 \gamma_r^2 k_{\beta,r}) = K_r^2 / (2(2 + K_r^2))$. Using Eqs.~(\ref{eq:periodaverageexp})-(\ref{eq:ponderomotiveidentity}) as well as the definition Eq.~(\ref{eq:cndefinition}) of the $\mathcal{C}_n$ functions discussed in Appendix~\ref{app:cn} and the integral Eq.~(\ref{eq:cnintegral1}), we compute the following (where $l \in \mathbb{Z}$)

\begin{figure*}[t]
    \centering
    \includegraphics[width=\columnwidth]{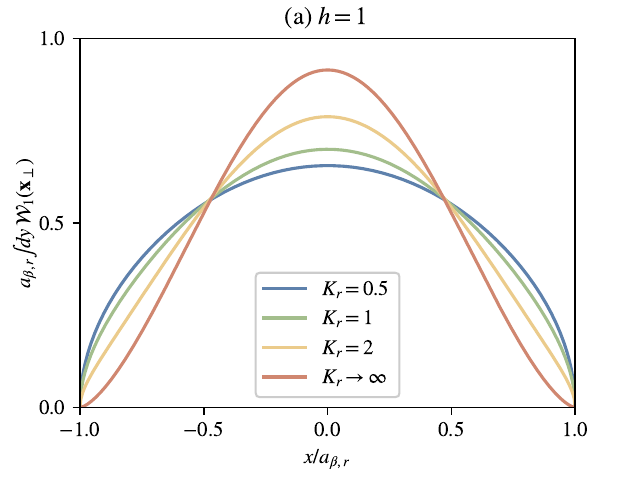}
    \includegraphics[width=\columnwidth]{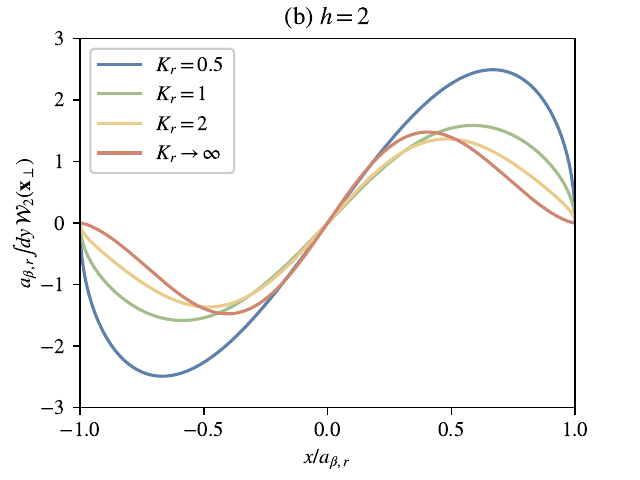}
    \includegraphics[width=\columnwidth]{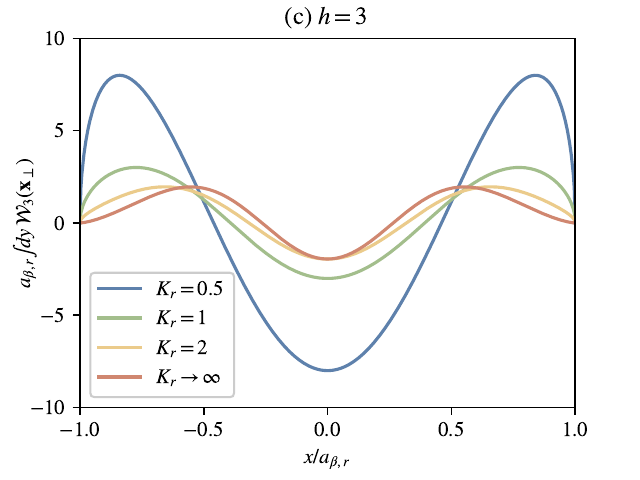}
   \includegraphics[width=\columnwidth]{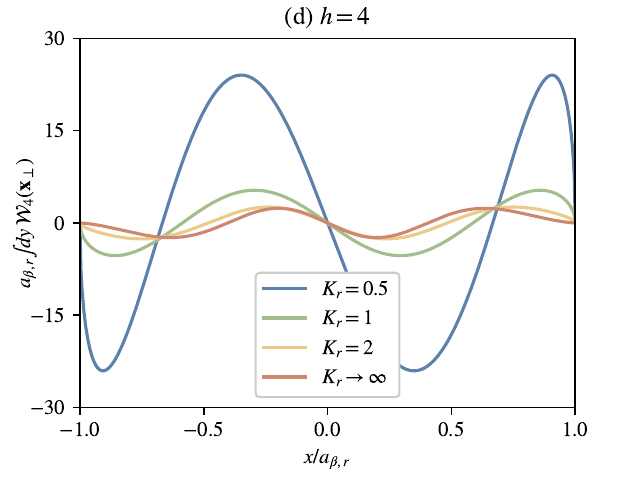}
    \caption{Plots of $a_{\beta,r} \int dy \, \mathcal{W}_h(\bm{x}_{\perp})$ for various values of $K_r$ and $h$.}
   \label{fig:w}
\end{figure*}

\begin{widetext}

\begin{equation} \label{eq:bigperiodaverage}
\begin{split}
&\overline{\delta^2(\bm{x}_{\perp} - \bm{x}_{\perp,j}) e^{i l (k_{\beta,r} z + \vartheta_j)} e^{\pm i \nu k_{1,r} \zeta_j} e^{\pm i \Delta \nu k_{\beta,r} z}} \\
&=\overline{\delta(x - a_{\beta,j} \cos(k_{\beta,r} z + \vartheta_j)) \delta(y) e^{i l (k_{\beta,r} z + \vartheta_j)} e^{\pm i \left(\nu \theta_j - h k_{\beta,r} z + \nu \xi_j \sin(2 (k_{\beta,r} z + \vartheta_j)) \right)}} \\
&=e^{\pm i \nu \theta_j} e^{\pm i h \vartheta_j} \delta(y) \int \frac{dk_x}{2 \pi} e^{i k_x x} \overline{e^{-i k_x  a_{\beta,j} \cos(k_{\beta,r} z + \vartheta_j)} e^{\pm i \nu \xi_j \sin(2(k_{\beta,r} z + \vartheta_j))} e^{i (l \mp h) (k_{\beta,r} z + \vartheta_j)}} \\
&=e^{\pm i \nu \theta_j} e^{\pm i h \vartheta_j} \delta(y) \int \frac{dk_x}{2 \pi} e^{i k_x x} \sum_{n,m \in \mathbb{Z}} (-i)^n J_n(k_x a_{\beta,j}) J_m(\nu \xi_j) \overline{e^{i (l \mp h \pm n \pm 2m) (k_{\beta,r} z + \vartheta_j)}} \\
&=e^{\pm i \nu \theta_j} e^{\pm i h \vartheta_j}  \delta(y) \int \frac{dk_x}{2 \pi} e^{i k_x x} \sum_{\substack{n \in \mathbb{Z}\\ h - n \mp l \; \mathrm{even}}} (-i)^n J_n(k_x a_{\beta,j}) J_{\frac{h - n \mp l}{2}}(\nu \xi_j) \\
&=e^{\pm i \nu \theta_j} e^{\pm i h \vartheta_j}  \delta(y) \sum_{\substack{n \in \mathbb{Z}\\ h - n \mp l \; \mathrm{even}}} J_{\frac{h - n \mp l}{2}}(\nu \xi_j) \left[\frac{1}{a_{\beta,j}} \int \frac{d \kappa}{2 \pi i^n} J_n(\kappa) e^{i \kappa \left(\frac{x}{a_{\beta,j}}\right)}\right]\\
&= e^{\pm i \nu \theta_j} e^{\pm i h \vartheta_j}   \sum_{\substack{n \in \mathbb{Z}\\ h - n \mp l \; \mathrm{even}}} J_{\frac{h - n \mp l}{2}}(\nu \xi_j) \frac{1}{a_{\beta,j}} \mathcal{C}_n\left(\frac{x}{a_{\beta,j}}\right) \delta(y) \\ 
&\simeq e^{\pm i \nu \theta_j} e^{\pm i h \vartheta_j}   \sum_{\substack{n \in \mathbb{Z}\\ h - n \mp l \; \mathrm{even}}} J_{\frac{h - n \mp l}{2}}(h \xi_r) \frac{1}{a_{\beta,r}} \mathcal{C}_n\left(\frac{x}{a_{\beta,r} }\right) \delta(y). \\ 
\end{split}
\end{equation}
\end{widetext}

\noindent In Eq.~(\ref{eq:bigperiodaverage}) we use a number of assumptions justified in the main text. We assume $\theta_j$ and $\vartheta_j$ are slowly varying over one betatron period, which allows $e^{\pm i \nu \theta_j}$ and $e^{\pm i h \vartheta_j}$ to be taken out of the period average in the third line. We assume $\eta, \delta, \Delta \nu \ll 1$ and then ignore subleading order terms, which allows us to replace the particle $j$ values of $\xi$ and $a_{\beta}$ with the reference particle values in the last line. The main result of this section follows directly from Eq.~(\ref{eq:bigperiodaverage}) and is

\begin{equation} \label{eq:fancyperiodaverage}
\begin{split}
&\overline{\delta^2(\bm{x}_{\perp} - \bm{x}_{\perp,j}) \sin(k_{\beta,r} z + \vartheta_j) e^{\pm i \nu k_{1,r} \zeta_j} e^{\pm i \Delta \nu k_{\beta,r} z} } \\
&\simeq \mp \frac{i}{2} e^{\pm i \nu \theta_j}e^{\pm i h \vartheta_j} [\mathrm{JJ}]_h \mathcal{W}_h(\bm{x}_{\perp})
\end{split}
\end{equation}

\noindent where we have defined

\begin{equation} \label{eq:wdefinition}
\mathcal{W}_h(\bm{x}_{\perp}) = \sum_{\substack{n \in \mathbb{Z}\\ h - n \; \mathrm{odd}}}  \frac{[\mathrm{JJ}]_{h - n}}{[\mathrm{JJ}]_h} \frac{1}{a_{\beta,r}} \mathcal{C}_n\left(\frac{x}{a_{\beta,r}}\right) \delta(y)
\end{equation}

\noindent and

\begin{equation} \label{eq:jjdefinition}
[\mathrm{JJ}]_n = J_{\frac{n - 1}{2}}(h \xi_r) - J_{\frac{n + 1}{2}}(h \xi_r)
\end{equation}

\noindent where $n \in \mathbb{Z}$. Plots of the $x$ part of $\mathcal{W}_h(\bm{x}_{\perp})$ for $h = 1\dots4$ are shown in Fig.~\ref{fig:w}. $\mathcal{W}_h(\bm{x}_{\perp})$ obeys the normalization condition 

\begin{equation}
\int d^2 \bm{x}_{\perp} \mathcal{W}_h(\bm{x}_{\perp}) = \begin{cases} 0 & h \; \mathrm{even} \\ 1 & h \; \mathrm{odd} \end{cases}.
\end{equation}

\section{\label{app:cn} $\mathcal{C}_n$ Functions}

For integer $n$ we define the function $\mathcal{C}_n$ as

\begin{equation} \label{eq:cndefinition}
\mathcal{C}_n(x) \equiv \begin{cases}
\frac{T_n(x)}{\pi \sqrt{1 - x^2}} & |x| < 1 \\
0 & |x| \geq 1.
\end{cases}
\end{equation}

\begin{figure}[b]
    \includegraphics[width=\columnwidth]{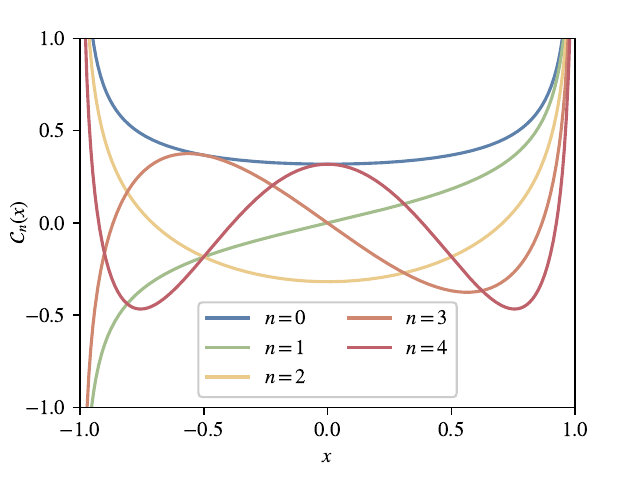}
    \caption{Plot of the first five $\mathcal{C}_n$ functions discussed in Appendix~\ref{app:cn}.}
    \label{fig:chebyshev}
\end{figure}

\noindent The first five $\mathcal{C}_n$ are plotted in Fig.~\ref{fig:chebyshev}. $\mathcal{C}_n$ are orthogonal functions with respect to the weight $w(x) = \sqrt{1 - x^2}$. These functions satisfy the identity $\mathcal{C}_{-n}(x) = \mathcal{C}_n(x)$. $\mathcal{C}_n$ is an even function for even $n$ and is an odd function for odd $n$. We note

\begin{equation}
\begin{split}
&\int_{-\infty}^{\infty} dx \, \mathcal{C}_n(x) = \int_{-1}^1 dx \, \mathcal{C}_n(x) \\
&= \frac{1}{\pi} \int_{-1}^1 dx\, \frac{T_n(x) T_0(x)}{\sqrt{1 - x^2}} = \delta_{n,0}
\end{split}
\end{equation}

\noindent where the last integral was evaluated using the Chebyshev polynomial orthogonality identity \cite[\href{https://dlmf.nist.gov/18.3.T1}{Table 18.3.1}]{NIST:DLMF}. We now evaluate the integral

\begin{equation} \label{eq:cnintegral1}
\begin{split}
&\int_{-\infty}^{\infty} \frac{d\kappa}{2 \pi i^n} e^{i \kappa x} J_n(\kappa) \\
&= \int_{-\infty}^{\infty} \frac{d\kappa}{2 \pi i^n} e^{i \kappa x} \left[\frac{i^{-n}}{\pi} \int_0^{\pi} d\theta \, e^{i \kappa \cos(\theta)} \cos(n \theta)\right] \\
&= \frac{(-1)^n}{\pi} \int_0^{\pi} d\theta \, \cos(n \theta) \int_{-\infty}^{\infty} \frac{d\kappa}{2 \pi} e^{i \kappa (x + \cos(\theta))} \\
&= \frac{(-1)^n}{\pi} \int_0^{\pi} d\theta \, \cos(n \theta) \delta(x + \cos(\theta)) \\
&= \frac{(-1)^n}{\pi} \int_{-1}^1 dy \, \frac{\cos(n \cos^{-1}(y)) \delta(x + y)}{\sqrt{1 - y^2}} \\
&= \begin{cases}
\frac{(-1)^n \cos(n (\cos^{-1}(x) + \pi))}{\pi \sqrt{1 - x^2}} & |x| < 1 \\
0 & |x| \geq 1
\end{cases} \\
&= \begin{cases}
\frac{\cos(n \cos^{-1}(x))}{\pi \sqrt{1 - x^2}} & |x| < 1 \\
0 & |x| \geq 1
\end{cases} \\
&= \begin{cases}
\frac{T_n(x)}{\pi \sqrt{1 - x^2}} & |x| < 1 \\
0 & |x| \geq 1
\end{cases} \\
&= \mathcal{C}_n(x)
\end{split} 
\end{equation}

\noindent where we have made use of  \cite[\href{https://dlmf.nist.gov/10.9.E2}{(10.9.2)},\href{https://dlmf.nist.gov/18.5.E1}{(18.5.1)}]{NIST:DLMF}.

\section{
\label{app:slowlyvaryingenvelopeapproximation} Slowly Varying Envelope Approximation}

The field $a(\bm{x}_{\perp}, \zeta, z)$ can be written as a Fourier transform

\begin{equation} \label{eq:fieldfouriertransform}
a(\bm{x}_{\perp}, \zeta, z) = \int d \nu \, e^{i \nu k_{1,r} \zeta} \tilde{a}(\bm{x}_{\perp}, \nu, z)
\end{equation}

\noindent where $\nu \equiv k / k_{1,r}$ is the spatial frequency normalized to the fundamental. Because $a(\bm{x}_{\perp}, \zeta, z)$ is a physical field and cannot have an imaginary component, the spectrum must satisfy $\tilde{a}^{*}(\bm{x}_{\perp}, -\nu, z) = \tilde{a}(\bm{x}_{\perp}, \nu, z)$. This means that 

\begin{equation} \label{eq:fieldfouriertransformreal}
a(\bm{x}_{\perp}, \zeta, z) = \int_0^{\infty} d \nu \, e^{i \nu k_{1,r} \zeta} \tilde{a}(\bm{x}_{\perp}, \nu, z) + \mathrm{c.c.}.
\end{equation}

\noindent One of the main assumptions used in this paper is that the radiation field spectrum consists of distinct peaks at frequencies that are integer harmonics of the fundamental. With this assumption we define

\begin{equation} \label{eq:mathcaladefinition}
\mathcal{A}_h(\bm{x}_{\perp}, \nu, z) \equiv \begin{cases}
e^{-i \Delta \nu k_{\beta,r} z} \tilde{a}(\bm{x}_{\perp}, \nu, z) & \nu \approx h \\
0 & \nu \not \approx h
\end{cases}
\end{equation}

\noindent where $\Delta \nu \equiv \nu -h$ and $|\Delta \nu| \ll 1$. $\mathcal{A}_h(\bm{x}_{\perp}, \nu, z)$ contains only the $h$-th peak of the spectrum of the field. Using Eqs.~(\ref{eq:fieldfouriertransformreal}) and (\ref{eq:mathcaladefinition}) and defining the ponderomotive phase coordinate $\theta(\zeta, z) \equiv k_{\beta,r} z + k_{1,r} \zeta$ we rewrite the field as

\begin{equation} \label{eq:slowlyvaryingsum}
\begin{split}
& a(\bm{x}_{\perp}, \zeta, z) \\
&= \int_0^{\infty} d \nu \,  e^{i \nu k_{1,r} \zeta} \tilde{a}(\bm{x}_{\perp}, \nu, z) + \mathrm{c.c.} \\
&= \int_0^{\infty} d \nu \, e^{i \nu k_{1,r} \zeta} \left[\sum_{h \in \mathbb{Z}} e^{i \Delta \nu k_{\beta,r} z}  \mathcal{A}_h(\bm{x}_{\perp}, \nu, z)\right] + \mathrm{c.c.} \\
&= \sum_{h \in \mathbb{N}^{+}} \left[\int_{\nu \approx h} d \nu \, e^{i \Delta \nu \theta} \mathcal{A}_h(\bm{x}_{\perp}, \nu, z) \right] e^{i h k_{1,r} \zeta} + \mathrm{c.c.} \\
&= \sum_{h \in \mathbb{N}^{+}} a_{\mathrm{sv},h}(\bm{x}_{\perp}, \theta, z) e^{i h k_{1,r} \zeta} + \mathrm{c.c.} \\
\end{split}
\end{equation}

\noindent which is a sum of slowly varying (independent of $\zeta$) envelope functions 

\begin{equation}
a_{\mathrm{sv},h}(\bm{x}_{\perp}, \theta, z) \equiv \int_{\nu \approx h} d \nu \, e^{i \Delta \nu \theta} \mathcal{A}_h(\bm{x}_{\perp}, \nu, z)
\end{equation}

\noindent multiplying carrier waves at integer harmonics of the fundamental---which is the slowly varying envelope approximation. Thus the assumption that the field spectrum consists of distinct narrow peaks at integer harmonics of the fundamental directly leads to the SVEA.

\section{\label{app:energy} Energy Conservation of the Maxwell-Klimontovich Equations}

We show that the Maxwell-Klimontovich equations in Eq.~(\ref{eq:mkharmonic}) conserve energy. The energy in the electromagnetic field per unit transverse area is given by

\begin{equation}
\begin{split}
&\mathcal{U}_{\perp, \mathrm{EM}}(\bm{x}_{\perp}, z) = \frac{I_A}{8 \pi} \int d\zeta \left(|\bm{E}|^2 + |\bm{B}|^2\right) \\
&\simeq \frac{I_A}{4 \pi} \int d\zeta \left|\frac{\partial}{\partial \zeta} a(\bm{x}_{\perp}, \zeta, z)\right|^2 \\
&= \frac{I_A}{4 \pi}  \int d\zeta \left(\int d\nu \, i \nu k_{1,r} \tilde{a}(\bm{x}_{\perp}, \nu, z) e^{i \nu k_{1,r} \zeta}\right) \\
&\times \left(\int d\nu' \,  (-i) \nu' k_{1,r} \tilde{a}^{*}(\bm{x}_{\perp}, \nu', z) e^{-i \nu' k_{1,r} \zeta}\right) \\
&= k_{1,r} I_A \int_0^{\infty} d\nu \, \nu^2 \left|\tilde{a}(\bm{x}_{\perp}, \nu, z)\right|^2. \\
\end{split}
\end{equation}

\noindent The total energy in the electromagnetic field is

\begin{equation}
\begin{split}
\mathcal{E}_{\mathrm{EM}}(z) &= \int d^2 \bm{x}_{\perp} \mathcal{U}_{\perp, \mathrm{EM}}(\bm{x}_{\perp}, z) \\
&= k_{1,r} I_A \int_0^{\infty} d\nu \,  \nu^2 \int d^2 \bm{x}_{\perp}\left|\tilde{a}(\bm{x}_{\perp}, \nu, z)\right|^2.
\end{split}
\end{equation}

\noindent The change in this quantity with respect to $z$ is

\begin{widetext}

\begin{equation}
\begin{split}
&\frac{d \mathcal{E}_{\mathrm{EM}}}{d z} = k_{1,r} I_A \int_0^{\infty} d\nu \nu^2 \int d^2 \bm{x}_{\perp} \tilde{a}^{*}(\bm{x}_{\perp}, \nu, z) \frac{\partial}{\partial z} \tilde{a}(\bm{x}_{\perp}, \nu, z) + \mathrm{c.c.} \\
&= k_{1,r} I_A \sum_{h \in \mathbb{N}^{+}} \int_{\nu \approx h} d\nu \, \nu^2 \int d^2 \bm{x}_{\perp} \mathcal{A}_h^{*}(\bm{x}_{\perp}, \nu, z) \left[\frac{\partial}{\partial z} + i \Delta \nu k_{\beta,r}\right] \mathcal{A}_h(\bm{x}_{\perp}, \nu, z) + \mathrm{c.c.} \\
&= k_{1,r} I_A \sum_{h \in \mathbb{N}^{+}} \int_{\nu \approx h} d\nu \, \nu^2 \int d^2 \bm{x}_{\perp} \mathcal{A}_h^{*}(\bm{x}_{\perp}, \nu, z) \left[\frac{\partial}{\partial z} + i \Delta \nu k_{\beta,r} - \frac{i}{2 \nu k_{1,r}} \bm{\nabla}_{\perp}^2\right] \mathcal{A}_h(\bm{x}_{\perp}, \nu, z) + \mathrm{c.c.} \\
&= -\sum_{h \in \mathbb{N}^{+}} \int_{\nu \approx h} d\nu \, \frac{\nu k_{1,r} K_r [\mathrm{JJ}]_h N_{\lambda_{1,r}}}{4 \pi \gamma_r} \int d^2 \bm{x}_{\perp} \mathcal{A}_h^{*}(\bm{x}_{\perp}, \nu, z)  \mathcal{W}_h(\bm{x}_{\perp}) \iiiint d\eta d\delta d\vartheta d\theta \, e^{-i \nu \theta} e^{-i h \vartheta} f(\eta, \delta, \vartheta, \theta, z) + \mathrm{c.c.} \\
&= -\frac{N_{\lambda_{1,r}} \gamma_r}{2 \pi} \iiiint d\eta d\delta d\vartheta d\theta \left[\sum_{h \in \mathbb{N}^{+}} \int_{\nu \approx h} d\nu \, \frac{\nu k_{1,r} K_r [\mathrm{JJ}]_h}{2 \gamma_r^2} e^{i h \vartheta} e^{i \nu \theta} \int d^2\bm{x}_{\perp} \mathcal{W}_h(\bm{x}_{\perp}) \mathcal{A}_h(\bm{x}_{\perp}, \nu, z) + \mathrm{c.c.} \right] f(\eta, \delta, \vartheta, \theta, z) \\
&= \frac{N_{\lambda_{1,r}} \gamma_r}{2 \pi} \iiiint d\eta d\delta d\vartheta d\theta \, \eta \left[\sum_{h \in \mathbb{N}^{+}} \int_{\nu \approx h} d\nu \, \frac{\nu k_{1,r} K_r [\mathrm{JJ}]_h}{2 \gamma_r^2} e^{i h \vartheta} e^{i \nu \theta}\int d^2\bm{x}_{\perp} \mathcal{W}_h(\bm{x}_{\perp}) \mathcal{A}_h(\bm{x}_{\perp}, \nu, z) + \mathrm{c.c.} \right] \frac{\partial}{\partial \eta} f(\eta, \delta, \vartheta, \theta, z) \\
&= \frac{N_{\lambda_{1,r}} \gamma_r}{2 \pi} \iiiint d\eta d\delta d\vartheta d\theta \, \eta \eta' \frac{\partial}{\partial \eta} f(\eta, \delta, \vartheta, \theta, z) \\
&= \frac{N_{\lambda_{1,r}} \gamma_r}{2 \pi} \iiiint d\eta d\delta d\vartheta d\theta \, \eta \left[\eta' \left(\frac{\partial}{\partial \eta} + \frac{1 + K_r^2}{2 K_r^2} \frac{\partial}{\partial \delta}\right) + 2 k_{\beta,r} \left(\eta - \frac{K_r^2}{2 + K_r^2} \delta\right) \frac{\partial}{\partial \theta} - \frac{1}{2} k_{\beta,r} \eta \frac{\partial}{\partial \vartheta}\right] f(\eta, \delta, \vartheta, \theta, z) \\
&= -\frac{N_{\lambda_{1,r}} \gamma_r}{2 \pi} \iiiint d\eta d\delta d\vartheta d\theta \, \eta \frac{\partial}{\partial z} f(\eta, \delta, \vartheta, \theta, z) \\
&= -\frac{d}{dz} \sum_{j = 1}^{N_e} \gamma_j
\end{split}
\end{equation}

\end{widetext}

\noindent which is exactly the negative of the change in the kinetic energy of the electrons in the bunch. Thus the sum of the energy in the field and the kinetic energy in the particles in the bunch is constant, as expected, and so energy is conserved.

\section{\label{app:vankampennormalmodes} van Kampen Normal Mode Expansion}

In this Appendix we follow existing 3D FEL theory by applying a van Kampen normal mode expansion \cite{huang2001,kim2017} to solve the initial value problem of Eq.~(\ref{eq:simplifiedmv}). For this subsection only we let the dependence of quantities on $\nu$, $h$, $\hat{\eta}$, $\hat{\delta}$, and $\hat{\bm{x}}_{\perp}$ be implicit in our notation. Further, in this section we assume $\hat{f}_0$ is independent of $\hat{z}$. Additionally we define 
\begin{equation} \label{eq:variousdefinitions}
\begin{split}
\mathcal{D} &\equiv \hat{\Delta \nu} - \mathcal{F}_D^{-1} \hat{\bm{\nabla}}_{\perp}^2  \\
\mathcal{G} &\equiv \int d\vartheta \left(\frac{\partial \hat{f}_0}{\partial \hat{\eta}} + \frac{1 + K_r^2}{2 K_r^2} \frac{\partial \hat{f}_0}{\partial \hat{\delta}}\right) \\
\Xi &\equiv \left(\nu - \frac{h}{4}\right)\hat{\eta} - \nu \frac{K_r^2}{2 + K_r^2} \hat{\delta} \\
R_J &\equiv \frac{[\mathrm{JJ}]_h^2}{[\mathrm{JJ}]_1^2}
\end{split}
\end{equation}

\noindent which will simplify notation moving forward. We define a Hilbert space with state vectors $\Psi(\hat{z}) \equiv \begin{pmatrix}
\hat{\mathcal{A}}(\hat{z}) & \mathcal{F}(\hat{z})
\end{pmatrix}^{\intercal}$ and inner product 
\begin{equation} \label{eq:innerproduct}
\langle \Psi_1, \Psi_2 \rangle \equiv \int d^2 \hat{\bm{x}}_{\perp} \hat{\mathcal{A}}_1\hat{\mathcal{A}}_2+ \int d\hat{\eta} \,d\hat{\delta} \,\mathcal{F}_1\mathcal{F}_2.    
\end{equation}

\noindent On this space, we define the linear operator $\bm{\mathcal{M}}$ by its action on an arbitrary state vector $\Psi(\hat{z})$:
\begin{equation}
\bm{\mathcal{M}}  \Psi(\hat{z}) \equiv \begin{pmatrix}
\hat{\mathcal{D}} \hat{\mathcal{A}}(\hat{z}) - \frac{i R_J}{2 \nu \mathcal{I}} \hat{\mathcal{W}} \int d\hat{\eta} \,d\hat{\delta} \, \mathcal{F}(\hat{z}) \\
-2 \pi i \nu \mathcal{G} \int d^2\hat{\bm{x}}_{\perp} \hat{\mathcal{W}} \hat{\mathcal{A}}(\hat{z}) + \Xi \mathcal{F}(\hat{z})
\end{pmatrix}.
\end{equation}

\noindent We have chosen $\bm{\mathcal{M}}$ so that Eq.~(\ref{eq:simplifiedmv}) is equivalent to 
\begin{equation} \label{eq:schrodinger}
i \frac{\partial}{\partial \hat{z}} \Psi(\hat{z})  =  \bm{\mathcal{M}} \Psi(\hat{z})
\end{equation}

\noindent which is of the same form as the Schr\"{o}dinger equation. Unlike the Hamiltonian operator in the Schr\"{o}dinger equation, however, $\bm{\mathcal{M}}$ is not necessarily Hermitian. The action of $\bm{\mathcal{M}}^{\dagger}$ on a state vector can be computed using the definition of the Hermitian conjugate $\langle \Psi_1, \bm{\mathcal{M}} \Psi_2 \rangle = \langle \bm{\mathcal{M}}^{\dagger} \Psi_1, \Psi_2 \rangle$ and Eq.~(\ref{eq:innerproduct}), which together yield
\begin{equation}
\bm{\mathcal{M}}^{\dagger}  \Psi(\hat{z}) \equiv \begin{pmatrix}
\hat{\mathcal{D}} \hat{\mathcal{A}}(\hat{z}) - 2 \pi i \nu  R_J \hat{\mathcal{W}} \int d\hat{\eta} \, d\hat{\delta} \, \mathcal{G} \mathcal{F}(\hat{z}) \\
- \frac{i }{2 \nu \mathcal{I}} \int d^2 \hat{\bm{x}}_{\perp} \hat{\mathcal{W}} \hat{\mathcal{A}}(\hat{z}) + \Xi \mathcal{F}(\hat{z})
\end{pmatrix}.
\end{equation}

We now introduce the normal modes $\Psi_{\ell}(\hat{z})$, which are solutions to Eq.~(\ref{eq:schrodinger}) of the form $\Psi_{\ell}(\hat{z}) = e^{-i \mu_{\ell} \hat{z}} \Psi_{\ell}$, where $\Psi_{\ell} = \begin{pmatrix} \hat{\mathcal{A}}_{\ell} & \mathcal{F}_{\ell} \end{pmatrix}^{\intercal}$ is independent of $\hat{z}$. Equation~(\ref{eq:schrodinger}) is now simply the eigenvalue equation $\bm{\mathcal{M}} \Psi_{\ell} = \mu_{\ell} \Psi_{\ell}$. We define $\Psi_{\ell}^{\dagger}$ and $\mu_{\ell}^{\dagger}$ through the adjoint eigenvalue equation $\bm{\mathcal{M}}^{\dagger} \Psi_{\ell}^{\dagger} = \mu_{\ell}^{\dagger} \Psi_{\ell}^{\dagger}$. Writing out the two eigenvalue equations for each of the two vector components gives four equations, two of which can be solved to give
\begin{equation} \label{eq:fs}
\begin{split}
\mathcal{F}_{\ell} &= -2 \pi i \nu \frac{\mathcal{G}}{\mu_{\ell} - \Xi} \int d^2\hat{\bm{x}}_{\perp} \hat{\mathcal{W}} \hat{\mathcal{A}}_{\ell} \\
\mathcal{F}_{\ell}^{\dagger} &= -\frac{i }{2 \nu \mathcal{I}} \frac{1}{\mu_{\ell}^{\dagger} - \Xi} \int d^2\hat{\bm{x}}_{\perp} \hat{\mathcal{W}} \hat{\mathcal{A}}_{\ell}^{\dagger}
\end{split}
\end{equation}

\noindent which can be substituted into the other two equations to get the dispersion relations for $\hat{\mathcal{A}}_{\ell}$ and $\hat{\mathcal{A}}_{\ell}^{\dagger}$:
\begin{equation} \label{eq:as}
\begin{split}
\left[\mu_{\ell} - \hat{\mathcal{D}}\right] \hat{\mathcal{A}}_{\ell} - \pi \mathcal{V}(\mu_{\ell}) \hat{\mathcal{W}} \int d^2 \bm{x}_{\perp} \hat{\mathcal{W}} \hat{\mathcal{A}}_{\ell} &= 0 \\
\left[\mu_{\ell}^{\dagger} - \hat{\mathcal{D}}\right] \hat{\mathcal{A}}_{\ell}^{\dagger} - \pi \mathcal{V}(\mu_{\ell}^{\dagger}) \hat{\mathcal{W}} \int d^2 \bm{x}_{\perp} \hat{\mathcal{W}} \hat{\mathcal{A}}_{\ell}^{\dagger} &= 0
\end{split}
\end{equation}

\noindent where
\begin{equation} \label{eq:vdefinition}
\mathcal{V}(\mu) \equiv -\frac{R_J}{\mathcal{I}} \iint d\hat{\eta} \,d\hat{\delta} \frac{\mathcal{G}}{\mu - \Xi}.
\end{equation}

\noindent Since the two dispersion relations are identical, we can set $\hat{\mathcal{A}}_{\ell}^{\dagger} = \hat{\mathcal{A}}_{\ell}$ and $\mu_{\ell}^{\dagger} = \mu_{\ell}$. As in FEL theory \cite{kim1986}, we assume the set of eigenvectors $\{\Psi_{\ell}\}$ is complete and the eigenvalues $\mu_{\ell}$ are not degenerate. From non-degeneracy, 
\begin{equation}
\begin{split}
0 &= \langle \Psi_{\ell}^{\dagger}, \bm{\mathcal{M}} \Psi_m \rangle - \langle  \bm{\mathcal{M}}^{\dagger} \Psi_{\ell}^{\dagger}, \Psi_m \rangle \\
&= \langle \Psi_{\ell}^{\dagger}, \mu_m \Psi_m \rangle - \langle  \mu_{\ell}^{\dagger} \Psi_{\ell}^{\dagger}, \Psi_m \rangle \\
&= (\mu_m - \mu_{\ell}) \langle \Psi_{\ell}^{\dagger}, \Psi_m \rangle \\
&\Longrightarrow \langle \Psi_{\ell}^{\dagger}, \Psi_m \rangle = \delta_{\ell,m} \langle \Psi_{\ell}^{\dagger}, \Psi_m \rangle
\end{split}
\end{equation}

\noindent which, along with completeness, allows us to write down the general solution to the initial value problem
\begin{equation} \label{eq:vankampenexpansion}
\Psi(\hat{z}) = \sum_{\ell} \frac{\langle \Psi_{\ell}^{\dagger}, \Psi(0)\rangle}{\langle \Psi_{\ell}^{\dagger}, \Psi_{\ell}\rangle} \Psi_{\ell} e^{-i \mu_{\ell} \hat{z}}.
\end{equation}

\noindent To use this equation, one would solve the first dispersion relation in Eq.~(\ref{eq:as}) for $\{\hat{\mathcal{A}}_{\ell}\}$ and $\{\mu_{\ell}\}$. From these, $\{\mathcal{F}_{\ell}\}$ and $\{\mathcal{F}_{\ell}^{\dagger}\}$ can then be obtained from Eq.~(\ref{eq:fs}), noting that $\hat{\mathcal{A}}_{\ell}^{\dagger} = \hat{\mathcal{A}}_{\ell}$. Next, the sets of vectors $\{\Psi_{\ell}\}$ and $\{\Psi_{\ell}^{\dagger}\}$ can be constructed, and using the inner product Eq.~(\ref{eq:innerproduct}) and $\Psi(0) = \begin{pmatrix} \hat{\mathcal{A}}(0) & \mathcal{F}(0) \end{pmatrix}^{\intercal}$, Eq.~(\ref{eq:vankampenexpansion}) can be used to compute $\Psi(\hat{z}) = \begin{pmatrix} \hat{\mathcal{A}}(\hat{z}) & \mathcal{F}(\hat{z}) \end{pmatrix}^{\intercal}$, which is the solution to the initial value problem of Eq.~(\ref{eq:simplifiedmv}).

\section{\label{app:computingVX} Computing $\mathcal{X}_h(\hat{z}, \hat{z}')$ and $\mathcal{V}(\mu)$}

In this Appendix we compute, for an electron bunch background distribution $\hat{f}_0$ with Gaussian energy detuning $\hat{\eta}$ and undulator parameter $\hat{\delta}$ distributions, the functions $\mathcal{X}_h(\hat{z}, \hat{z}')$ defined in Eq.~(\ref{eq:Xdefinition}) and appearing in the ICL initial value problem Eq.~(\ref{eq:ivp}) and $\mathcal{V}(\mu)$ defined in Eq.~(\ref{eq:vdefinition2}) and appearing in the dispersion relation Eq.~(\ref{eq:dispersion}). After this we take the limit as the spread in $\hat{\eta}$ and $\hat{\delta}$ goes to zero to obtain $\mathcal{X}(\hat{z}, \hat{z}')$ and $\mathcal{V}(\mu)$ for a ``cold'' bunch. We assume that the background beam distribution $\hat{f}_0$ has no dependence on $\hat{z}$ and that there are no correlations between the betatron phase detuning $\vartheta$ and $\hat{\eta}$ or $\hat{\delta}$. With these assumptions the background beam distribution can be written 

\begin{equation} \label{eq:correlationcondition}
\hat{f}_0(\hat{\eta}, \hat{\delta}, \vartheta, \hat{z}) = \hat{f}_{0, \hat{\eta}\hat{\delta}}(\hat{\eta}, \hat{\delta}) \hat{f}_{0, \vartheta}(\vartheta)
\end{equation}

\noindent where we choose the normalization such that $\iint d\hat{\eta} d\hat{\delta} \hat{f}_{0, \hat{\eta}\hat{\delta}}(\hat{\eta}, \hat{\delta}) = 1$ and $\int d\vartheta \hat{f}_{0, \vartheta}(\vartheta) = 1$. Because the $\vartheta$ integration in Eqs.~(\ref{eq:Xdefinition}) and (\ref{eq:vdefinition2}) can be carried out using this normalization condition and Eq.~(\ref{eq:correlationcondition}), the physics is independent of the $\vartheta$ distribution. For a Gaussian distribution in $\hat{\eta}$ and $\hat{\delta}$,

\begin{equation} \label{eq:f0Gaussian}
\hat{f}_{0, \hat{\eta}\hat{\delta}}(\hat{\eta}, \hat{\delta}) = \frac{e^{-\frac{(\hat{\eta} - \hat{\eta}_0)^2}{2 \sigma_{\hat{\eta}}^2}}}{\sqrt{2 \pi} \sigma_{\hat{\eta}}} \times \frac{e^{-\frac{(\hat{\delta} - \hat{\delta}_0)^2}{2 \sigma_{\hat{\delta}}^2}}}{\sqrt{2 \pi} \sigma_{\hat{\delta}}}.
\end{equation}

First we compute $\mathcal{X}_h(\hat{z}, \hat{z}')$ using Eq.~(\ref{eq:Xdefinition}). Plugging in Eqs.~(\ref{eq:correlationcondition}) and (\ref{eq:f0Gaussian}) and making the substitutions $\hat{\eta} \equiv \hat{\eta}_0 + u \sigma_{\hat{\eta}}$ and $\hat{\delta} \equiv \hat{\delta}_0 + v \sigma_{\hat{\delta}}$ we obtain

\begin{figure*}
\centering
\includegraphics[width=\columnwidth]{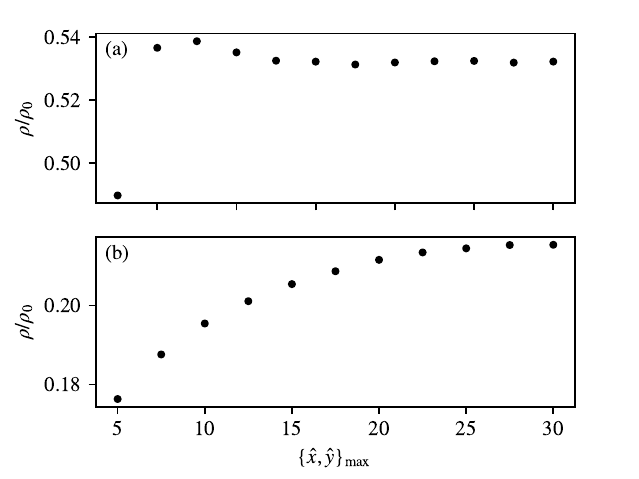}
\includegraphics[width=\columnwidth]{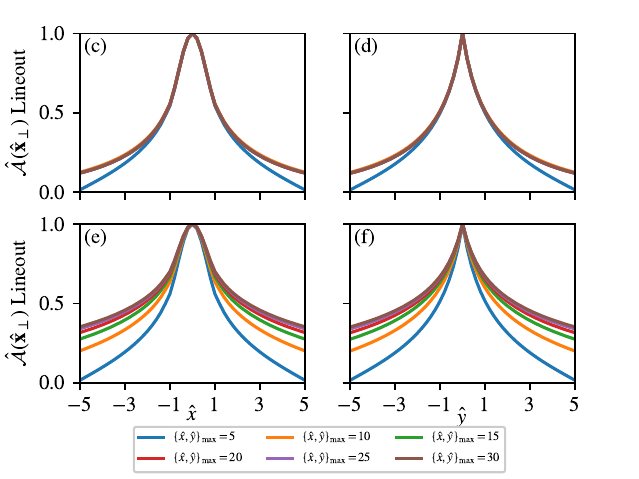}
\caption{Simulation box size convergence scan. Left plots show relative gain parameter $\rho / \rho_0$ as a function of $\{\hat{x}, \hat{y}\}_{\mathrm{max}}$ for $\rho_0 = 0.01$ (a) and $\rho_0 = 0.001$ (b). $\{\hat{x}, \hat{y}\}_{\mathrm{max}} = \hat{x}_{\mathrm{max}} = \hat{y}_{\mathrm{max}} = -\hat{x}_{\mathrm{min}} = -\hat{y}_{\mathrm{min}}$ is half the side length of the square transverse area over which the algorithm solves Eq.~(\ref{eq:ivpb}). Right plots show $x$ (c,e) and $y$ (d,f) lineouts of the transverse radiation profiles for $\rho_0 = 0.01$ (c,d) and $\rho_0 = 0.001$ (e,f). Colors within these plots indicate different values of $\{\hat{x}, \hat{y}\}_{\mathrm{max}}$.}
\label{fig:xmaxscan}
\end{figure*}

\begin{equation}
\begin{split}
&\mathcal{X}_h(\hat{z}, \hat{z}') = -\frac{e^{i (\hat{z} - \hat{z}') \left[-\left(\frac{3}{4}h + \Delta \nu \right) \hat{\eta}_0 + (h + \Delta \nu)\frac{K_r^2}{2 + K_r^2} \hat{\delta}_0\right]}}{2 \mathcal{I}} \frac{[\mathrm{JJ}]_h^2 }{[\mathrm{JJ}]_1^2} \times \\
&\times \iint du dv \left(\frac{u}{\sigma_{\hat{\eta}}} + \frac{1 + K_r^2}{2 K_r^2} \frac{v}{\sigma_{\hat{\delta}}}\right) e^{-\frac{1}{2} (u^2 + v^2)} \times \\
&\times e^{i (\hat{z} - \hat{z}') \left[-\left(\frac{3}{4}h + \Delta \nu\right)\sigma_{\hat{\eta}} u + (h + \Delta \nu) \frac{K_r^2}{2 + K_r^2} \sigma_{\hat{\delta}} v\right]}
\end{split}
\end{equation}

\noindent which can be integrated to give

\begin{equation} \label{eq:xGaussian}
\mathcal{X}_h(\hat{z}, \hat{z}') = i \pi A (\hat{z} - \hat{z}') e^{-i (\hat{z} - \hat{z}') \mu_0} e^{-\frac{1}{2} (\hat{z} - \hat{z}')^2 \Sigma^2}
\end{equation}

\noindent where

\begin{equation} \label{eq:vxeqdefs}
\begin{split}
A &\equiv \left(h + \frac{2 (3 + K_r^2)}{4 + K_r^2} \Delta \nu\right) \frac{[\mathrm{JJ}]_h^2 }{[\mathrm{JJ}]_1^2} \\
\mu_0 &\equiv \left(\frac{3}{4} h + \Delta \nu\right)\hat{\eta}_0 - (h + \Delta \nu) \frac{K_r^2}{2 + K_r^2} \hat{\delta}_0 \\
\Sigma &\equiv \sqrt{\Sigma_{\hat{\eta}}^2 + \Sigma_{\hat{\delta}}^2} \\
\Sigma_{\hat{\eta}} &\equiv \left(\frac{3}{4}h + \Delta \nu\right) \sigma_{\hat{\eta}} \\
\Sigma_{\hat{\delta}} &\equiv(h + \Delta \nu) \frac{K_r^2}{2 + K_r^2}  \sigma_{\hat{\delta}}. \\
\end{split}
\end{equation}

Next we compute $\mathcal{V}(\mu)$ in an analogous way by plugging Eqs.~(\ref{eq:correlationcondition}) and (\ref{eq:f0Gaussian}) into Eq.~(\ref{eq:vdefinition2}) and following the same steps to get

\begin{equation} \label{eq:uvintegral}
\mathcal{V}(\mu) = \frac{A}{2 \pi} \iint du dv \frac{e^{-\frac{u^2 + v^2}{2}}}{\left(\Delta \mu -  u \Sigma_{\hat{\eta}} + v \Sigma_{\hat{\delta}}\right)^2}
\end{equation}

\noindent where $\Delta \mu \equiv \mu - \mu_0$. This can be further simplified by making the transformation $u = \frac{\Sigma}{2 \Sigma_{\hat{\eta}}} (w + x)$, $v = \frac{\Sigma}{2 \Sigma_{\hat{\delta}}} (w - x)$ and performing the $w$ integral to give 

\begin{equation} \label{eq:veq1}
\mathcal{V}(\mu) = \frac{A}{(\Delta \mu)^2} I\left(\frac{\Sigma}{\Delta \mu}\right)
\end{equation}

\noindent where

\begin{figure*}
\centering
\includegraphics[width=\columnwidth]{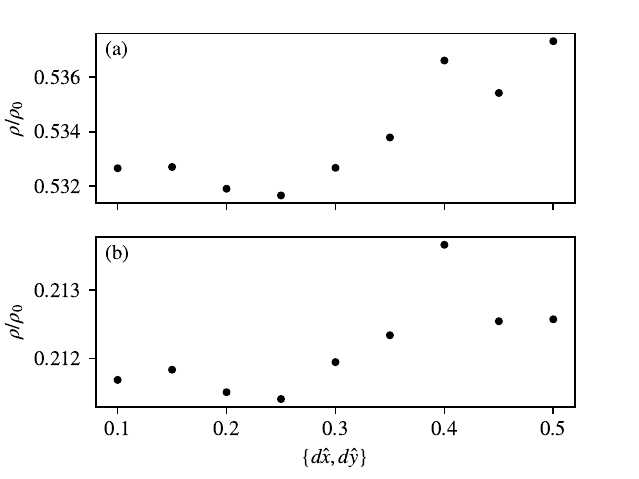}
\includegraphics[width=\columnwidth]{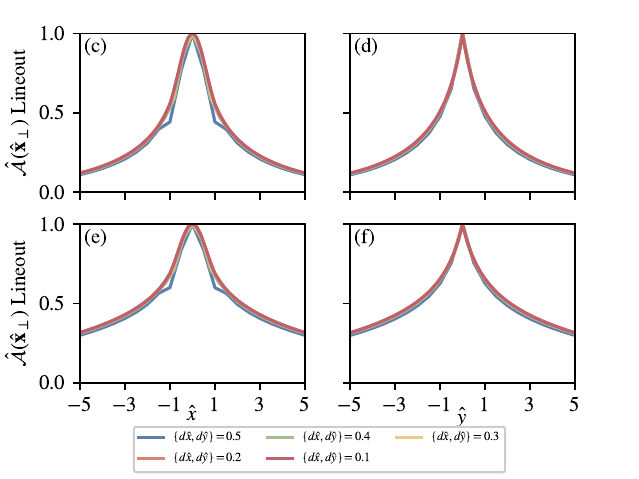}
\caption{Transverse grid size convergence scan. Left plots show relative gain parameter $\rho / \rho_0$ as a function of transverse cell size $\{d\hat{x}, d\hat{y}\}$ for $\rho_0 = 0.01$ (a) and $\rho_0 = 0.001$ (b). Right plots show $x$ (c,e) and $y$ (d,f) lineouts of the transverse radiation profiles for $\rho_0 = 0.01$ (c,d) and $\rho_0 = 0.001$ (e,f). Colors within these plots indicate different values of $\{d\hat{x}, d\hat{y}\}$. }
\label{fig:dxscan}
\end{figure*}

\begin{equation} \label{eq:veq2}
I(y) = \frac{1}{\sqrt{2 \pi}} \int dx \frac{e^{-\frac{1}{2} x^2}}{\left(1 - x y\right)^2}.
\end{equation}

\noindent While it is possible to write the above in terms of complex error functions, we find the given form to be more useful.

Now that we have derived Eqs.~(\ref{eq:xGaussian}) and (\ref{eq:veq1}), which give the $\mathcal{X}(\hat{z}, \hat{z}')$ and $\mathcal{V}(\mu)$ functions, respectively, for a bunch with Gaussian distributions in $\hat{\eta}$ and $\hat{\delta}$, we take the limit $\sigma_{\hat{\eta}}, \sigma_{\hat{\delta}} \rightarrow 0$ to compute the functions for a ``cold'' bunch $\hat{f}_{0, \hat{\eta}\hat{\delta}}(\hat{\eta}, \hat{\delta}) = \delta(\hat{\eta} - \hat{\eta}_0) \delta(\hat{\delta} - \hat{\delta}_0)$:

\begin{equation}
\mathcal{X}(\hat{z}, \hat{z}') = i \pi A (\hat{z} - \hat{z}') e^{-i (\hat{z} - \hat{z}') \mu_0} 
\end{equation}

\begin{equation}
\mathcal{V}(\mu) = A / (\Delta \mu)^2
\end{equation}

\section{\label{app:convergence} Crank-Nicolson Code Convergence}

The Crank-Nicolson code discussed in Section~\ref{sec:numericalalgorithm} depends on a number of nonphysical parameters: the location of the transverse boundary $\hat{x}_{\mathrm{max}} (= \hat{y}_{\mathrm{max}} = -\hat{x}_{\mathrm{min}} = -\hat{y}_{\mathrm{min}})$, the transverse cell size $d\hat{x} (= d\hat{y})$, the parameter $\mu (=\mu_x = \mu_y)$ which the code uses to calculate the $\hat{z}$ step size, and the standard deviations $\sigma_{\hat{x},\mathrm{initial}}$ and $\sigma_{\hat{y}, \mathrm{initial}}$ of the initial Gaussian radiation field. Here we demonstrate the convergence (or, in the case of $\sigma_{\hat{x},\mathrm{initial}}$ and $\sigma_{\hat{y}, \mathrm{initial}}$,  insensitivity) of the simulation results with respect to these nonphysical parameters. We use the physical parameters $h = 1$, $K_r \rightarrow \infty$, $\hat{\Delta \nu} = 0$, $\Sigma = 0$, and $\rho_0 = 0.01$ from the ``base'' simulation from Section~\ref{sec:numericalresults}. Because we find that some quantities converge more slowly for smaller $\rho$, we also include convergence scans for a second simulation with these same parameters except $\rho_0 = 0.001$. For the nonphysical parameters, we start with $\hat{x}_{\mathrm{max}} = 20$, $d\hat{x} = 0.2$, $\mu = 0.5$, and $\sigma_{\hat{x}, \mathrm{initial}} = \sigma_{\hat{y}, \mathrm{initial}} = 1$ and vary each individually.

For the first convergence scan we vary the value of $\hat{x}_{\mathrm{max}}$. The results, shown in Fig.~\ref{fig:xmaxscan}, demonstrate clear convergence as the simulation box size increases. For $\rho_0 = 0.01$ ($\rho_0 = 0.001$), the value $\hat{x}_{\mathrm{max}} = 20$ used in Section~\ref{sec:numericalresults} and in other convergence scans gives a value of $\rho$ with a relative error of $0.053\%$ ($1.8\%$) compared to the $\hat{x}_{\mathrm{max}} = 30$ simulation. We note faster convergence of the lineout for $\rho_0 = 0.01$ than for $\rho_0 = 0.001$. We suspect this is due to the wider beam in the latter case, although even in that case the lineout for $\hat{x}_{\mathrm{max}} = 20$ is only marginally different from $\hat{x}_{\mathrm{max}} = 30$.

\begin{figure*}
\centering
\includegraphics[width=\columnwidth]{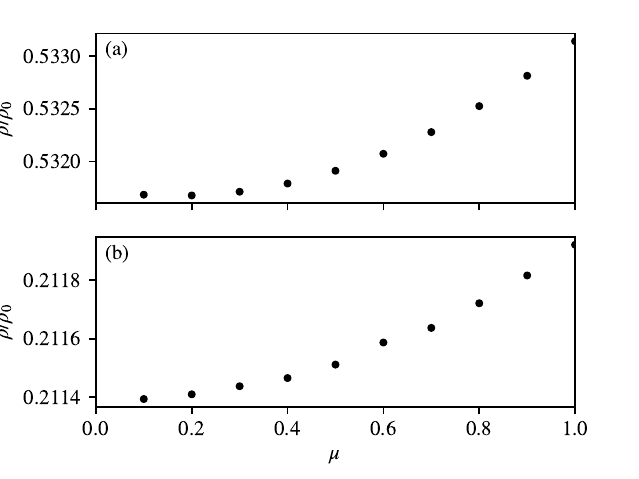}
\includegraphics[width=\columnwidth]{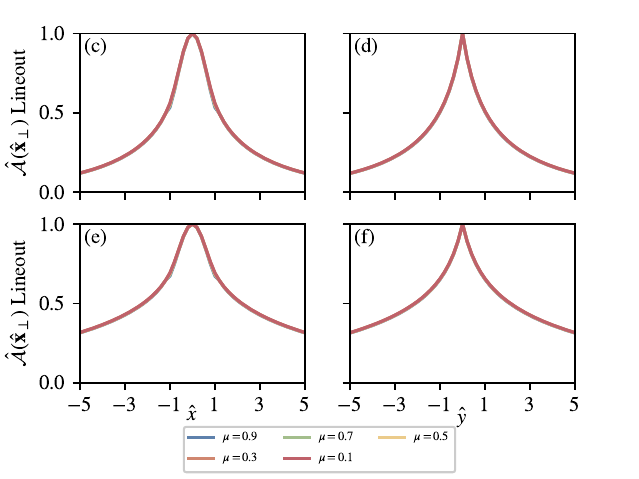}
\caption{Normalized $\hat{z}$ step size convergence scan. Left plots show relative gain parameter $\rho / \rho_0$ as a function of $\mu = d\hat{z} / (2 \mathcal{F}_D d\hat{x}^2) = d\hat{z} / (2 \mathcal{F}_D d\hat{y}^2)$ for $\rho_0 = 0.01$ (a) and $\rho_0 = 0.001$ (b). Right plots show $x$ (c,e) and $y$ (d,f) lineouts of the transverse radiation profiles for $\rho_0 = 0.01$ (c,d) and $\rho_0 = 0.001$ (e,f). Colors within these plots indicate different values of $\mu$.}
\label{fig:muscan}
\end{figure*}

The next scan is over $d\hat{x}$ and is shown in Fig.~\ref{fig:dxscan}. These results show relatively straightforward convergence with the value of $\rho$ having only a $0.14\%$ ($0.085\%$) error at the value $d\hat{x} = 0.2$ used in Section~\ref{sec:numericalresults} and other convergence scans compared to the more accurate $d\hat{x} = 0.1$. 

The next convergence scan is over $\mu$ and the results are shown in Fig.~\ref{fig:muscan}. Again this shows straightforward convergence with the value $\mu = 0.5$ used in Section~\ref{sec:numericalresults} and other convergence scans yielding a value of $\rho$ with a relative error of only $0.043\%$ ($0.056\%$) when compared to the more accurate simulation with $\mu = 0.1$. There is almost no noticeable difference in the lineout shapes for simulations with different values of $\mu$.

\begin{figure*}
\centering
\includegraphics[width=\columnwidth]{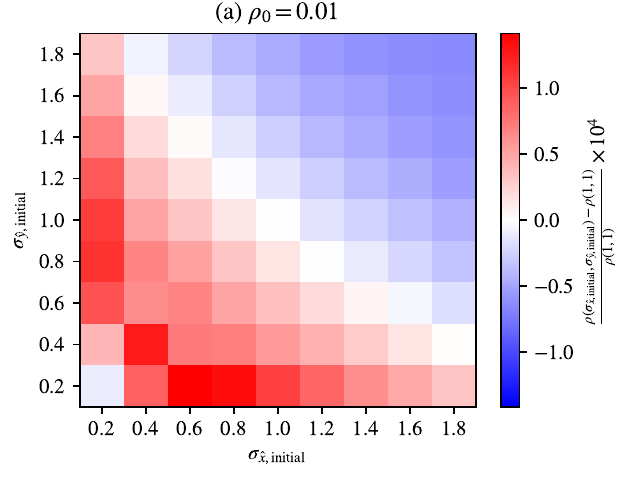}
\includegraphics[width=\columnwidth]{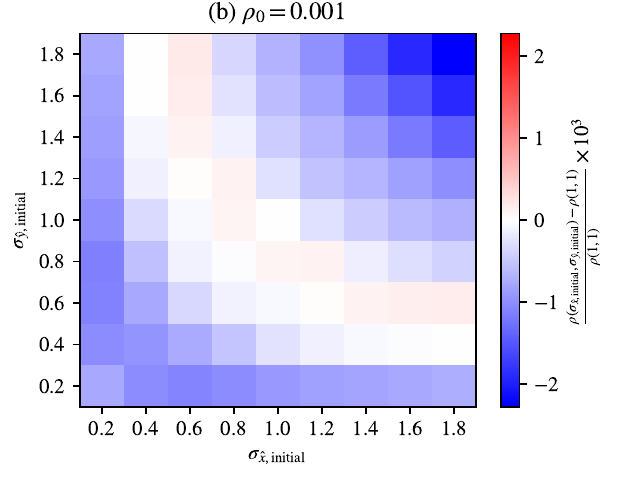}
\caption{Sensitivity of $\rho$ to the $x$ and $y$ sizes of the Gaussian seed radiation field. Plots show the relative error of $\rho(\sigma_{\hat{x}, \mathrm{initial}}, \sigma_{\hat{y}, \mathrm{initial}})$ relative to $\rho(\sigma_{\hat{x}, \mathrm{initial}} = 1, \sigma_{\hat{y}, \mathrm{initial}}=1)$ used in the ``base'' simulation for $\rho_0 = 0.01$ (a) and $\rho_0 = 0.001$ (b). Note the $10^4$ (a) and $10^3$ (b) scaling factor in the colorbar.}
\label{fig:sigmainitial}
\end{figure*}

While not technically a convergence scan, we ran a number of simulations with different values of $\sigma_{\hat{x},\mathrm{initial}}$ and $\sigma_{\hat{y},\mathrm{initial}}$ with the resulting values of $\rho$ shown in Fig.~\ref{fig:sigmainitial}. These results show that, as expected, the ICL gain parameter is only barely sensitive to the size of the seed radiation field with differences of at most a few tenths of a percent between simulations. 

To summarize, we have demonstrated the convergence of two representative ICL simulations with respect to the nonphysical parameters $\hat{x}_{\mathrm{max}}$, $d\hat{x}$, and $\mu$, and the insensitivity of these simulations with respect to the nonphysical parameters $\sigma_{\hat{x},\mathrm{initial}}$ and $\sigma_{\hat{y},\mathrm{initial}}$. This is a good indication that this algorithm behaves as we expect and the nonphysical parameters used in Section~\ref{sec:numericalresults} are adequate to accurately resolve the relevant physics.

\section{\label{app:optimization} Bunch Distribution Optimization}

The normalized phase space distribution functions for the matched offset Gaussian, optimally mismatched offset Gaussian, and phase space annular sector bunches are 

\begin{equation} \label{eq:distributionfunctions}
\begin{split}
&f_{\mathrm{MOG}}(\tilde{\bm{x}}_{\perp}, \tilde{\bm{p}}_{\perp}) = \frac{e^{-\frac{(\tilde{x} - \tilde{\mu})^2 + \tilde{y}^2 +  \tilde{p}_x^2 + \tilde{p}_y^2}{2 \tilde{\sigma}^2}}}{4 \pi^2 \tilde{\sigma}^4} \\
&f_{\mathrm{OMOG}}(\tilde{\bm{x}}_{\perp}, \tilde{\bm{p}}_{\perp}) = \frac{e^{-\frac{(\tilde{x} - \tilde{\mu})^2}{2 \tilde{\sigma}_x^2} -\frac{\tilde{y}^2}{2 \tilde{\sigma}_y^2}-\frac{\tilde{p}_x^2}{2 \tilde{\sigma}_{p_x}^2}-\frac{\tilde{p}_y^2}{2 \tilde{\sigma}_{p_y}^2}}}{4 \pi^2 \tilde{\sigma}_x \tilde{\sigma}_y \tilde{\sigma}_{p_x} \tilde{\sigma}_{p_y}} \\
&f_{\mathrm{PSAS}}(\tilde{\bm{x}}_{\perp}, \tilde{\bm{p}}_{\perp}) = \frac{1}{\Theta \tilde{\mu} \tilde{w}} \mathrm{rect}\left(\frac{\sqrt{\tilde{x}^2 + \tilde{p}_x^2} - \tilde{\mu}}{\tilde{w}}\right) \times \\
&\times \mathrm{rect}\left(\frac{\mathrm{atan2}(\tilde{p}_x, \tilde{x})}{\Theta}\right) \frac{e^{-\frac{\tilde{y}^2}{2 \tilde{\sigma}_y^2}-\frac{\tilde{p}_y^2}{2 \tilde{\sigma}_{p_y}^2}}}{2 \pi \tilde{\sigma}_y \tilde{\sigma}_{p_y}} 
\end{split}
\end{equation}

\noindent where a tilde indicates a normalized quantity where positions are normalized to $a_{\beta,r}$ and momenta to $K_r$. Note that we are assuming these bunches are monoenergetic. Computing moments of these distribution functions allows us to obtain the mean radiation wavelength for each case

\begin{equation} \label{eq:lambda1b}
\begin{split}
&\lambda_{1,b} = \langle \lambda_1 \rangle = \left \langle \frac{\lambda_{\beta,r}}{2 \gamma_r^2} \left(1 + K_r^2 \frac{\tilde{x}^2 + \tilde{y}^2 + \tilde{p}_x^2 + \tilde{p}_y^2}{2}\right) \right \rangle \\
&= \begin{cases}
\frac{\lambda_{\beta,r}}{2 \gamma_r^2}\left(1 + \frac{K_r^2 (\tilde{\mu}^2 + 4 \tilde{\sigma}^2)}{2}\right) & \mathrm{MOG}\\
\frac{\lambda_{\beta,r}}{2 \gamma_r^2}\left(1 + \frac{K_r^2 (\tilde{\mu}^2 + \tilde{\sigma}_x^2 + \tilde{\sigma}_y^2 + \tilde{\sigma}_{p_x}^2 + \tilde{\sigma}_{p_y}^2)}{2}\right) & \mathrm{OMOG}\\\frac{\lambda_{\beta,r}}{2 \gamma_r^2}\left(1 + \frac{K_r^2 \left(\tilde{\mu}^2 +  \frac{\tilde{w}^2}{4} + \tilde{\sigma}_y^2 + \tilde{\sigma}_{p_y}^2\right)}{2}\right) & \mathrm{PSAS}
\end{cases}
\end{split}
\end{equation}

\noindent as well as the wavelength spread

\begin{equation} \label{eq:sigmalambda1b}
\begin{split}
&\sigma_{\lambda_1,b} = \sqrt{\langle \lambda_1^2 \rangle - \langle \lambda_1 \rangle^2} \\
&= \begin{cases}
\frac{\lambda_{\beta,r} K_r^2}{2 \gamma_r^2} \sqrt{\tilde{\sigma}^2 (\tilde{\mu}^2 + 2 \tilde{\sigma}^2)} & \mathrm{MOG} \\
\frac{\lambda_{\beta,r} K_r^2}{2 \gamma_r^2} \sqrt{\tilde{\mu}^2 \tilde{\sigma}_x^2 + \frac{\tilde{\sigma}_x^4 + \tilde{\sigma}_y^4 + \tilde{\sigma}_{p_x}^4 + \tilde{\sigma}_{p_y}^4}{2}} & \mathrm{OMOG} \\
\frac{\lambda_{\beta,r} K_r^2}{2 \gamma_r^2} \sqrt{\frac{\tilde{\mu}^2 \tilde{w}^2}{12} + \frac{\tilde{\sigma}_y^4 + \tilde{\sigma}_{p_y}^4}{2}} & \mathrm{PSAS} 
\end{cases}.
\end{split}
\end{equation}

\noindent Next we compute the emittance for each of the three cases. Note that 2D emittances with a tilde are normalized to $a_{\beta,r} K_r$ and 4D emittances with a tilde are normalized to $a_{\beta,r}^2 K_r^2$. For the $x$ and $y$ dimensions of the two Gaussian cases as well as the $y$ dimension of the phase space annular sector case, there is no difference between the RMS emittance $\tilde{\epsilon}_{n,x, \mathrm{RMS}} \equiv \sqrt{\mathrm{det}(\mathrm{cov}(\tilde{x}, \tilde{p}_x))}$ and the geometric emittance $\tilde{\epsilon}_{n,x, \mathrm{geometric}} \equiv \tilde{A}/\pi$ where $\tilde{A}$ is the area of the distribution in normalized phase space (for Gaussian bunch distributions $\tilde{A}$ is the area of an ellipse with semiaxes equal to $\tilde{\sigma}_x$ and $\tilde{\sigma}_{p_x}$). For the $x$ dimension of the phase space annular sector case, however, the RMS emittance

\begin{equation} \label{eq:annularsectorrms}
\begin{split}
&\tilde{\epsilon}_{n,x, \mathrm{RMS}} = \frac{\tilde{\mu}^2}{24 \Theta} \Bigg[\left(4 + \frac{\tilde{w}^2}{\tilde{\mu}^2}\right) \left(1 - \mathrm{sinc}(\Theta)\right) \times \\
&\times \Bigg( 9 \Theta^2 \left(4 + \frac{\tilde{w}^2}{\tilde{\mu}^2}\right) \left(1 + \mathrm{sinc}(\Theta)\right) + \\
&+ \left(12 + \frac{\tilde{w}^2}{\tilde{\mu}^2}\right)^2 (\cos(\Theta) - 1)\Bigg)  \Bigg]^{\frac{1}{2}}
\end{split}
\end{equation}

\noindent is different from the geometric emittance $\tilde{\epsilon}_{n, x, \mathrm{geometric}} = \Theta \tilde{\mu} \tilde{w} / \pi$. We use the geometric instead of the RMS emittance here because the geometric emittance goes to zero when $\tilde{w} \rightarrow 0$ while the RMS emittance does not. The $x$ and $y$ emittances in each case are

\begin{equation}
\tilde{\epsilon}_{n,x}, \tilde{\epsilon}_{n,y} = \left \{ \begin{array}{lll}
 \tilde{\sigma}^2, &  \tilde{\sigma}^2  & \mathrm{MOG} \\
 \tilde{\sigma}_x \tilde{\sigma}_{p_x},& \tilde{\sigma}_y \tilde{\sigma}_{p_y}  & \mathrm{OMOG} \\
\Theta \tilde{\mu} \tilde{w} / \pi, &  \tilde{\sigma}_y \tilde{\sigma}_{p_y}  & \mathrm{PSAS} \\
\end{array} \right. .
\end{equation}

\noindent In each case, the distribution functions must obey the resonance condition $\lambda_{1,b} = \lambda_{1,r} = \frac{\lambda_{\beta,r}}{2 \gamma_r^2} \left(1 + \frac{K_r^2}{2}\right)$. Using Eq.~(\ref{eq:lambda1b}) this allows us to solve for the offset $\tilde{\mu}$:

\begin{equation} \label{eq:muparam}
\tilde{\mu} = \begin{cases}
\sqrt{1 - 4\tilde{\sigma}^2}
 & \mathrm{MOG} \\
\sqrt{1 - \tilde{\sigma}_x^2 - \tilde{\sigma}_y^2 - \tilde{\sigma}_{p_x}^2 - \tilde{\sigma}_{p_y}^2} & \mathrm{OMOG} \\
\sqrt{1 - \tilde{\sigma}_y^2 - \tilde{\sigma}_{p_y}^2 - \frac{\tilde{w}^2}{4}} & \mathrm{PSAS} \\
\end{cases}.
\end{equation}

Our goal is to, for each of the three cases and for a given wavelength spread $\Delta \lambda_1$, find the parameters that maximize the emittance. Because there are two emittances ($x$ and $y$), we choose the 4D emittance as the quantity to maximize

\begin{equation} \label{eq:4Demittances}
\begin{split}
&\tilde{\epsilon}_{n, \mathrm{4D}} = \tilde{\epsilon}_{n,x} \tilde{\epsilon}_{n,y} \\
&=\begin{cases}
\tilde{\sigma}^4 & \mathrm{MOG} \\
\tilde{\sigma}_x \tilde{\sigma}_y \tilde{\sigma}_{p_x}\tilde{\sigma}_{p_y}& \mathrm{OMOG} \\
\frac{\Theta \tilde{w} \tilde{\sigma}_y \tilde{\sigma}_{p_y}}{\pi} \sqrt{1 - \tilde{\sigma}_y^2 - \tilde{\sigma}_{p_y}^2 - \frac{\tilde{w}^2}{4}}
 & \mathrm{PSAS} \\
\end{cases}.
\end{split}
\end{equation}

\noindent This maximization is subject to the constraint that the wavelength spread of the bunch $\sigma_{\lambda_1,b}$ is equal to the given wavelength spread $\Delta \lambda_1$. To simplify the math we use the equivalent constraint $\sigma_{\lambda_1,b}^2 = (\Delta \lambda_1)^2$ during the calculation. With the help of Eqs.~(\ref{eq:sigmalambda1b}) and (\ref{eq:muparam}) this constraint is

\begin{equation} \label{eq:deltaconstraints}
\Delta^2 = \begin{cases}
\tilde{\sigma}^2\left(1 - 2\tilde{\sigma}^2\right) & \mathrm{MOG} \\
\begin{split}
&\frac{1}{2} \left(\tilde{\sigma}_{p_x}^4 + \tilde{\sigma}_{p_y}^4 + \tilde{\sigma}_y^4 - \tilde{\sigma}_x^4\right) + \\
&+\tilde{\sigma}_{x}^2\left(1 - \tilde{\sigma}_y^2 - \tilde{\sigma}_{p_x}^2 - \tilde{\sigma}_{p_y}^2\right)
\end{split} & \mathrm{OMOG} \\
\begin{split}&\frac{1}{2} (\tilde{\sigma}_y^4 + \tilde{\sigma}_{p_y}^4) + \\
&+\frac{\tilde{w}^2}{12} (1 - \tilde{\sigma}_y^2 - \tilde{\sigma}_{p_y}^2) -\frac{\tilde{w}^4}{48} \\
\end{split} & \mathrm{PSAS} \\
\end{cases}
\end{equation}

\noindent where

\begin{equation} 
\Delta \equiv \frac{1 + \frac{K_r^2}{2}}{K_r^2} \frac{\Delta \lambda_1}{\lambda_{1,r}}.
\end{equation}

We now consider each of the three cases individually. First, for the matched offset Gaussian distribution, no optimization needs to be performed as Eq.~(\ref{eq:deltaconstraints}) fixes the only free parameter $\tilde{\sigma}$. For this case, the parameters in terms of wavelength spread are

\begin{equation} \label{eq:matchedsolution}
\begin{split}
\tilde{\mu} &= \left(1 - 8 \Delta^2\right)^{\frac{1}{4}} \\
\tilde{\sigma} &= \frac{1}{2} \sqrt{1 - \sqrt{1 - 8 \Delta^2}}
\end{split}
\end{equation}

\noindent and the emittances are

\begin{equation} \label{eq:matchedsolution2}
\tilde{\epsilon}_{n,x} = \tilde{\epsilon}_{n,y} =  \frac{1}{4} \left(1 - \sqrt{1 - 8 \Delta^2}\right).
\end{equation}

\noindent To obtain the actual emittance constraint we write $\tilde{\epsilon}_{n,x}$ and $\tilde{\epsilon}_{n,y}$ in Eq.~(\ref{eq:matchedsolution2}) to leading order ($\Delta \sim \rho \ll 1$ unless $K_r$ is small, but as discussed in Section~\ref{sec:maxwellklimontovich} the small $K_r$ ICL is infeasible due to diffraction) and then use the wavelength spread constraint $\Delta \lambda_1 / \lambda_{1,r} \lesssim \rho$ to write, in non-normalized units,

\begin{equation} \label{eq:matchedemittanceconstraint}
\epsilon_{n,x},\epsilon_{n,y} \lesssim \frac{\gamma \lambda_1}{\pi} \frac{1 + \frac{K^2}{2}}{K^2} \rho^2.
\end{equation}

Next we consider the more complicated optimally mismatched case. We solved this optimization problem using Lagrange multipliers in a computer algebra system. While there is an exact solution that can be written in terms of radicals, due to its excessive complexity, we write the solution as a power series in $\Delta$. For an optimally mismatched bunch, the bunch parameters in terms of $\Delta$ are 

\begin{equation} \label{eq:mismatchedparameters}
\begin{split}
&\tilde{\mu} \simeq 1 - \frac{3}{\sqrt{10}} \Delta - \frac{77}{100} \Delta^2 - \frac{3}{4} \sqrt{\frac{5}{2}} \Delta^3 + O\left(\Delta^4\right) \\
&\tilde{\sigma}_x  \simeq \sqrt{\frac{2}{5}} \Delta + \frac{12}{25} \Delta^2 + \frac{28}{25} \sqrt{\frac{2}{5}} \Delta^3 + O\left(\Delta^4\right) \\
&\tilde{\sigma}_y = \tilde{\sigma}_{p_x} = \tilde{\sigma}_{p_y} \simeq \left(\frac{2}{5}\right)^{\frac{1}{4}} \Delta^{\frac{1}{2}} + \frac{1}{5 \times 2^{\frac{1}{4}} \times 5^{\frac{3}{4}}} \Delta^{\frac{3}{2}} + \\
&+ \frac{39}{250 \times 2^{\frac{3}{4}} \times 5^{\frac{1}{4}}} \Delta^{\frac{5}{2}} + O\left(\Delta^{\frac{7}{2}}\right)
\end{split}
\end{equation}

\noindent and the emittances are

\begin{equation} \label{eq:mismatchedparameters2}
\begin{split}
&\tilde{\epsilon}_{n,x} \simeq \left(\frac{2}{5}\right)^{\frac{3}{4}} \Delta^{\frac{3}{2}} + \frac{13}{25} \left(\frac{2}{5}\right)^{\frac{1}{4}}\Delta^{\frac{5}{2}} + \\
&+ \frac{623}{250 \times 2^{\frac{1}{4}} \times 5^{\frac{3}{4}}} \Delta^{\frac{7}{2}} + \mathrm{O}\left(\Delta^{\frac{9}{2}}\right)\\
&\tilde{\epsilon}_{n,y} \simeq \sqrt{\frac{2}{5}} \Delta + \frac{2}{25}  \Delta ^2 + \frac{4}{25} \sqrt{\frac{2}{5}} \Delta^3 + \mathrm{O}\left(\Delta^4\right).
\end{split}
\end{equation}

\noindent In the same way as before, we compute the emittance constraint, which is, in non-normalized units,

\begin{equation} \label{eq:mismatchedemittanceconstraint}
\begin{split}
\epsilon_{n,x} &\lesssim \frac{\gamma \lambda_1}{\pi} \left(\frac{2}{5}\right)^{\frac{3}{4}} \left(\frac{1 + \frac{K^2}{2}}{K^2}\right)^{\frac{1}{2}} \rho^{\frac{3}{2}} \\
\epsilon_{n,y} &\lesssim \frac{\gamma \lambda_1}{\pi} \sqrt{\frac{2}{5}} \rho.
\end{split}
\end{equation}

Finally we consider the phase space annular sector case. In this case we treat $\Theta$ as fixed rather than one of the parameters to optimize. Again we solved this optimization problem using Lagrange multipliers, although this time the exact solution is far simpler and so we don't compute a power series. The optimal parameters are

\begin{equation}
\begin{split}
\tilde{w} &= \sqrt{2\left(1 - \sqrt{2} \Delta - \sqrt{1 - 2 \sqrt{2} \Delta - 4 \Delta^2}\right)} \\
\tilde{\mu} &= \sqrt{\frac{1 - \sqrt{2} \Delta + \sqrt{1 - 2 \sqrt{2} \Delta - 4 \Delta^2}}{2}} \\
\tilde{\sigma}_y &= \tilde{\sigma}_{p_y} = \frac{\sqrt{\Delta}}{2^{\frac{1}{4}}}
\end{split}
\end{equation}

\noindent and the corresponding emittances are 

\begin{equation}
\begin{split}
\tilde{\epsilon}_{n, x} &= \frac{\sqrt{6} \Theta \Delta}{\pi} \\
\tilde{\epsilon}_{n, y} &= \frac{\Delta}{\sqrt{2}}.
\end{split}
\end{equation}

\noindent Finally, the emittance constraint in non-normalized units is

\begin{equation}
\begin{split}
\epsilon_{n,x} &\lesssim \frac{\gamma \lambda_1}{\pi} \frac{\sqrt{6} \Theta}{\pi} \rho \\
\epsilon_{n,y} &\lesssim \frac{\gamma \lambda_1}{\pi} \frac{1}{\sqrt{2}}\rho.
\end{split}
\end{equation}

\bibliography{refs}

\end{document}